\pdfoutput=1
\documentclass[12pt,a4paper]{article}
\usepackage{ifthen} % for conditional statements
\newboolean{pdflatex}
\setboolean{pdflatex}{true} % False for eps figures 

\newboolean{articletitles}
\setboolean{articletitles}{true} % False removes titles in references

\newboolean{uprightparticles}
\setboolean{uprightparticles}{false} %True for upright particle symbols

\def\paperauthors{LHCb RICH group} % Leave as is for PAPER and CONF
\def\paperasciititle{Performance of the LHCb RICH detectors during LHC Run 2} % Set ASCII title here
\def\papertitle{Performance of the LHCb RICH detectors during LHC~Run~2} % Latex formatted title
\def\paperkeywords{{High Energy Physics}, {LHCb}} % Comma separated list
\def\papercopyright{\the\year\ CERN for the benefit of the LHCb collaboration} % new since 9/Apr/2018
\def\paperlicence{CC-BY-4.0 licence}
\def\paperlicenceurl{https://creativecommons.org/licenses/by/4.0/}

\usepackage[top=1in, bottom=1.25in, left=1in, right=1in]{geometry}

\usepackage{microtype}
\usepackage{lineno}  % for line numbering during review
\usepackage{xspace} % To avoid problems with missing or double spaces after
\usepackage{caption} %these three command get the figure and table captions automatically small

\usepackage{graphicx}  % to include figures (can also use other packages)
\usepackage{color}
\usepackage{colortbl}
\graphicspath{{./figs/}} % Make Latex search fig subdir for figures
\DeclareGraphicsExtensions{.pdf,.PDF,png,.PNG}

\usepackage{amsmath} % Adds a large collection of math symbols
\usepackage{amssymb}
\usepackage{amsfonts}
\usepackage{upgreek} % Adds in support for greek letters in roman typeset

\newcommand*\patchAmsMathEnvironmentForLineno[1]{%
\expandafter\let\csname old#1\expandafter\endcsname\csname #1\endcsname
\expandafter\let\csname oldend#1\expandafter\endcsname\csname
end#1\endcsname
 \renewenvironment{#1}%
   {\linenomath\csname old#1\endcsname}%
   {\csname oldend#1\endcsname\endlinenomath}%
}
\newcommand*\patchBothAmsMathEnvironmentsForLineno[1]{%
  \patchAmsMathEnvironmentForLineno{#1}%
  \patchAmsMathEnvironmentForLineno{#1*}%
}
\AtBeginDocument{%
\patchBothAmsMathEnvironmentsForLineno{equation}%
\patchBothAmsMathEnvironmentsForLineno{align}%
\patchBothAmsMathEnvironmentsForLineno{flalign}%
\patchBothAmsMathEnvironmentsForLineno{alignat}%
\patchBothAmsMathEnvironmentsForLineno{gather}%
\patchBothAmsMathEnvironmentsForLineno{multline}%
\patchBothAmsMathEnvironmentsForLineno{eqnarray}%
}

\usepackage{hyperxmp}

\usepackage[pdftex,
            pdfauthor={\paperauthors},
            pdftitle={\paperasciititle},
            pdfkeywords={\paperkeywords},
            pdfcopyright={Copyright (C) \papercopyright},
            pdflicenseurl={\paperlicenceurl}]{hyperref}

\usepackage[colorinlistoftodos,textsize=scriptsize]{todonotes}

\usepackage[all]{hypcap} % Internal hyperlinks to floats.

\usepackage{xspace} 
\usepackage{upgreek}

\def\lhcb   {\mbox{LHCb}\xspace}

\def\richone {RICH1\xspace}
\def\richtwo {RICH2\xspace}

\def\MagUp {\mbox{\em Mag\kern -0.05em Up}\xspace}

\ifthenelse{\boolean{uprightparticles}}%
{

 \def\Ppi         {\ensuremath{\uppi}\xspace}

 \def\PDelta      {\ensuremath{\Delta}\xspace}                 
 \def\PXi         {\ensuremath{\Xi}\xspace}                 
 \def\PLambda     {\ensuremath{\Lambda}\xspace}                 
 \def\PSigma      {\ensuremath{\Sigma}\xspace}                 
 \def\POmega      {\ensuremath{\Omega}\xspace}                 
 \def\PUpsilon    {\ensuremath{\Upsilon}\xspace}

 \def\PB      {\ensuremath{\mathrm{B}}\xspace}                 
                  
 \def\PD      {\ensuremath{\mathrm{D}}\xspace}

 \def\PK      {\ensuremath{\mathrm{K}}\xspace}

 \def\Pb      {\ensuremath{\mathrm{b}}\xspace}                 
 \def\Pc      {\ensuremath{\mathrm{c}}\xspace}

 \def\Pi      {\ensuremath{\mathrm{i}}\xspace}

 \def\Pp      {\ensuremath{\mathrm{p}}\xspace}

 \def\thebaroffset{0.0em}
}
{

 \def\Ppi         {\ensuremath{\pi}\xspace}

 \mathchardef\PDelta="7101
 \mathchardef\PXi="7104
 \mathchardef\PLambda="7103
 \mathchardef\PSigma="7106
 \mathchardef\POmega="710A
 \mathchardef\PUpsilon="7107
                  
 \def\PB      {\ensuremath{B}\xspace}                 
                  
 \def\PD      {\ensuremath{D}\xspace}

 \def\PK      {\ensuremath{K}\xspace}

 \def\Pb      {\ensuremath{b}\xspace}                 
 \def\Pc      {\ensuremath{c}\xspace}

 \def\Pi      {\ensuremath{i}\xspace}

 \def\Pp      {\ensuremath{p}\xspace}

 \def\thebaroffset{0.18em}
}
\newcommand{\offsetoverline}[2][\thebaroffset]{\kern #1\overline{\kern -#1 #2}}%

\makeatletter
\ifcase \@ptsize \relax% 10pt
  \newcommand{\miniscule}{\@setfontsize\miniscule{4}{5}}% \tiny: 5/6
\or% 11pt
  \newcommand{\miniscule}{\@setfontsize\miniscule{5}{6}}% \tiny: 6/7
\or% 12pt
  \newcommand{\miniscule}{\@setfontsize\miniscule{5}{6}}% \tiny: 6/7
\fi
\makeatother

\DeclareRobustCommand{\optbar}[1]{\shortstack{{\miniscule (\rule[.5ex]{1.25em}{.18mm})}
  \\ [-.7ex] $#1$}}

\def\cquark    {{\ensuremath{\Pc}}\xspace}

\def\bquark    {{\ensuremath{\Pb}}\xspace}

\def\pion   {{\ensuremath{\Ppi}}\xspace}

\def\pipm   {{\ensuremath{\pion^\pm}}\xspace}

\def\kaon    {{\ensuremath{\PK}}\xspace}
\def\KorKbar {\kern \thebaroffset\optbar{\kern -\thebaroffset \PK}{}\xspace}

\def\Kpm     {{\ensuremath{\kaon^\pm}}\xspace}

\def\DorDbar {\kern \thebaroffset\optbar{\kern -\thebaroffset \PD}\xspace}

\def\BorBbar {\kern \thebaroffset\optbar{\kern -\thebaroffset \PB}\xspace}

\def\Y#1S{\ensuremath{\PUpsilon{(#1S)}}\xspace}

\def\proton      {{\ensuremath{\Pp}}\xspace}
\def\antiproton  {{\ensuremath{\overline \proton}}\xspace}

\def\LorLbar     {\kern \thebaroffset\optbar{\kern -\thebaroffset \PLambda}\xspace}

\def\to                 {\ensuremath{\rightarrow}\xspace}

\def\CP                {{\ensuremath{C\!P}}\xspace}

\def\AT#1     {\ensuremath{A_{\mathrm{T}}^{#1}}\xspace}           % 2

\def\C#1      {\ensuremath{\mathcal{C}_{#1}}\xspace}                       % 9
\def\Cp#1     {\ensuremath{\mathcal{C}_{#1}^{'}}\xspace}                    % 7
\def\Ceff#1   {\ensuremath{\mathcal{C}_{#1}^{\mathrm{(eff)}}}\xspace}        % 9  
\def\Cpeff#1  {\ensuremath{\mathcal{C}_{#1}^{'\mathrm{(eff)}}}\xspace}       % 7
\def\Ope#1    {\ensuremath{\mathcal{O}_{#1}}\xspace}                       % 2
\def\Opep#1   {\ensuremath{\mathcal{O}_{#1}^{'}}\xspace}                    % 7

\newcommand{\aunit}[1]{\ensuremath{\text{\,#1}}}       
\newcommand{\tev}{\aunit{Te\kern -0.1em V}\xspace}
\newcommand{\gev}{\aunit{Ge\kern -0.1em V}\xspace}
\newcommand{\mev}{\aunit{Me\kern -0.1em V}\xspace}
\newcommand{\kev}{\aunit{ke\kern -0.1em V}\xspace}
\newcommand{\ev}{\aunit{e\kern -0.1em V}\xspace}
\newcommand{\mevc}{\ensuremath{\aunit{Me\kern -0.1em V\!/}c}\xspace}
\newcommand{\gevc}{\ensuremath{\aunit{Ge\kern -0.1em V\!/}c}\xspace}
\newcommand{\mevcc}{\ensuremath{\aunit{Me\kern -0.1em V\!/}c^2}\xspace}
\newcommand{\gevcc}{\ensuremath{\aunit{Ge\kern -0.1em V\!/}c^2}\xspace}
\def\ns   {\ensuremath{\aunit{ns}}\xspace}

\def\gsim{{~\raise.15em\hbox{$>$}\kern-.85em
          \lower.35em\hbox{$\sim$}~}\xspace}
\def\lsim{{~\raise.15em\hbox{$<$}\kern-.85em
          \lower.35em\hbox{$\sim$}~}\xspace}

\def\sqs   {\ensuremath{\protect\sqrt{s}}\xspace}

\def\mrad{\aunit{mrad}}

\xspace

\def\tell1  {TELL1\xspace}
\def\ukl1   {UKL1\xspace}

\newcommand{\eg}{\mbox{\itshape e.g.}\xspace}

\usepackage{cite} % Allows for ranges in citations
\usepackage{mciteplus}

\usepackage{ifthen}
\newcommand{\displaycomments}{true} %switch off to make sure all comments are gone
\newcommand{\mike}[1]{\ifthenelse{\equal{\displaycomments}{true}}{\textcolor{blue}{(Mike: #1)}}{}}

\begin{document}

%%%%%%%%%%%%%%%%%%%%%%%%%
%%%%% Title     %%%%%%%%%
%%%%%%%%%%%%%%%%%%%%%%%%%
\renewcommand{\thefootnote}{\fnsymbol{footnote}}
\setcounter{footnote}{1}

% $Id: title-LHCb-PAPER.tex 122889 2018-08-17 17:59:55Z pkoppenb $
% ===============================================================================
% Purpose: LHCb-PAPER journal paper title page template
% Author: 
% Created on: 2010-09-25
% ===============================================================================

%%%%%%%%%%%%%%%%%%%%%%%%%
%%%%%  TITLE PAGE  %%%%%%
%%%%%%%%%%%%%%%%%%%%%%%%%
\begin{titlepage}
\pagenumbering{roman}

% Header ---------------------------------------------------
\vspace*{-1.5cm}
\centerline{\large EUROPEAN ORGANIZATION FOR NUCLEAR RESEARCH (CERN)}
\vspace*{1.5cm}
\noindent
\begin{tabular*}{\linewidth}{lc@{\extracolsep{\fill}}r@{\extracolsep{0pt}}}
\ifthenelse{\boolean{pdflatex}}% Logo format choice
{\vspace*{-1.5cm}\mbox{\!\!\!\includegraphics[width=.14\textwidth]{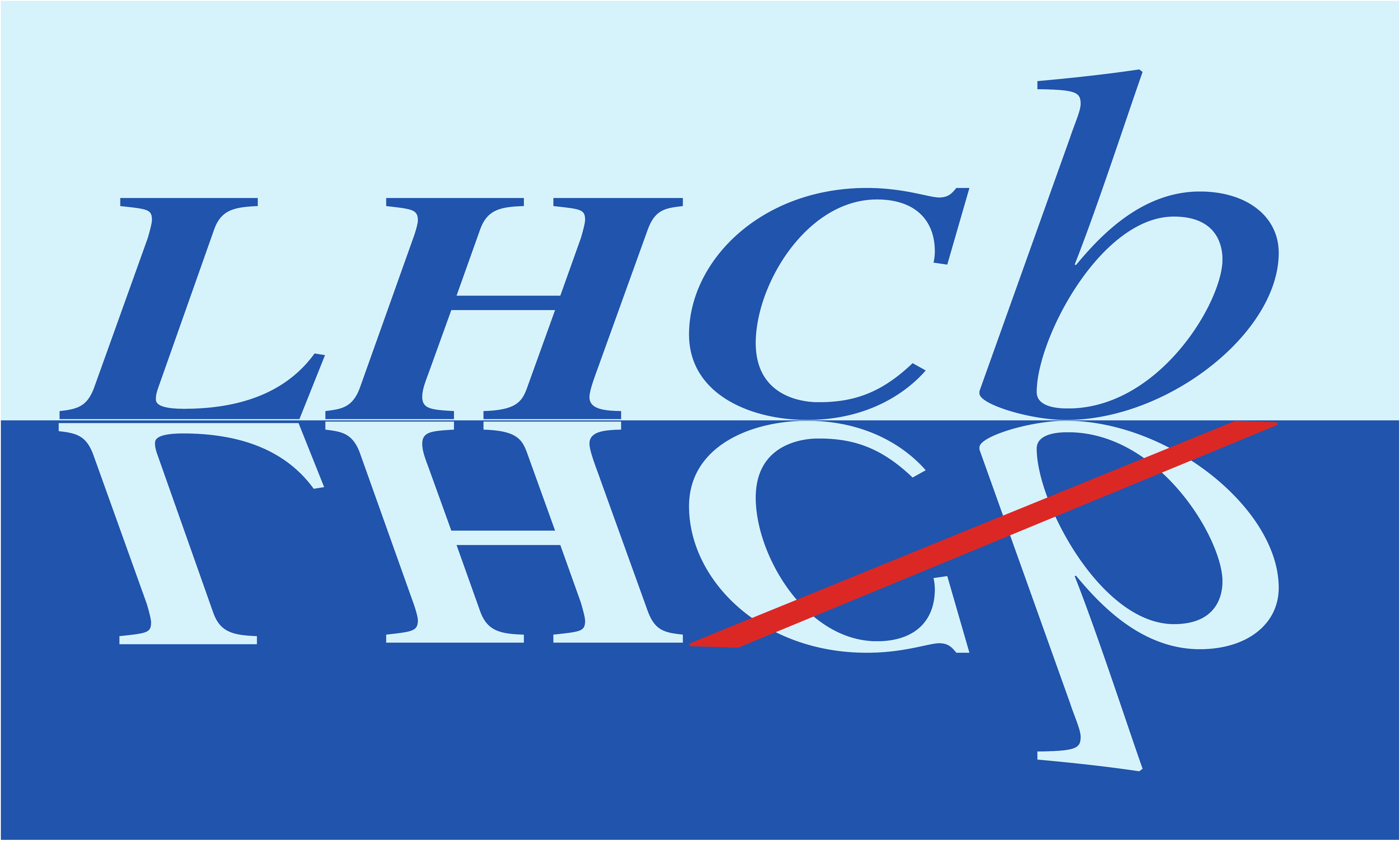}} & &}%
{\vspace*{-1.2cm}\mbox{\!\!\!\includegraphics[width=.12\textwidth]{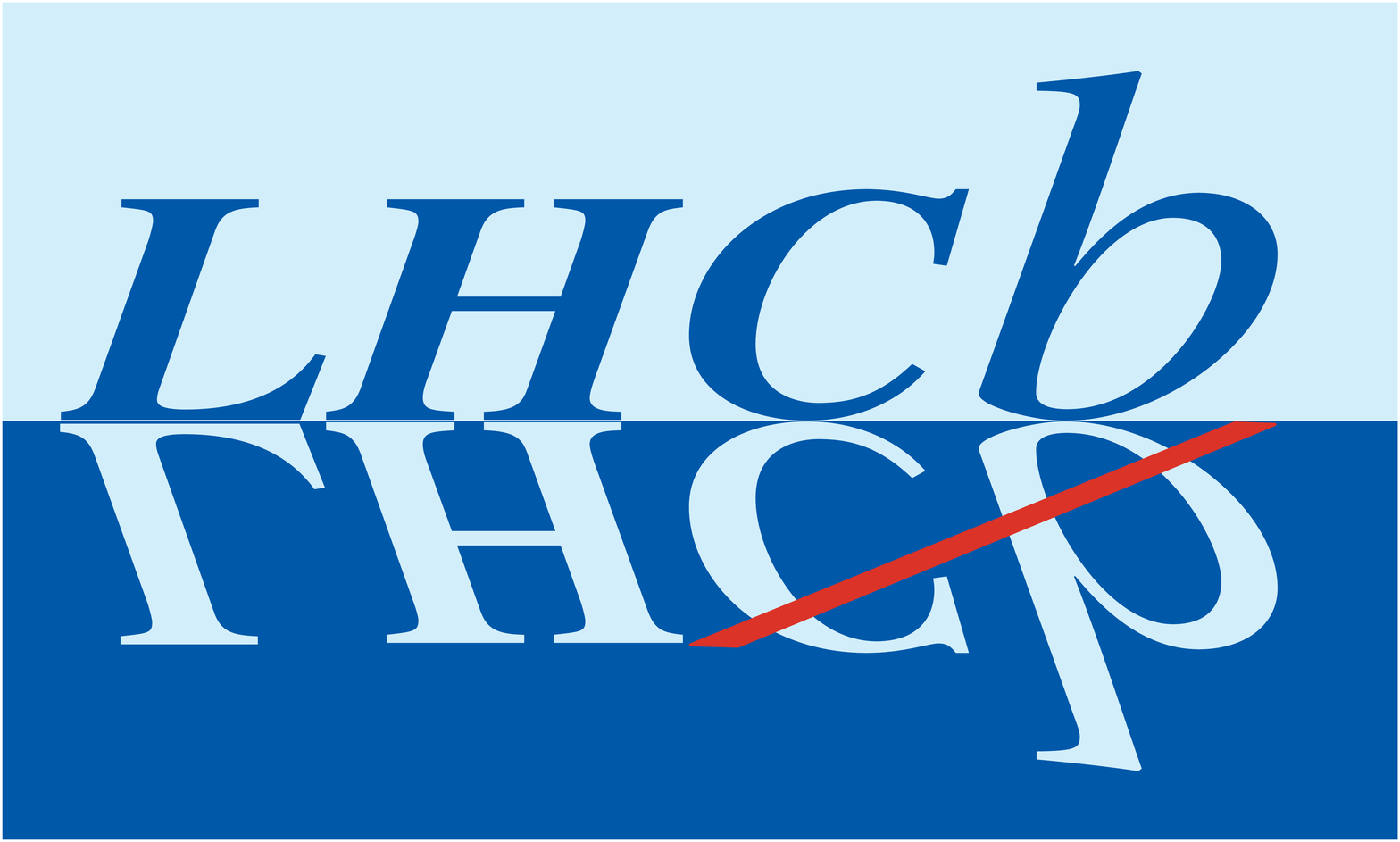}} & &}%
\\
% & & CERN-EP-20XX-ZZZ \\  % ID 
 & & LHCb-DP-2021-004 \\  % ID 
 & & \today \\ % Date - Can also hardwire e.g.: 23 March 2010
 & & \\
% not in paper \hline
\end{tabular*}

\vspace*{0.5cm}

% Title --------------------------------------------------
{\normalfont\bfseries\boldmath\huge
\begin{center}
% DO NOT EDIT HERE. Instead edit macro in main.tex to keep metadata correct
  \papertitle 
\end{center}
}

\vspace*{0.5cm}

% Authors -------------------------------------------------

\begin
{flushleft}
\small

R.~Calabrese$^{a,b}$,
M.~Fiorini$^{a,b}$,
E.~Luppi$^{a,b}$,
L.~Minzoni$^{a,b}$,
I.~Slazyk$^{a,b}$,
L.~Tomassetti$^{a,b}$,
M.~Bartolini$^{c,d}$,
R.~Cardinale$^{c,d}$,
F.~Fontanelli$^{c,d}$,
A.~Petrolini$^{c,d}$,
A.~Pistone$^{c,d,1}$,
M.~Calvi$^{e,f}$,
C.~Matteuzzi$^{e}$,
A.~Lupato$^{g,2}$,
G.~Simi$^{g}$,
M.~Kucharczyk$^{h}$,
B.~Malecki$^{h}$,
M.~Witek$^{h}$,
S.~Benson$^{i}$,
M.~Blago$^{i}$,
G.~Cavallero$^{i}$,
A.~Contu$^{i}$,
C.~D'Ambrosio$^{i}$,
C.~Frei$^{i}$,
T.~Gys$^{i}$,
J.~He$^{i}$,
D.~Piedigrossi$^{i}$,
K.~Wyllie$^{i}$,
M.~Adinolfi$^{j}$,
M.G.~Chapman$^{j}$,
J.~Dalseno$^{j,3}$,
S.~Harnew$^{j}$,
S.~Maddrell-Mander$^{j}$,
P.~Naik$^{j}$,
C.~Prouve$^{j,3}$,
J.~Rademacker$^{j}$,
A.~Solomin$^{j}$,
J.~Garra Tic\'{o}$^{k}$,
V.~Gibson$^{k}$,
C.R.~Jones$^{k}$,
S.~Tolk$^{k}$,
S.A.~Wotton$^{k}$,
S.~Easo$^{l}$,
A.~Papanestis$^{l}$,
F. F.~Wilson$^{l}$,
L.~Carson$^{m}$,
S.~Eisenhardt$^{m}$,
S.~Gambetta$^{m}$,
K.~Gizdov$^{m}$,
H.~Luo$^{m}$,
A.~Morris$^{m}$,
F.~Muheim$^{m}$,
M.~Pappagallo$^{m}$,
I.~Smith$^{m}$,
J.~Webster$^{m}$,
J.~Zonneveld$^{m}$,
S.~Karodia$^{n}$,
S.~Ogilvy$^{n}$,
P.~Sail$^{n}$,
F.J.P.~Soler$^{n}$,
P.~Spradlin$^{n}$,
M.~Traill$^{n}$,
S.~Baker$^{o}$,
D.~Clark$^{o}$,
I.~Clark$^{o}$,
K.~Ladhams$^{o}$,
M.~McCann\footnote[2]{Corresponding author.}$^{o}$,
R.D.~Moise$^{o,4}$,
D.~Nardini$^{o}$,
M.~Patel$^{o}$,
T.~Savidge$^{o}$,
E.~Smith$^{o,5}$,
S.~Stefkova$^{o,6}$,
D.~Websdale$^{o}$,
M.~Bjorn$^{p}$,
F.~Cheung$^{p}$,
R.~Gao$^{p}$,
N.~Harnew$^{p}$,
D.~Hill$^{p}$,
S.~Malde$^{p}$,
A.~Nandi$^{p}$,
S.~Topp-Jorgensen$^{p}$.\bigskip

{\footnotesize \it

$^{a}$ INFN Sezione di Ferrara, Ferrara, Italy\\
$^{b}$ Universit{\`a} di Ferrara, Ferrara, Italy\\
$^{c}$ INFN Sezione di Genova, Genova, Italy\\
$^{d}$ Universit{\`a} di Genova, Genova, Italy\\
$^{e}$ INFN Sezione di Milano-Bicocca, Milano, Italy\\
$^{f}$ Universit{\`a} di Milano Bicocca, Milano, Italy\\
$^{g}$ Universita degli Studi di Padova, Universita e INFN, Padova, Padova, Italy\\
$^{h}$ Henryk Niewodniczanski Institute of Nuclear Physics  Polish Academy of Sciences, Krak{\'o}w, Poland\\
$^{i}$ European Organization for Nuclear Research (CERN), Geneva, Switzerland\\
$^{j}$ H.H. Wills Physics Laboratory, University of Bristol, Bristol, United Kingdom\\
$^{k}$ Cavendish Laboratory, University of Cambridge, Cambridge, United Kingdom\\
$^{l}$ STFC Rutherford Appleton Laboratory, Didcot, United Kingdom\\
$^{m}$ School of Physics and Astronomy, University of Edinburgh, Edinburgh, United Kingdom\\
$^{n}$ School of Physics and Astronomy, University of Glasgow, Glasgow, United Kingdom\\
$^{o}$ Imperial College London, London, United Kingdom\\
$^{p}$ Department of Physics, University of Oxford, Oxford, United Kingdom\\
\bigskip
$^{1}$ Now at Italian institute of Technology\\
$^{2}$ Now at Department of Physics and Astronomy, University of Manchester, Manchester, United Kingdom\\
$^{3}$ Now at Instituto Galego de F{\'\i}sica de Altas Enerx{\'\i}as (IGFAE), Universidade de Santiago de Compostela, Santiago de Compostela, Spain\\
$^{4}$ Now at I. Physikalisches Institut, RWTH Aachen University, Aachen, Germany\\
$^{5}$ Now at Physik-Institut, Universit{\"a}t Z{\"u}rich, Z{\"u}rich, Switzerland\\
$^{6}$ Now at Deutsches Elektronen-Synchrotron, Hamburg, Germany\\

}\end{flushleft}

\vspace{\fill}

% Abstract -----------------------------------------------
\begin{abstract}
  \noindent
  
  The performance of the ring-imaging Cherenkov detectors at the LHCb experiment is determined during the LHC Run 2 period between 2015 and 2018. The stability of the Cherenkov angle resolution and number of detected photons with time and running conditions is measured. The particle identification performance is evaluated with data and found to satisfy the requirements of the physics programme.
   
\end{abstract}

\vspace*{2.0cm}

\begin{center}
  Submitted to
  JInst. 
\end{center}

\vspace{\fill}

{\footnotesize 
% Edit macro in main.tex to keep metadata correct
\centerline{\copyright~\papercopyright. \href{\paperlicenceurl}{\paperlicence}.}}
\vspace*{2mm}

\end{titlepage}

%%%%%%%%%%%%%%%%%%%%%%%%%%%%%%%%
%%%%%  EOD OF TITLE PAGE  %%%%%%
%%%%%%%%%%%%%%%%%%%%%%%%%%%%%%%%

\clearpage

\renewcommand{\thefootnote}{\arabic{footnote}}
\setcounter{footnote}{0}

%%%%%%%%%%%%%%%%%%%%%%%%%%%%%%%%
%%%%%  Table of Content   %%%%%%
%%%%%%%%%%%%%%%%%%%%%%%%%%%%%%%%
%%%% Uncomment next 2 lines if desired
%\tableofcontents
%\cleardoublepage

%\input{todo}

%%%%%%%%%%%%%%%%%%%%%%%%%
%%%%% Main text %%%%%%%%%
%%%%%%%%%%%%%%%%%%%%%%%%%

\pagestyle{plain} % restore page numbers for the main text
\setcounter{page}{1}
\pagenumbering{arabic}

\section{Introduction}
\label{sec:Introduction}
 
The \lhcb experiment is designed for the precision study of decays of hadrons containing \bquark and \cquark quarks, with the aim of discovering new physics beyond the Standard Model of particle physics and measuring the parameters of the CKM matrix.
A key requirement for isolating specific decays is the identification of long-lived charged hadrons, i.e. \pipm, \Kpm, \proton/\antiproton and $d$/$\overline{d}$.
\lhcb uses a system of two ring-imaging Cherenkov (RICH) detectors to discriminate between these particles.

The performance of the RICH detectors was studied during the LHC Run~1 (2011--2012)~\cite{LHCb-DP-2012-003}. During Run~2 (2015--2018), the \lhcb experiment took data with a 25\ns $pp$-collision spacing, half that of Run~1, and an increased centre-of-mass energy, \sqs=13\tev, creating a more challenging environment with higher track multiplicities and larger detector occupancies. A system of automated real-time alignment and calibration measurements of the \lhcb detector was implemented for Run~2, allowing hadron identification to be used in the software trigger for the first time.

The RICH detectors are currently undergoing a major upgrade in preparation for the significantly increased data rates expected during LHC Run~3, with completely new sensors and readout chain, and substantially different optics allowed by removal of an aerogel radiator at the end of Run~1. This paper addresses the ultimate performance of the LHCb RICH detectors in their original configuration.

This paper describes the performance of the RICH detectors in LHC Run~2. In section \ref{sec:Detector} the \lhcb spectrometer and the RICH detectors are described. The calibrations and alignments of the RICH detectors are given in section \ref{sec:Calibration}. Measurements of the Cherenkov angle resolution are presented in section \ref{sec:Resolution}, and the number of detected photons per charged track in section \ref{sec:Yield}. The particle identification (PID) performance measured with several calibration channels is presented in section \ref{sec:Performance}, along with its implications on representative \lhcb physics analyses demonstrating the utility of the excellent hadron PID.

\section{Detector and reconstruction}
\label{sec:Detector}

\begin{figure}[tb]
  \begin{center}
    \includegraphics[width=\linewidth,page=1]{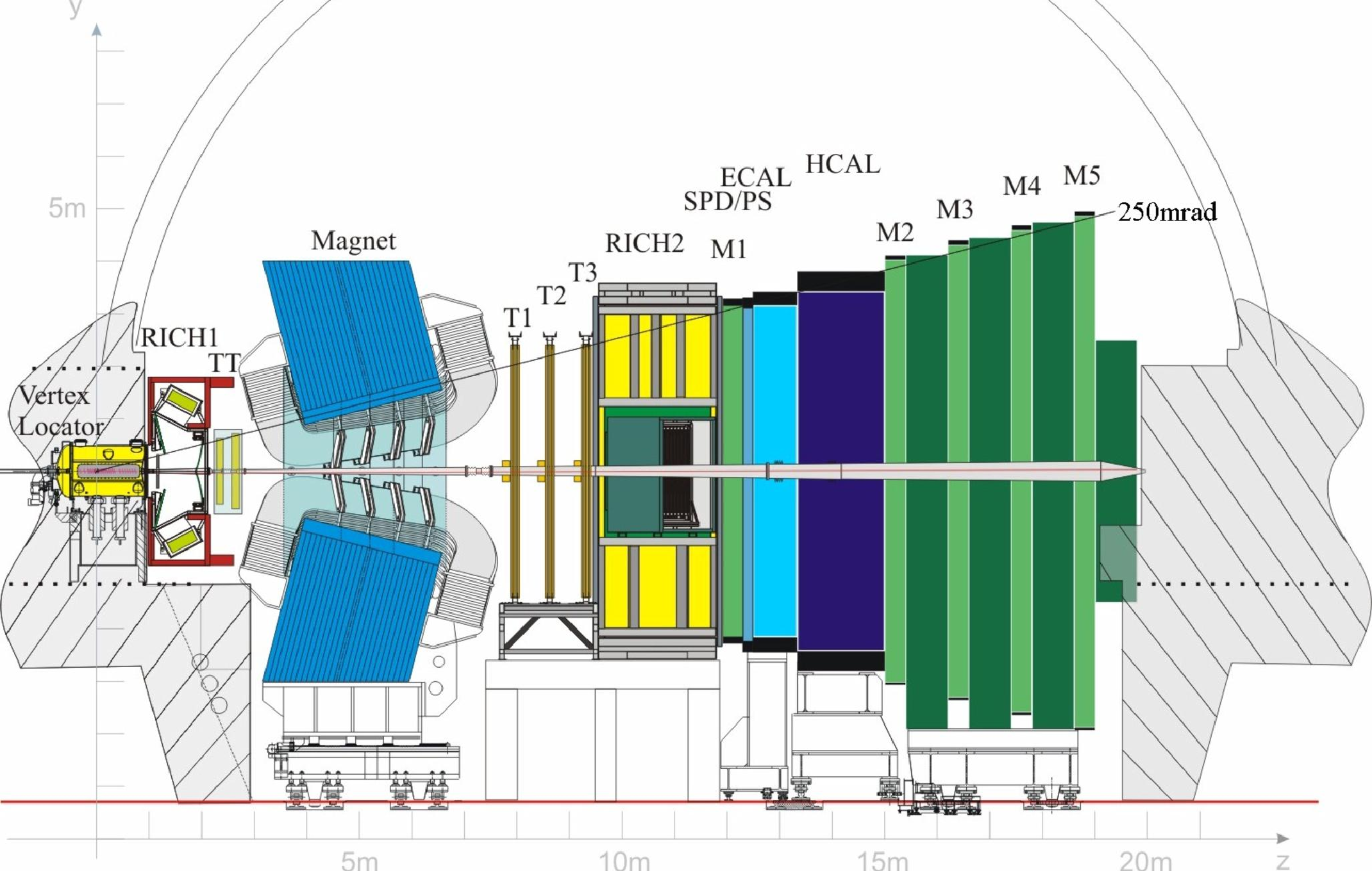}
    \vspace*{-0.5cm}
  \end{center}
  \caption{
    A view of the LHCb detector \cite{LHCb-TDR-009}. The proton-proton interaction point is on the extreme left at $z=0$~m.
    }
  \label{fig:lhcb_schematic}
\end{figure}

The LHCb 
detector~\cite{Alves:2008zz,LHCb-DP-2014-002} is a
single-arm forward spectrometer covering the pseudorapidity range ${2 < \eta < 5}$, designed for
the study of particles containing \bquark\ or \cquark\ quarks. A schematic view of the detector is given in figure~\ref{fig:lhcb_schematic}. The detector elements include: a silicon-strip vertex detector surrounding the $pp$ interaction
region that allows \cquark\ and \bquark\ hadrons to be identified from their characteristically long
flight distance; a tracking system that provides a measurement of the momentum, $p$, of charged
particles as they traverse a dipole magnetic field with a bending power of about $4{\mathrm{\,Tm}}$; and two RICH detectors that are able to discriminate between different species of charged particles. Photons, electrons and hadrons are additionally identified by a calorimeter system consisting of scintillating-pad and preshower detectors, an electromagnetic and a hadronic calorimeter. Muons are identified by a system composed of alternating layers of iron and multiwire proportional chambers. The magnetic field is periodically flipped between vertically up and down to reduce systematic uncertainties associated with opposing charges experiencing different parts of the detector, and left-right detector asymmetries.

\begin{figure}[tb]
  \begin{center}
    \includegraphics[width=0.49\linewidth,page=1]{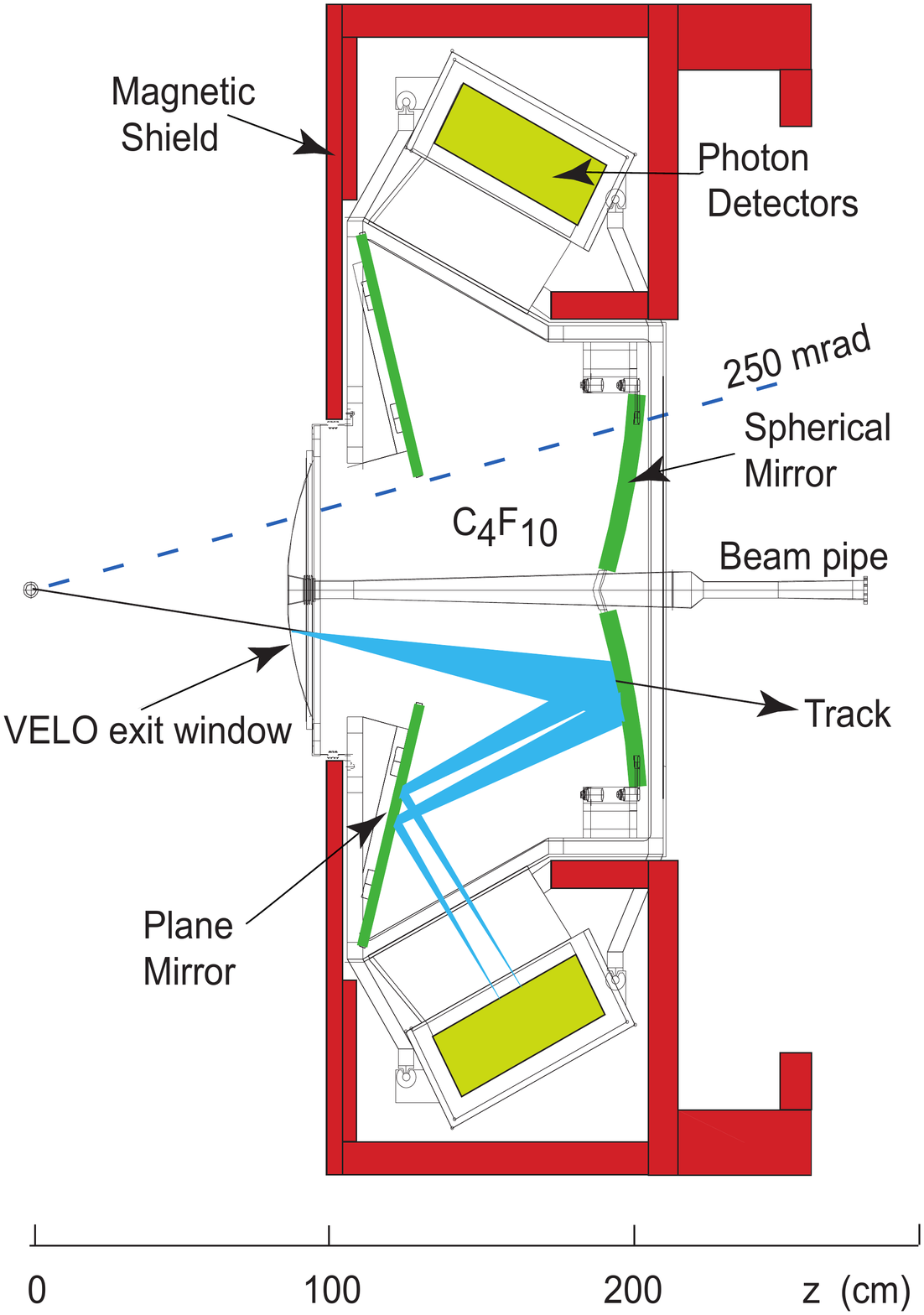}\put(-220,285){(a)}
    \includegraphics[width=0.49\linewidth,page=1]{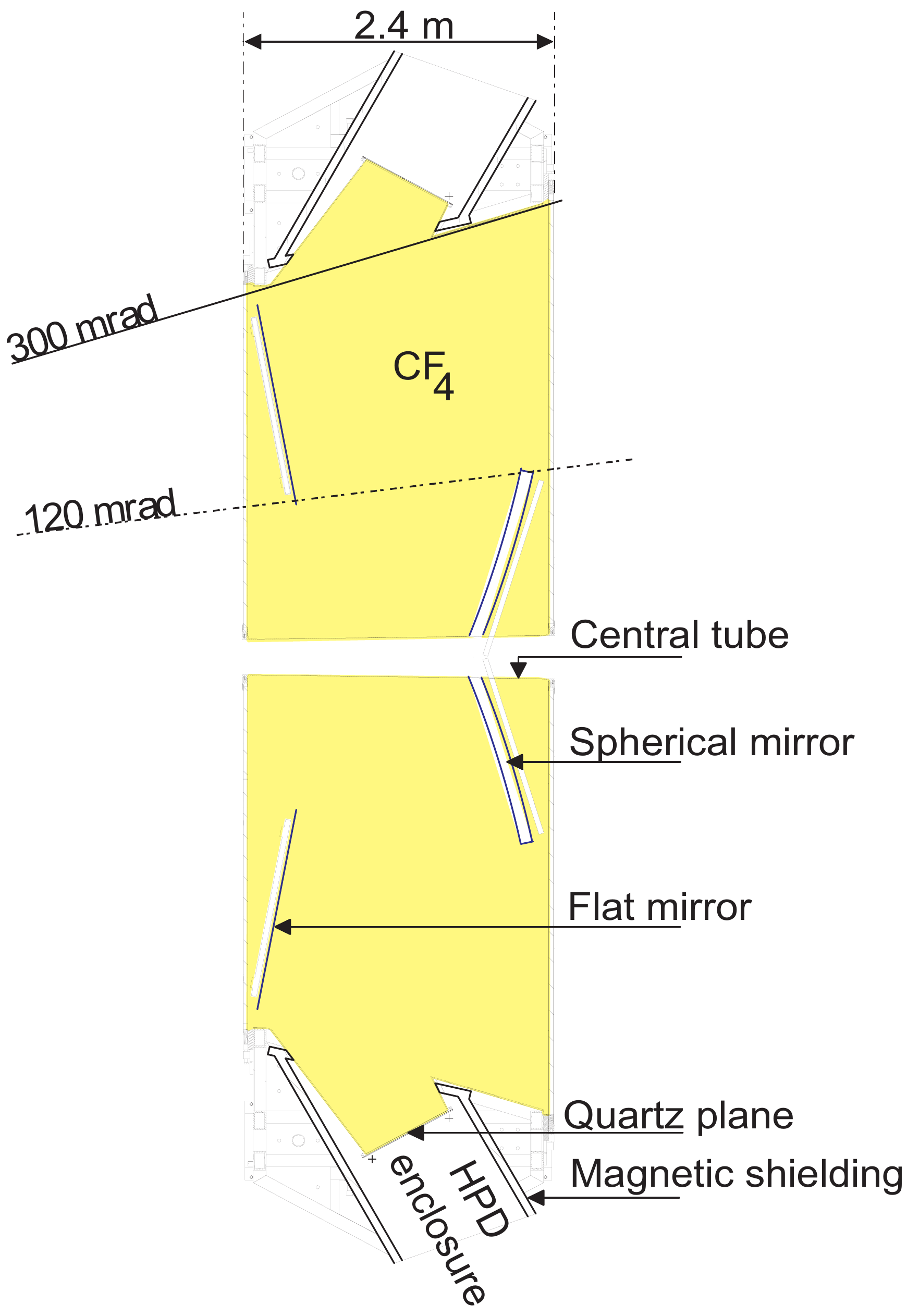}\put(-220,285){(b)}
    \vspace*{-0.5cm}
  \end{center}
  \caption{
    The optical system of (a) \richone and (b) \richtwo~\cite{LHCb-TDR-009}.}
  \label{fig:rich_schematic}
\end{figure}

The two RICH detectors are each configured with a single gas radiator for Run~2. Upstream of the dipole magnet is \richone, optimised for low-momentum particles using C$_4$F$_{10}$ as a radiator with refractive index 1.0014, and covering the full angular acceptance. Downstream of the magnet is \richtwo, optimised for a higher momentum range, using CF$_4$ as a radiator with refractive index 1.0005, and covering approximately a quarter of the solid-angular acceptance of the spectrometer, $3 < \eta < 5$.

Both RICH detectors use spherical mirrors to focus the cones of Cherenkov light emitted from a charged particle as rings onto an array of hybrid photon detectors (HPDs). An intermediate flat mirror between the spherical mirror and HPD array allows the HPDs to be located outside of the spectrometer acceptance. A schematic of the two RICH detectors is shown in figure~\ref{fig:rich_schematic}.

\begin{figure}[tb]
  \begin{center}
    \includegraphics[width=0.49\linewidth,page=1]{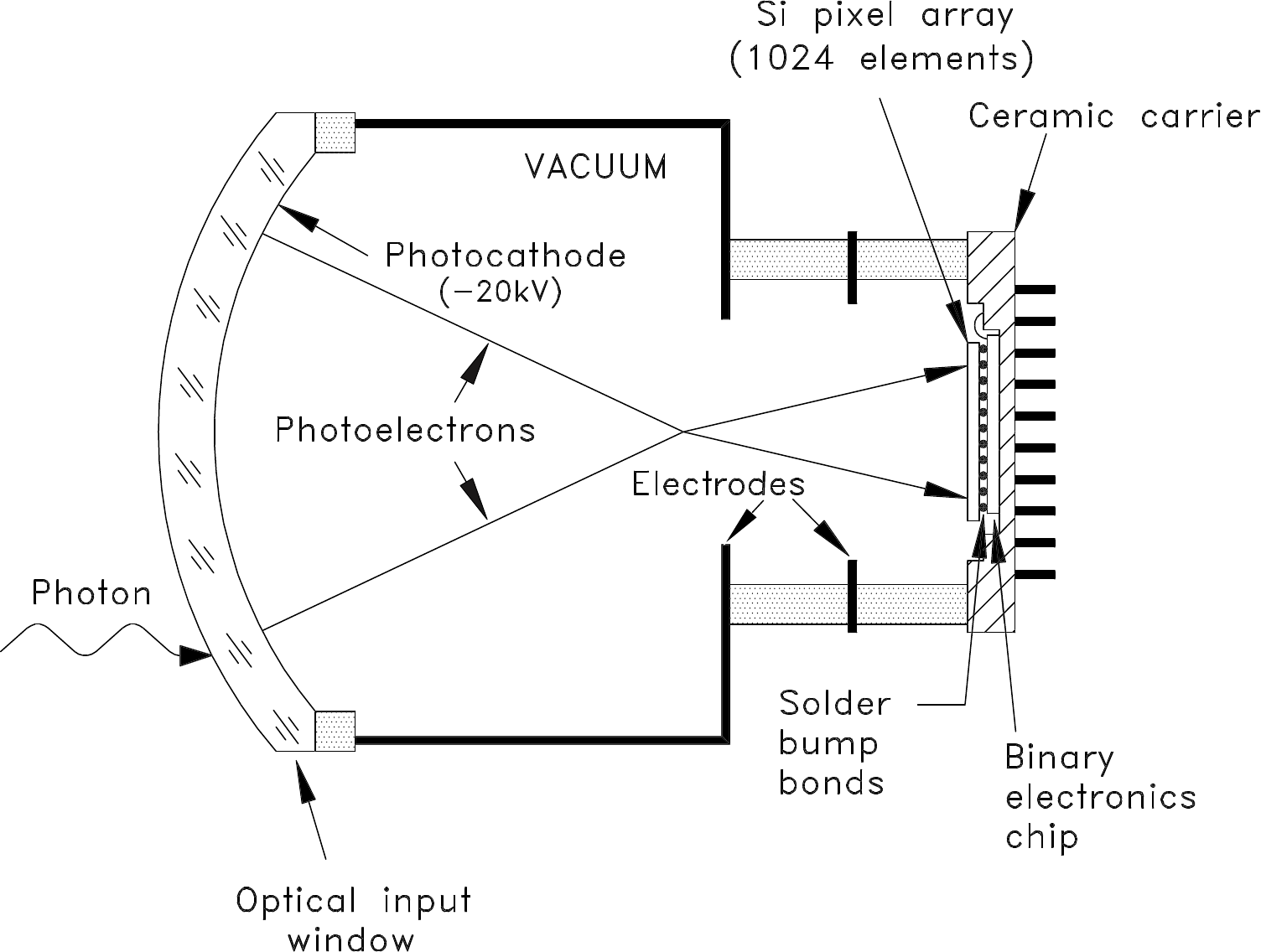}
    \vspace*{-0.5cm}
  \end{center}
  \caption{
    The focusing optics of an HPD~\cite{LHCb-TDR-009}. A photoelectron generated at the photocathode is accelerated to the silicon sensor chip.
    }
  \label{fig:hpd_schematic}
\end{figure}

The HPDs~\cite{ALEMIA200048} consist of a photocathode deposited on the inside of the quartz window of a vacuum tube containing acceleration optics that guide the electron produced at the photocathode to a silicon sensor where it is read out as a hit. A schematic of an HPD is shown in figure~\ref{fig:hpd_schematic}. The tube operates with a high voltage range of 16 to 20~kV applied between the anode and cathode. The high energy of the electron entering the silicon sensor allows the signal to be well isolated from the electronic noise pedestal, making a very efficient, almost noise free photon detector. In total, across the two RICH detectors there are approximately 500\,000 active channels and in a typical LHC event readout window of 25ns only 10 dark noise counts are observed. The trajectory of the photoelectron through the focusing optics of the HPD is affected by external electric and magnetic fields. While most of these effects are minimised by ample magnetic shielding placed around the HPDs, residual distortions are corrected in software.

Charged particles are probabilistically identified by the construction of a log-likelihood value from the measured hits on the detector plane and the expected hit patterns calculated from all the tracks measured through the detector~\cite{Forty:1996gj}. Each particle-type hypothesis is tried iteratively to find the configuration of particle hypotheses that minimises the global likelihood. With all other tracks under the hypothesis that minimises the global likelihood, a delta-log-likelihood ($\Delta$LL) value is computed for each track as the difference between the log-likelihoods when the track is given a specific hypothesis and that of the pion hypothesis. This value is used to discriminate between different hypotheses further along the analysis chain.

\section{Detector calibration}
\label{sec:Calibration}

To achieve optimal performance several detector effects need to be accounted for through various calibration schemes. These include the timing of the readout electronics relative to the transit of charged particles, the alignment of the mirrors, changes in the refractive index of the radiators, and the effects of electric and magnetic fields on the images of the HPD's photocathode on the silicon sensor. Each of these requires a careful calibration process, which is described and evaluated below. Further details of the calibration procedures, and their implementation can be found in ref.~\cite{richcalpaper}.

\subsection{Time alignment}
\label{subsec:TimeAlignment}

The difference between the arrival time of photons at an HPD's photocathode and the peak of the output from the silicon sensor depends on the sensor characteristics, particularly the leakage current. In addition, the arrival time of photons relative to a \proton\proton collision is dependent on the photon path length through the optical system to each HPD. These effects are compensated for by applying an offset of the trigger signal to each of the HPD's readout boards, L0 boards~\cite{ADINOLFI2007689}, relative to the global LHCb clock. Each L0 board reads out two HPDs,  so care is taken to select HPD pairs with similar characteristics.

\begin{figure}[tb]
  \begin{center}
    \includegraphics[width=0.49\linewidth,page=1]{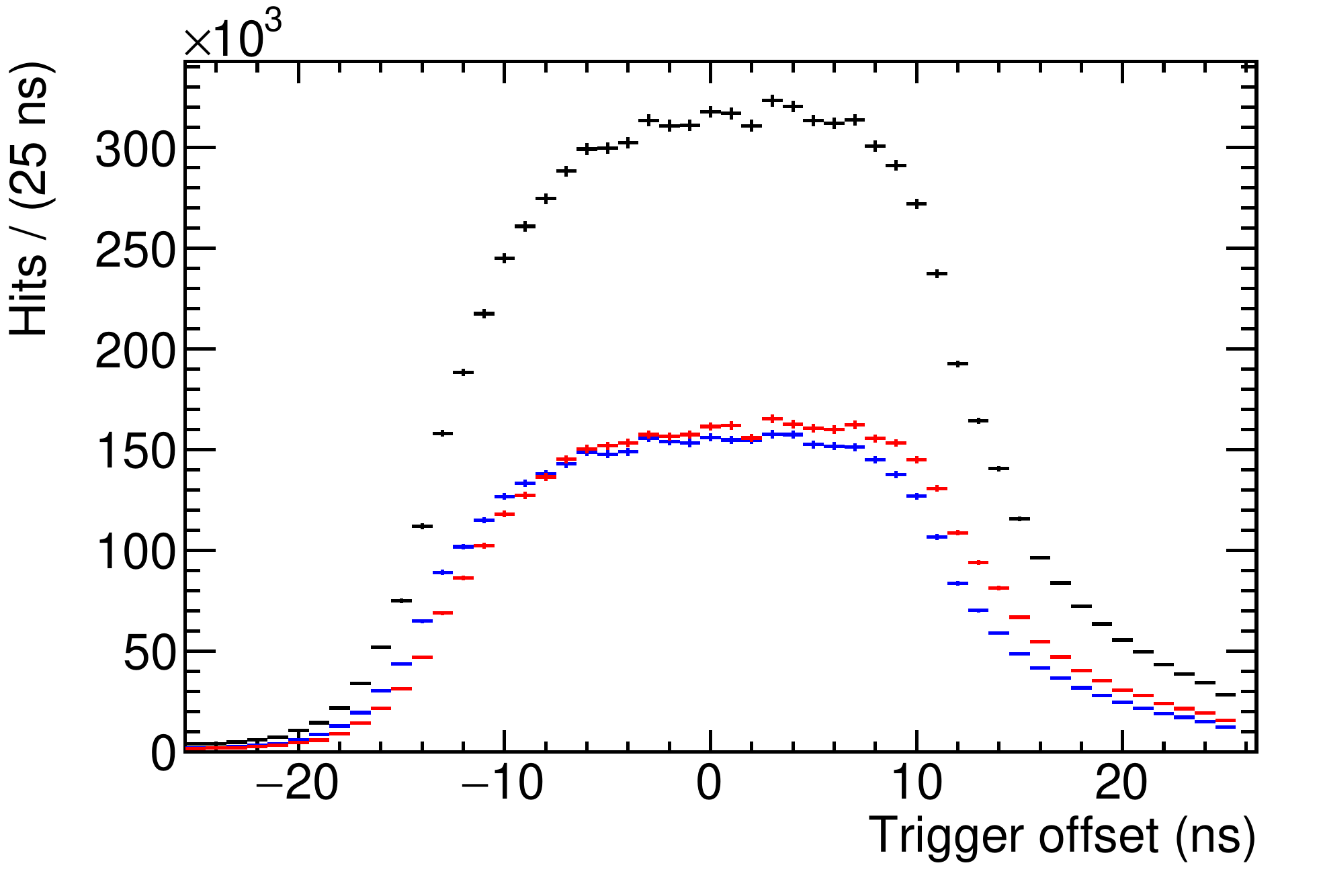}
  \end{center}
  \caption{
    The number of hits from two HPDs (blue and red points) read out by the same L0 board and the combination (black points) for data taken with differing trigger offsets in steps of 1\ns. Each step integrates the number of photons detected over 10\,000 bunch crossings.}
  \label{fig:timing_curve_example}
\end{figure}

To time align the HPDs, dedicated calibration data is taken when there is a very large bunch spacing in the LHC, typically with a single colliding bunch per orbit. The number of photons detected by each HPD is recorded with different values of time offset relative to the LHCb clock. The offset is adjusted in steps of 1\ns in the range $\pm25$\ns from a point seen to be in the correct 25\ns window. Typically 10\,000 events are taken per step. An example of such a scan for one pair of HPDs is shown in figure~\ref{fig:timing_curve_example}. The scan is performed without input from the tracking system, so the total photon count cannot be normalised to the number of tracks. Instead, the instantaneous luminosity as measured by the calorimeters is monitored during the scan to ensure it remains stable. For each pair of photon detectors a distribution of detected photons as a function of time is produced. This distribution is typically a top hat, and the centre point of the high region (defined as 90\% of maximum) is taken as the optimal timing. To further account for variation in the beam conditions, several scans are performed and the required time adjustment for each board computed from the average.

\begin{figure}[tb]
  \begin{center}
    \includegraphics[width=0.49\linewidth,page=1]{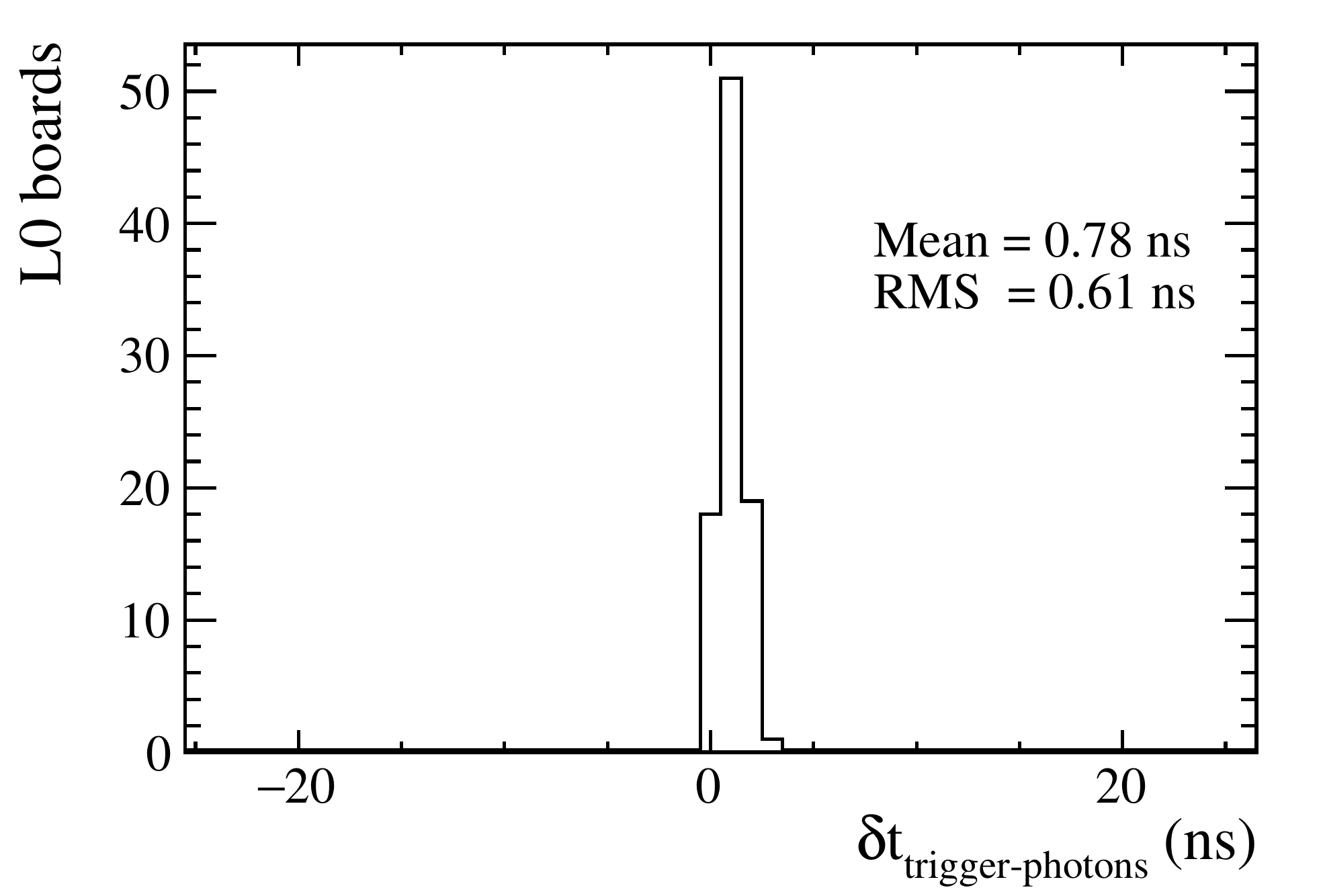}\put(-32,120){(a)}
    \includegraphics[width=0.49\linewidth,page=2]{figs/Fig5}\put(-32,120){(b)}
    \vspace*{-0.5cm}
  \end{center}
  \caption{
    Distribution of the residual time differences between the trigger to each L0 board and the central point of the photon signal from that board for (a) \richone and (b) \richtwo. Both detectors are aligned to better than 1\ns.}
  \label{fig:timing_scan}
\end{figure}

The distribution of the residual time differences between the trigger and the centre point of the photon signal is shown in figure~\ref{fig:timing_scan}. The mean and RMS of the L0 board timing distributions are well centred, giving an alignment better than 1\ns, comparable with the resolution of the timing data taken. The combined HPD response for each L0 board is typically a top-hat function with a width of 10\ns, making a 1\ns alignment sufficiently precise.

\subsection{Magnetic distortion correction}
\label{subsec:MDCS}

The HPDs are sensitive to magnetic fields, which alter the trajectory the photo-electrons take between the cathode and anode sensor. The LHCb dipole creates a large magnetic field that extends over the region including \richone with a peak value of about $60{\mathrm{\,mT}}$ at the location of the photon detectors. To attenuate this field, the HPDs are contained within a soft-iron box, and are wrapped in individual mu-metal shields. However they still experience a residual, predominantly axial, field of up to 2.5~mT. The distortion caused by this field is the largest effect that must be corrected in order to achieve the optimal resolution in \richone. A dedicated calibration system, which scans a grid of point light sources across the detector plane to map the magnetic distortions, is periodically run (typically once per field polarity per year) to derive corrections that are applied to the hit positions. A detailed description of the operation and performance of the correction system  can be found in ref.~\cite{BORGIA201444}. The magnetic distortion maps are one of the largest sources of differences in reconstruction performance between the magnetic field polarities in \richone.

\richtwo is in a region of much lower magnetic field, and the HPDs are aligned perpendicular to the predominant direction of the field. This leads to much smaller distortions, the time dependence of which can be ignored. Corrections for \richtwo were computed during the early detector commissioning and remain unchanged~\cite{Cardinale:2011eu}.

\begin{figure}[tb]
  \begin{center}
    \includegraphics[width=0.49\linewidth,page=1]{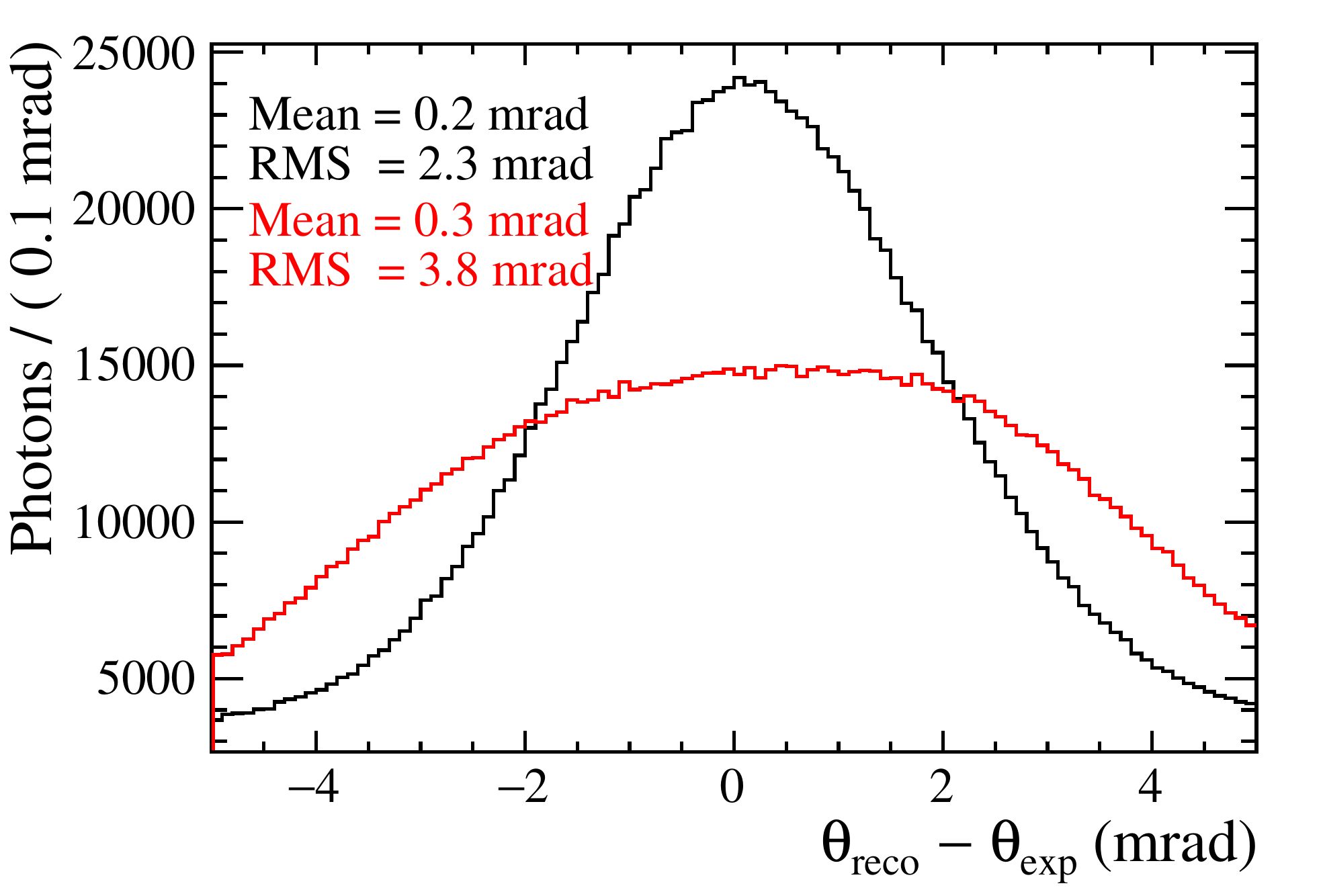}\put(-32,120){(a)} \includegraphics[width=0.49\linewidth,page=2]{figs/Fig6}\put(-32,120){(b)}
    \vspace*{-0.5cm}
  \end{center}
  \caption{
    The distribution of the difference of reconstructed and expected Cherenkov angles in (a) \richone and (b) \richtwo. The data is shown for reconstruction with corrections for the distortions caused by magnetic field applied (black) and ignored (red). A significant improvement is seen in the width of the distributions in \richone, and hence the Cherenkov angle resolutions, when applying the corrections.}
  \label{fig:MDCSRes}
\end{figure}

Figure~\ref{fig:MDCSRes} shows the change in angular resolution for both \richone and \richtwo without corrections for the magnetic field distortions, and when applying corrections. A large degradation of approximately a factor of five in the resolution is observed in \richone when no corrections are applied. Almost no degradation is seen in \richtwo.

\subsection{Mirror alignment}
\label{subsec:MirrorAlignment}

\richone contains 4 spherical mirrors and 16 flat mirrors whilst \richtwo contains 56 spherical mirrors and 40 flat mirrors. The relative orientation of the mirrors with respect to the tracking system and/or the HPDs could change with time through movements in the tracking stations, the detector support structures, or genuine movements of the mirrors. Such effects can be corrected using a data driven determination of the orientation of the mirrors. The alignment of the mirrors is monitored once per fill of the LHC and if found to change by more than 0.03~\mrad, the alignment used by the reconstruction is recomputed and updated.

The alignment of the mirrors is computed from the distribution of the photon hits as a function of the angular position around the tracks. If the orientation of the mirrors is perfectly known, the mean reconstructed Cherenkov angle would be constant as a function of the reconstructed azimuthal angle around the Cherenkov ring. Misalignments from the orientation used in the reconstruction introduce a sinusoidal dependency to the mean. The extent of this sinusoidal component is measured for spherical/flat mirror pairs and is used to correct mirror alignment. The process is performed iteratively until the corrections fall below a predefined threshold. A full description of the alignment procedure and its performance can be found in ref.~\cite{Prouve:2308521}. Throughout Run 2 the mirror alignments have been stable with very few updates required.

\subsection{Refractive index}
\label{subsec:RefIndex}

The refractive index, $n$, of the radiators used in the RICH detectors is a function of the pressure, temperature, and purity of the gases. The temperature and pressure are constantly monitored, but the purity of the gases is only measured periodically. A parameterisation of the refractive index as a function of pressure and temperature is used to approximate the actual refractive index, to which a correction factor, measured using data, is applied.

\begin{figure}[tb]
  \begin{center}
    \includegraphics[width=0.49\linewidth,page=1]{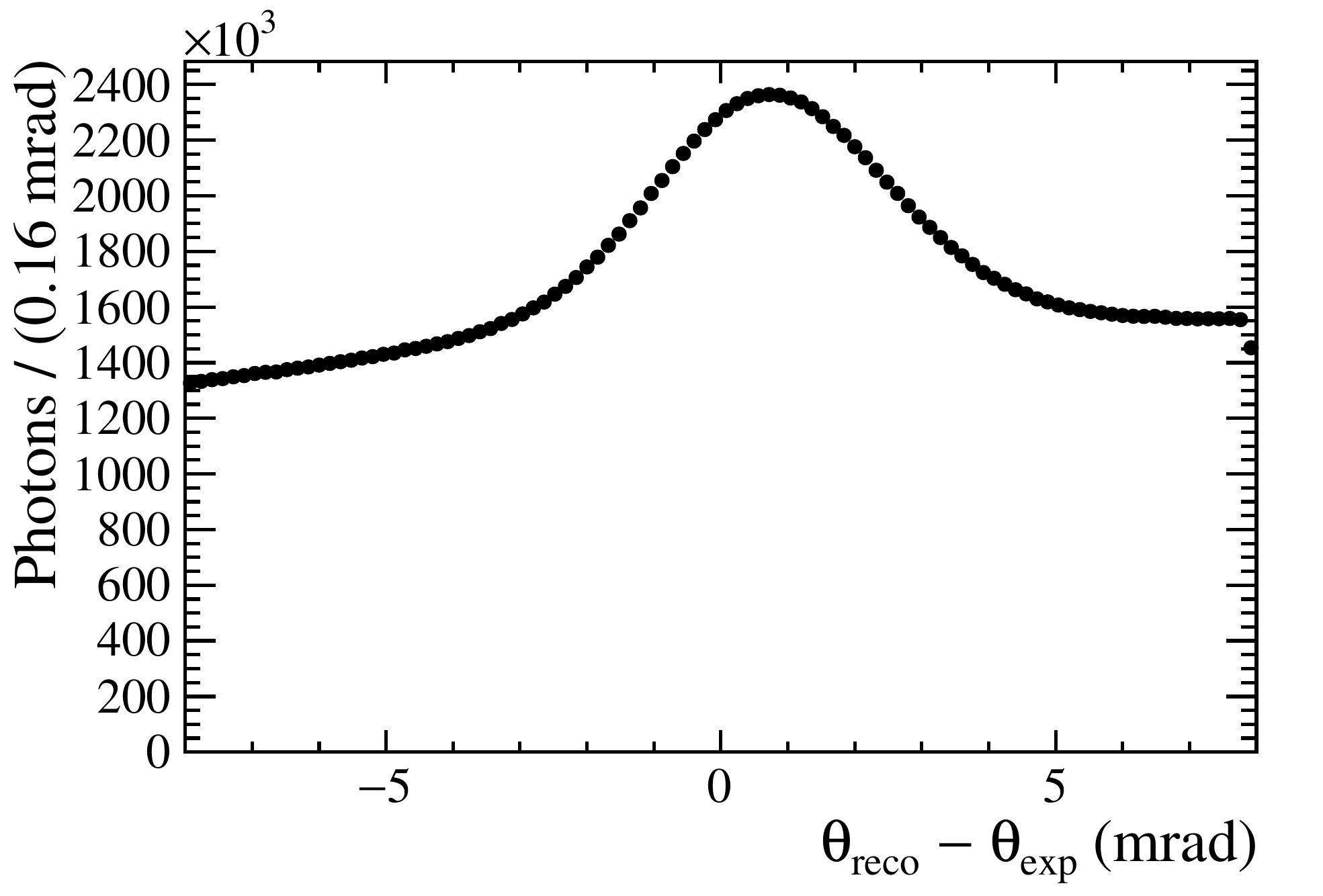}\put(-32,120){(a)}
    \includegraphics[width=0.49\linewidth,page=2]{figs/Fig7}\put(-32,120){(b)}
    \vspace*{-0.5cm}
  \end{center}
  \caption{
    Example distribution of photons detected in (a) \richone and (b) \richtwo from online monitoring data, from which the refractive index correction is computed. The correction is derived from the displacement of the mean of the Gaussian from zero.}
  \label{fig:ex_ckRefIndex}
\end{figure}

Figure \ref{fig:ex_ckRefIndex} shows the difference in the Cherenkov angle reconstructed from hits around a track and the expected angle calculated using only the temperature and pressure parameterisation of the refractive index. The refractive index must be monitored frequently and early in the reconstruction chain, so a high-statistics sample requiring little additional processing is used. The distribution is formed from a sample of high momentum tracks, for which the velocity approaches $c$ and the Cherenkov angle is approximately a function of $n$ only, collected over a one hour period. The peak representing the true Cherenkov photons is not centred at zero, indicating that the refractive index used in the calculation of the expected angle is different from the true value. This displacement is used to compute a corrective scale factor applied to the estimated refractive index in the reconstruction.

\begin{figure}[tb]
  \begin{center}
    \includegraphics[width=0.49\linewidth,page=9]{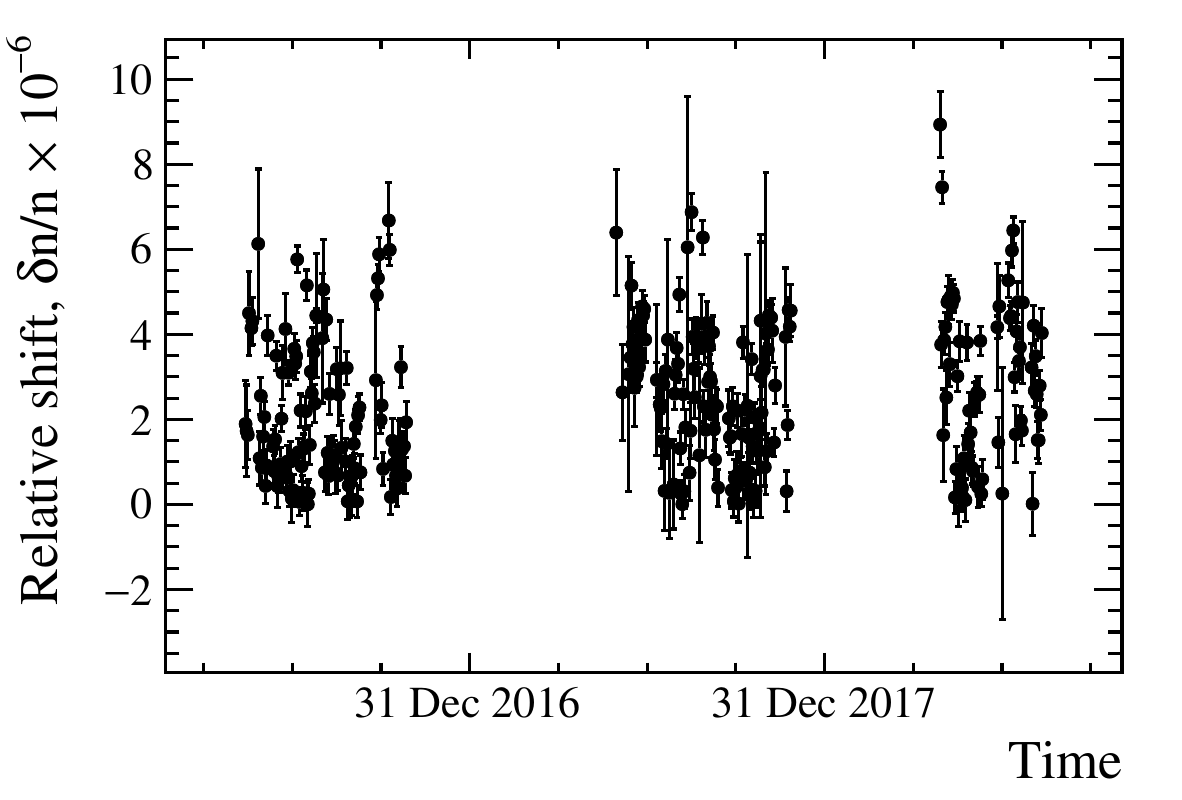}\put(-32,120){(a)}
    \includegraphics[width=0.49\linewidth,page=10]{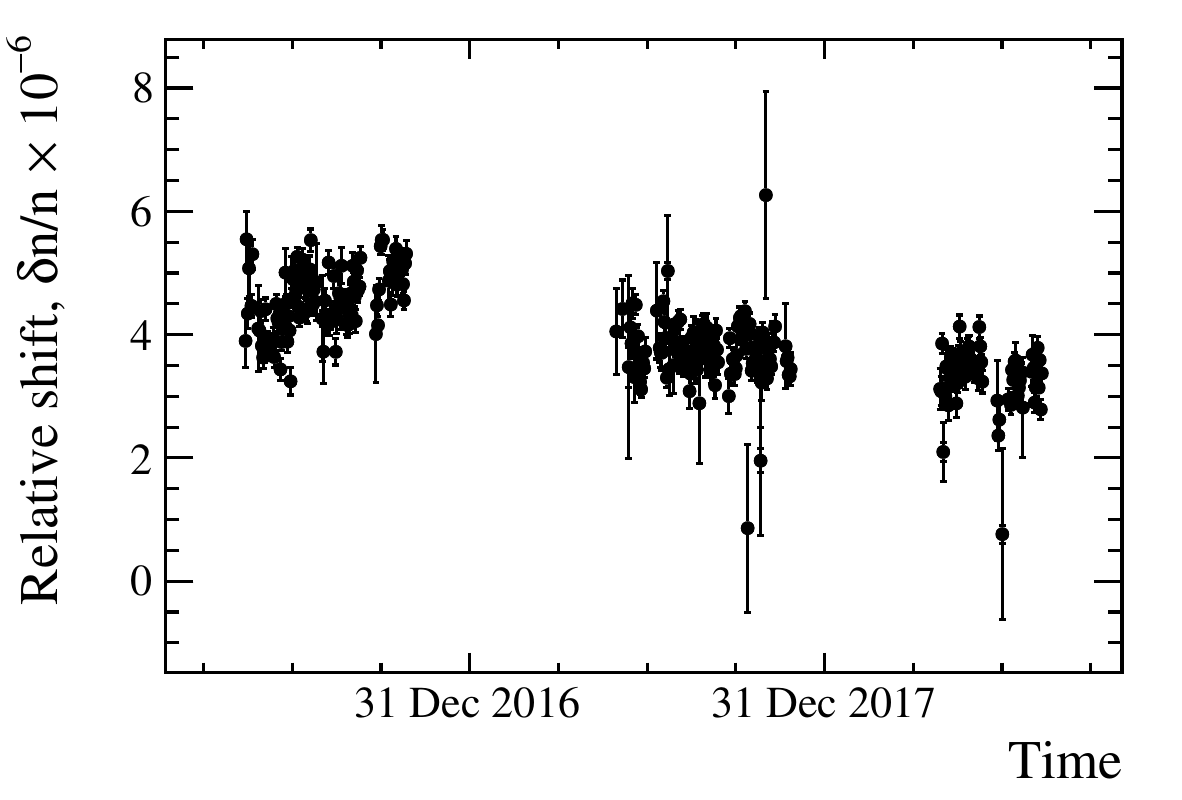}\put(-32,120){(b)}
    \vspace*{-0.5cm}
  \end{center}
  \caption{
    The relative difference in the refractive index used in the reconstruction after the calibration, and the true refractive index of the radiators in (a) \richone and (b) \richtwo evaluated using the peak position of low-multiplicity dimuon data as a function of time. Each point integrates data taken over a 24 hour period. The stability of the correction is excellent over several years.}
  \label{fig:ckRefIndexVTime}
\end{figure}

\begin{figure}[tb]
  \begin{center}
    \includegraphics[width=0.49\linewidth,page=1]{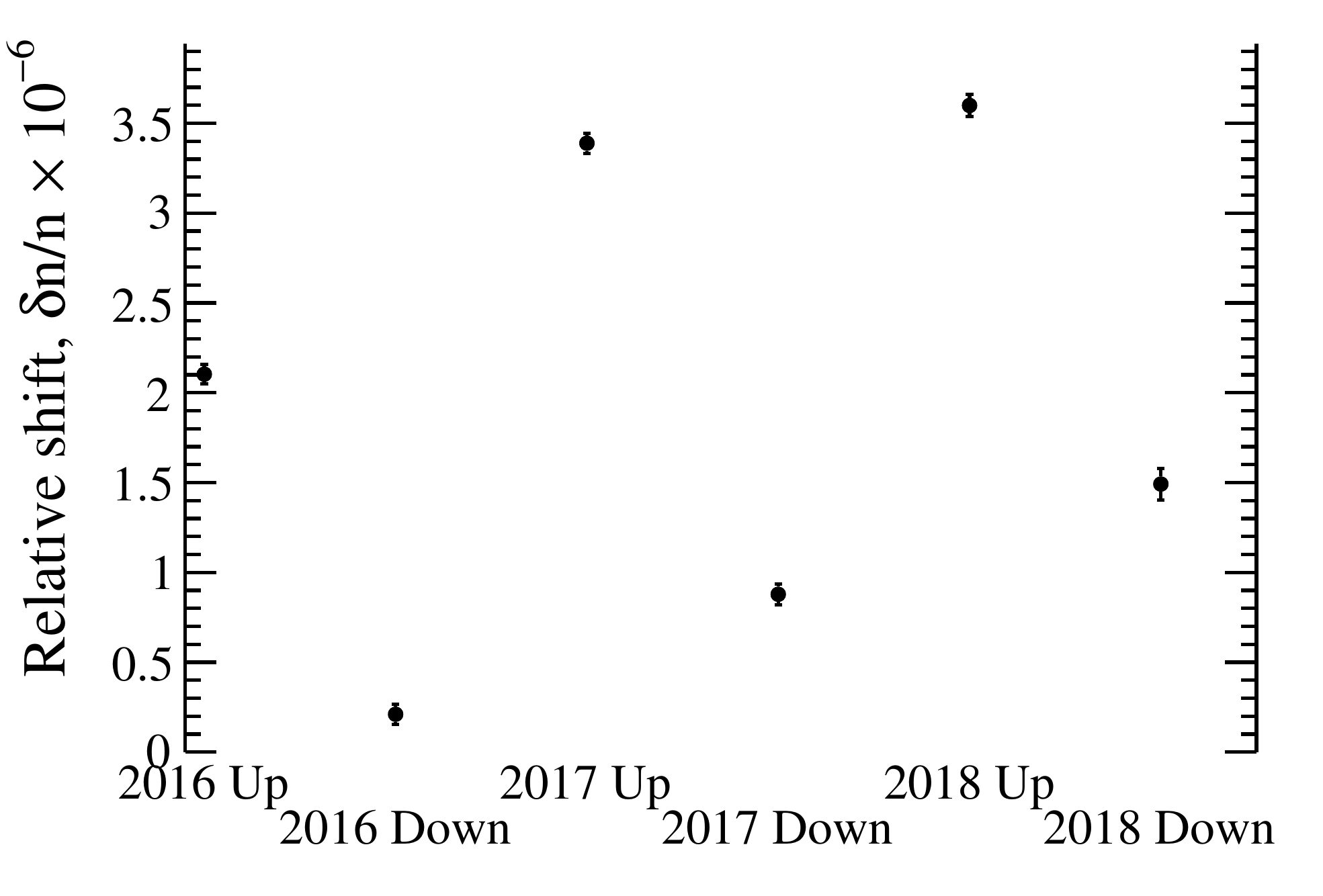}\put(-32,120){(a)}
    \includegraphics[width=0.49\linewidth,page=2]{figs/Fig8_9}\put(-32,120){(b)}\\
    \includegraphics[width=0.49\linewidth,page=5]{figs/Fig8_9}\put(-32,133){(c)}
    \includegraphics[width=0.49\linewidth,page=6]{figs/Fig8_9}\put(-32,133){(d)}
        \vspace*{-0.5cm}
  \end{center}
  \caption{
    The relative difference in the refractive index used in the reconstruction after the calibration, and the true refractive index of the radiators in (left) \richone and (right) \richtwo evaluated using the peak position of low-multiplicity dimuon data split by magnet state (a and b), and track charge (c and d).}
  \label{fig:ckRefIndexVPolarity}
\end{figure}

After correction, there is a residual difference between the refractive index used in the construction and the true refractive index. This difference is evaluated using the offset from zero of the signal peak of low-multiplicity dimuon data, selected only using information from the tracking and muon systems to avoid any bias from selecting tracks well identified in the RICH. The estimated residual difference as a function of time is shown in figure~\ref{fig:ckRefIndexVTime}. A small bias is observed in the refractive index after correction, but this results in angular deviations an order of magnitude smaller than those due to the HPD pixel size, and therefore is negligible. The stability of the correction is excellent over a period of several years. Figure~\ref{fig:ckRefIndexVPolarity} shows the residual difference split by magnet polarity, and track charge for each year. All differences are very small. The differences seen between magnet polarities follows the same pattern as seen in the resolutions determined in section \ref{sec:Resolution}, where the resolution can affect the determination of the calibration. The differences between years follows the same pattern as the photon yields determined in section \ref{sec:Yield}, which again can affect the ability to extract the calibration.

\begin{figure}[tb]
  \begin{center}
    \includegraphics[width=0.49\linewidth,page=1]{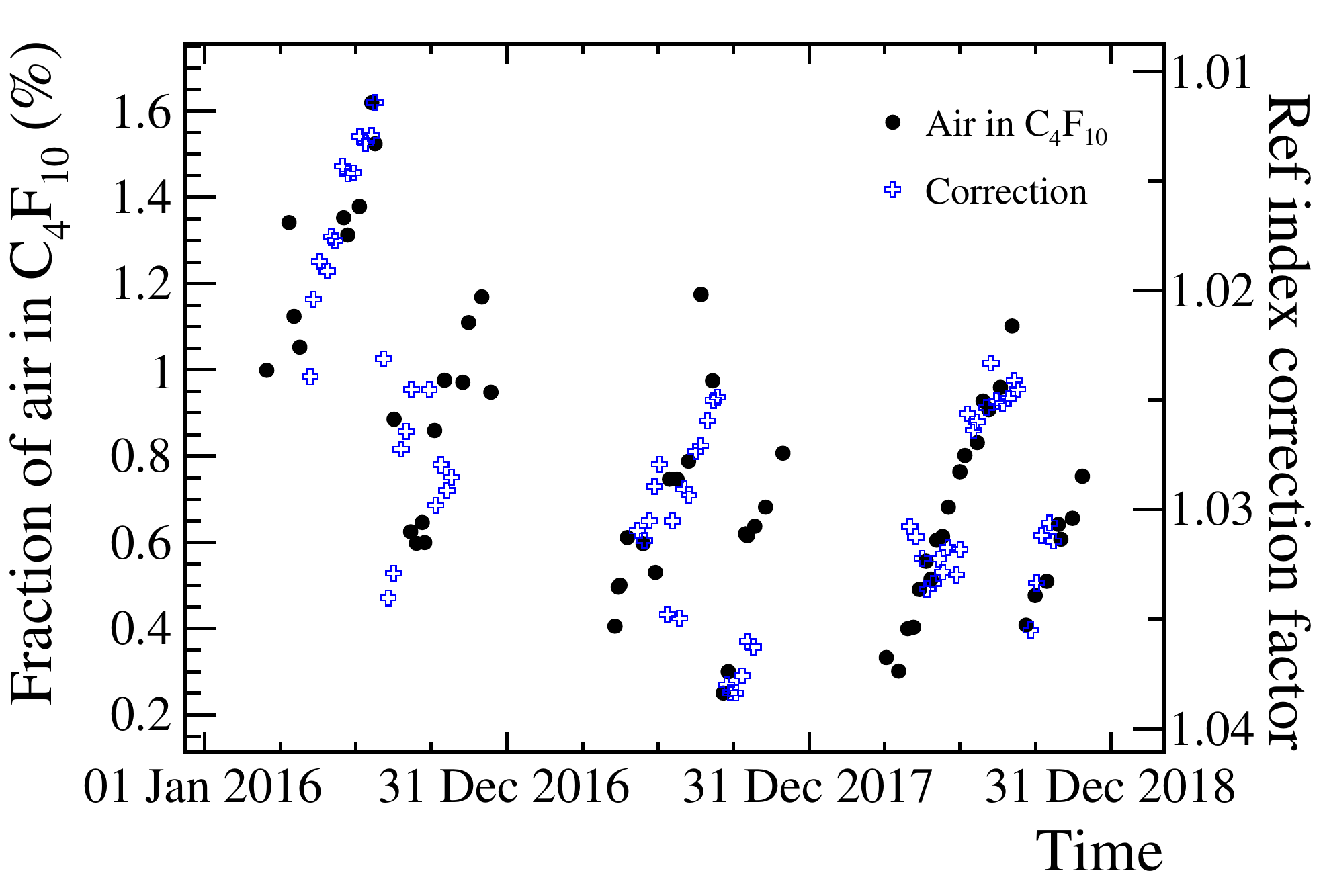}\put(-180,130){(a)}\hspace{5pt}\includegraphics[width=0.49\linewidth,page=12]{figs/Fig8_9}\put(-32,130){(b)}
        \vspace*{-0.5cm}
  \end{center}
  \caption{
    (a) The air contamination in \richone as a function of time (black circles), and the correction factor applied to the refractive index (blue crosses). The axis for the correction factor is reversed. (b) The relative difference in the refractive index used in the reconstruction after the calibration, and the true refractive index of the radiator in \richone for differing levels of contamination by air.}
  \label{fig:ckRefIndexVPurity}
\end{figure}

The gas system in \richone is known to have a slowly increasing contamination with air. The gas is regularly cleaned to keep the contamination below 2\,\%. The level of contamination and the associated correction factor are shown in figure~\ref{fig:ckRefIndexVPurity}, along with the corresponding effect on the residual of the refractive index determination for different levels of contamination. No significant bias is seen as a function of this contamination. The bias is not expected to be constant as a function of the contamination with air. The refractive index correction procedure was tuned using data taken over a short time period, therefore not exhibiting the full range of contamination. The air contamination in the gas radiator of \richtwo has been kept below 2\,\% over 10 years without the need to purify the gas.

\subsection{HPD photocathode image position}
\label{subsec:HPDImage}
Charging and temperature effects in the HPD can cause the image of the photocathode projected on to the silicon sensor to move slowly with time. The image the circular photocathode makes on the square pixel array from an accumulated set of events is sufficiently circular that it is well described by a centre point and radius. This image drift is compensated for by fitting a circle to the distribution of hits in an HPD, integrated over approximately one hour, to locate the projected centre point of the photocathode. The difference between this and the centre of the silicon sensor is applied as a correction to individual photon hits. Before fitting, the distribution is cleaned and smoothed to remove artefacts and to compensate for the non-uniform illumination across an HPD. A Sobel filter~\cite{sobelarticle} is applied to enhance the edge of the photocathode image.

\begin{figure}[tb]
  \begin{center}
    \includegraphics[width=0.32\linewidth,page=17]{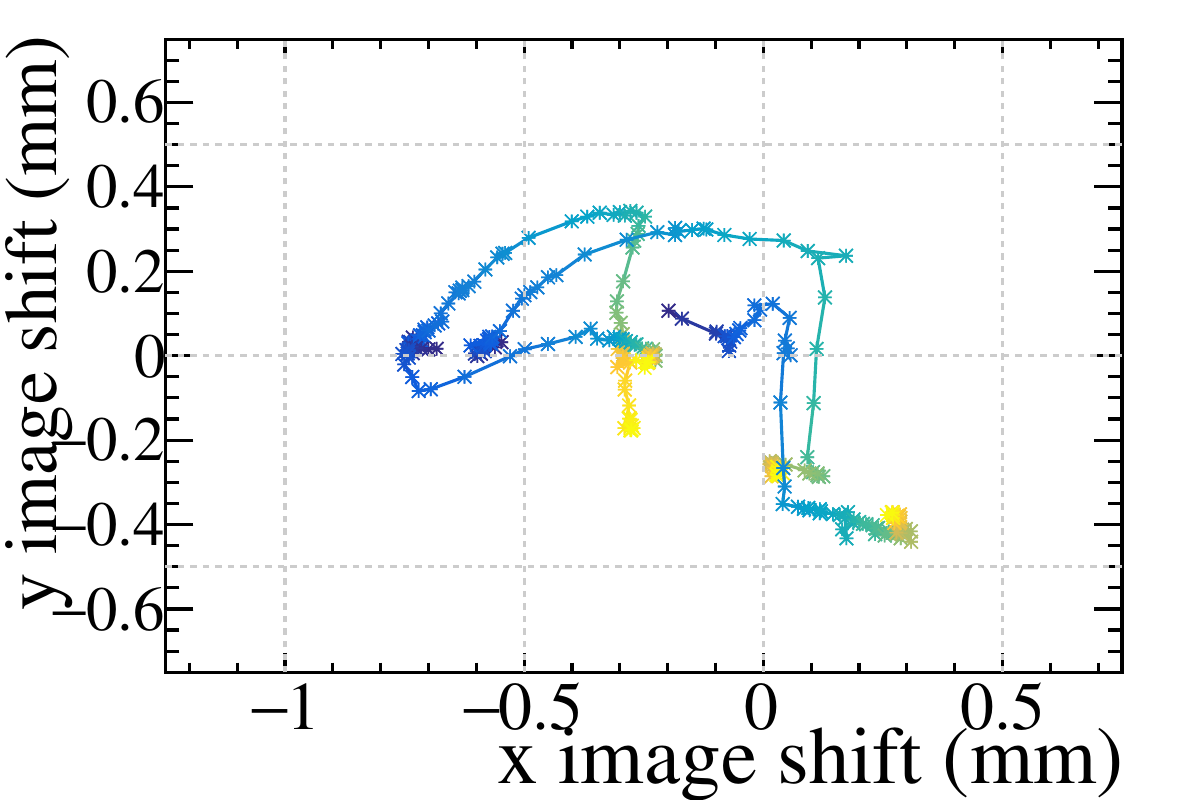}     \includegraphics[width=0.32\linewidth,page=34]{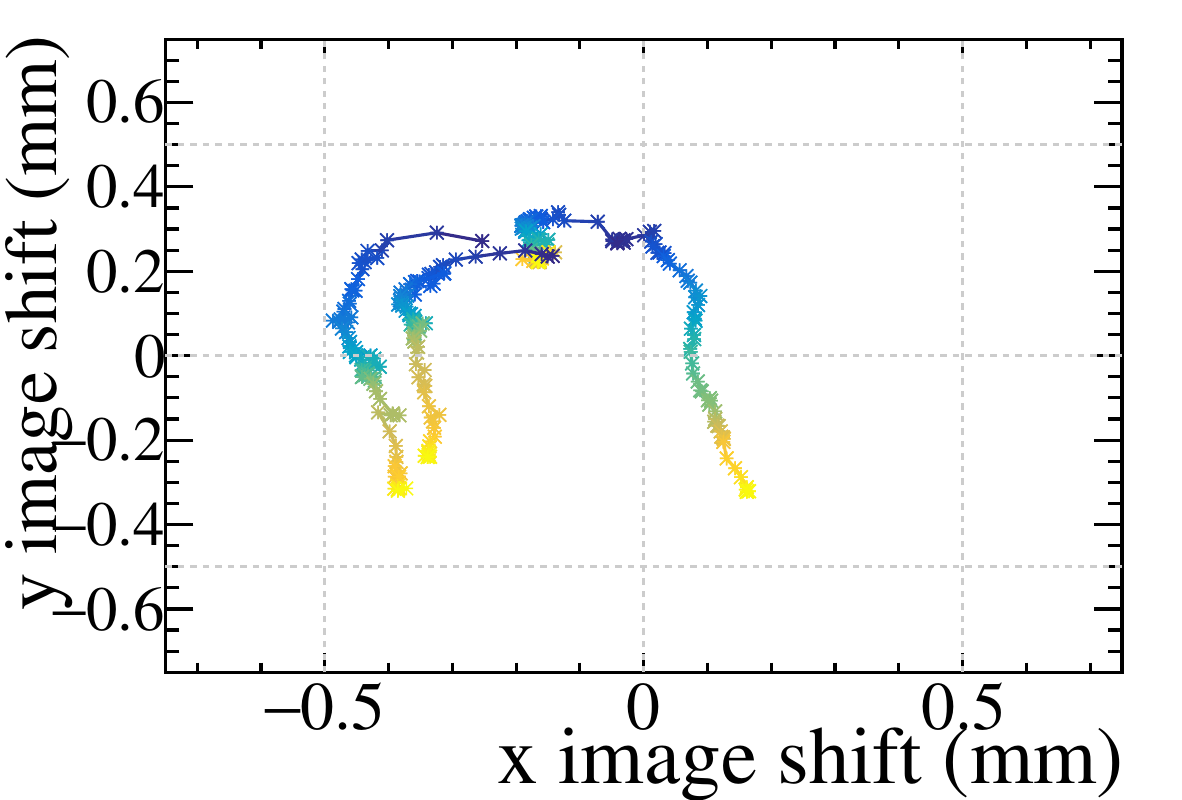}     \includegraphics[width=0.32\linewidth,page=36]{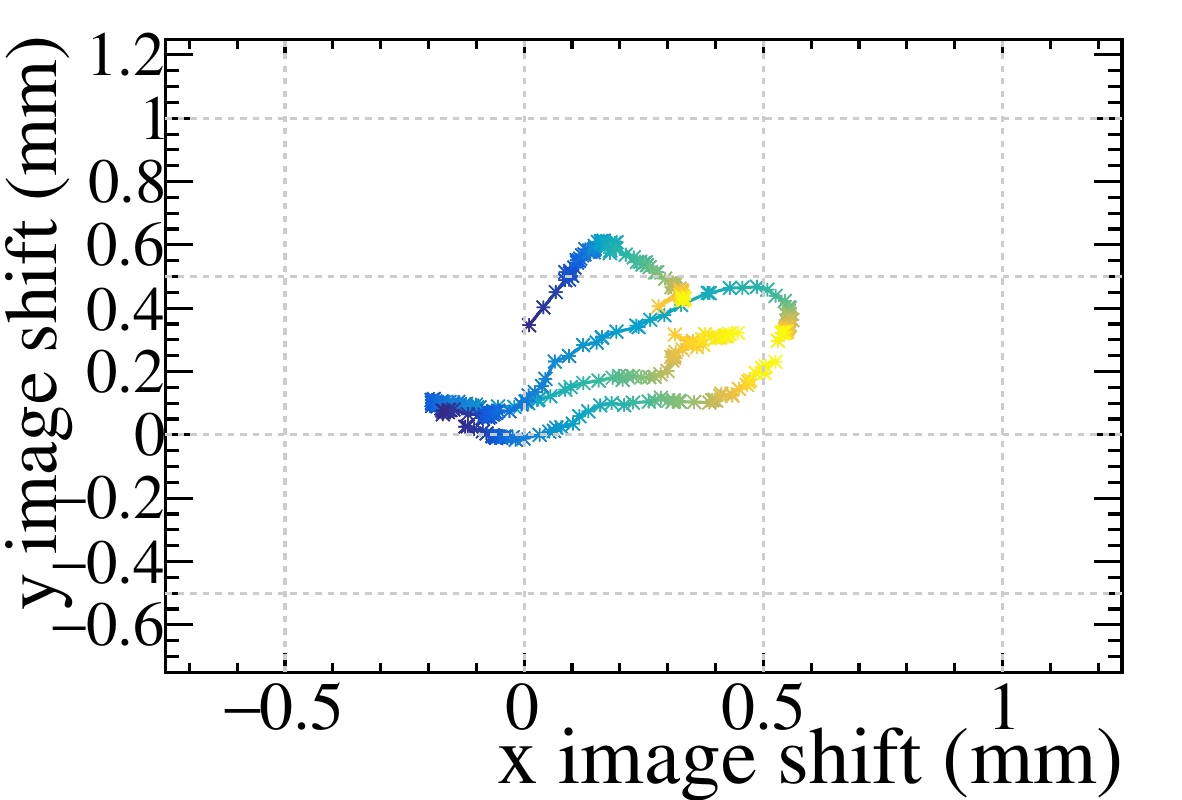}\\
    \includegraphics[width=0.32\linewidth,page=55]{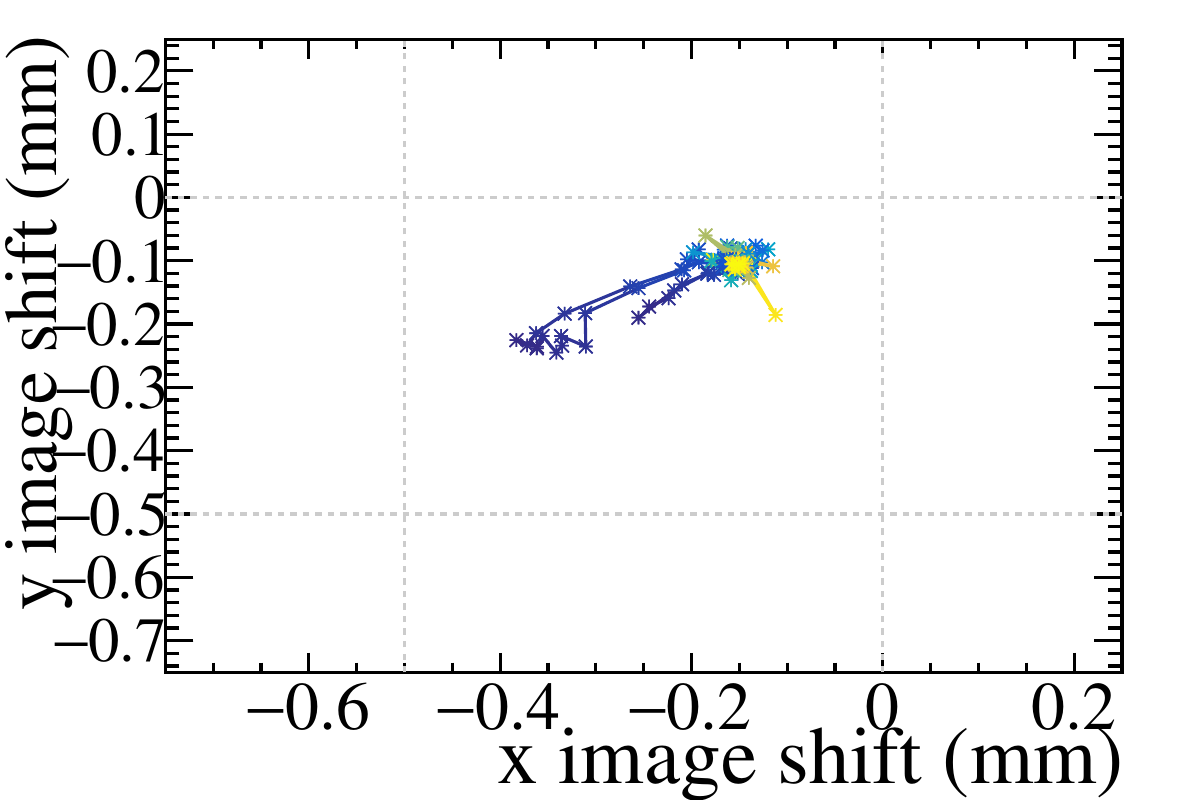}     \includegraphics[width=0.32\linewidth,page=61]{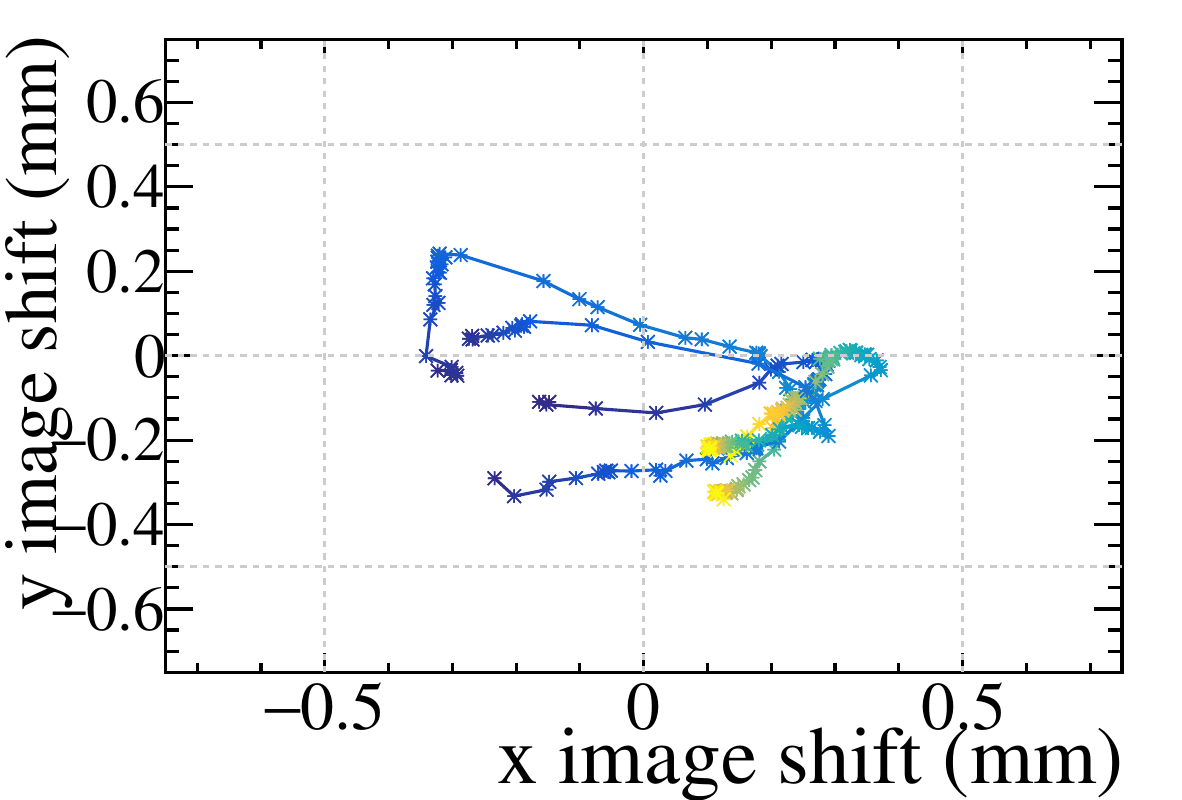}     \includegraphics[width=0.32\linewidth,page=75]{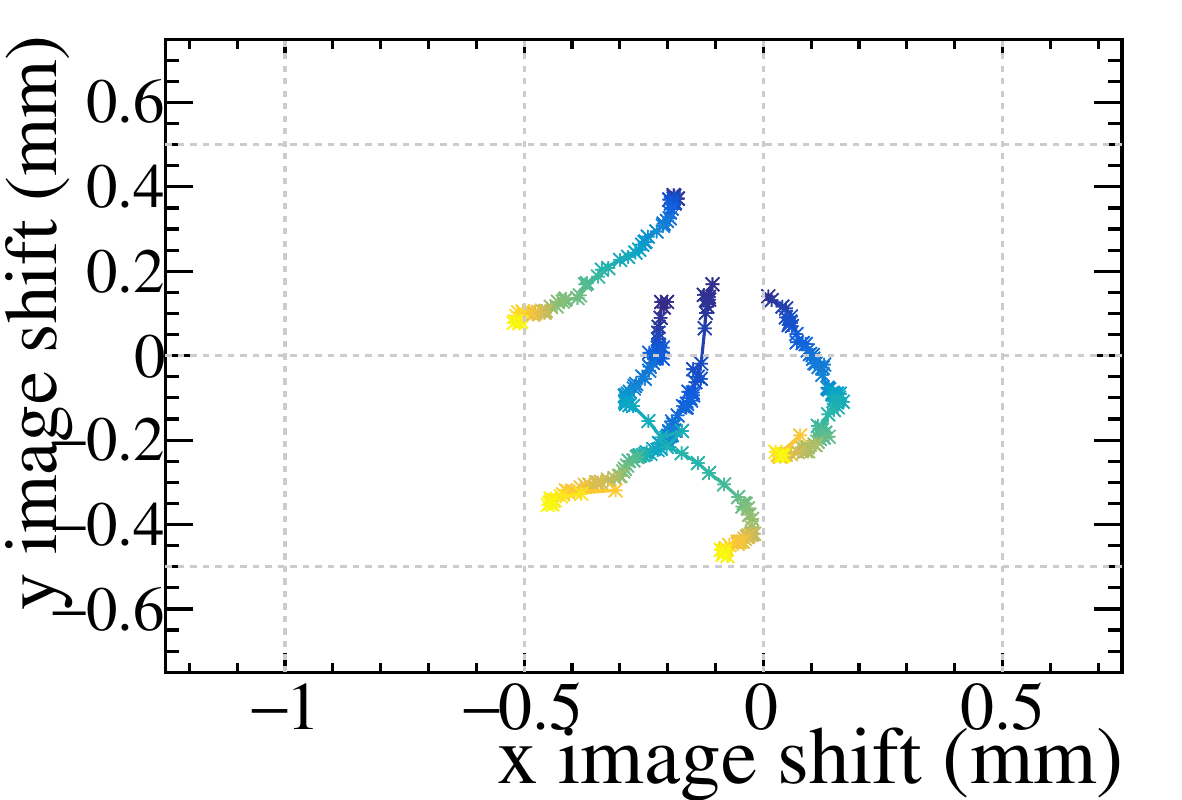}\\
    \includegraphics[width=0.32\linewidth,page=87]{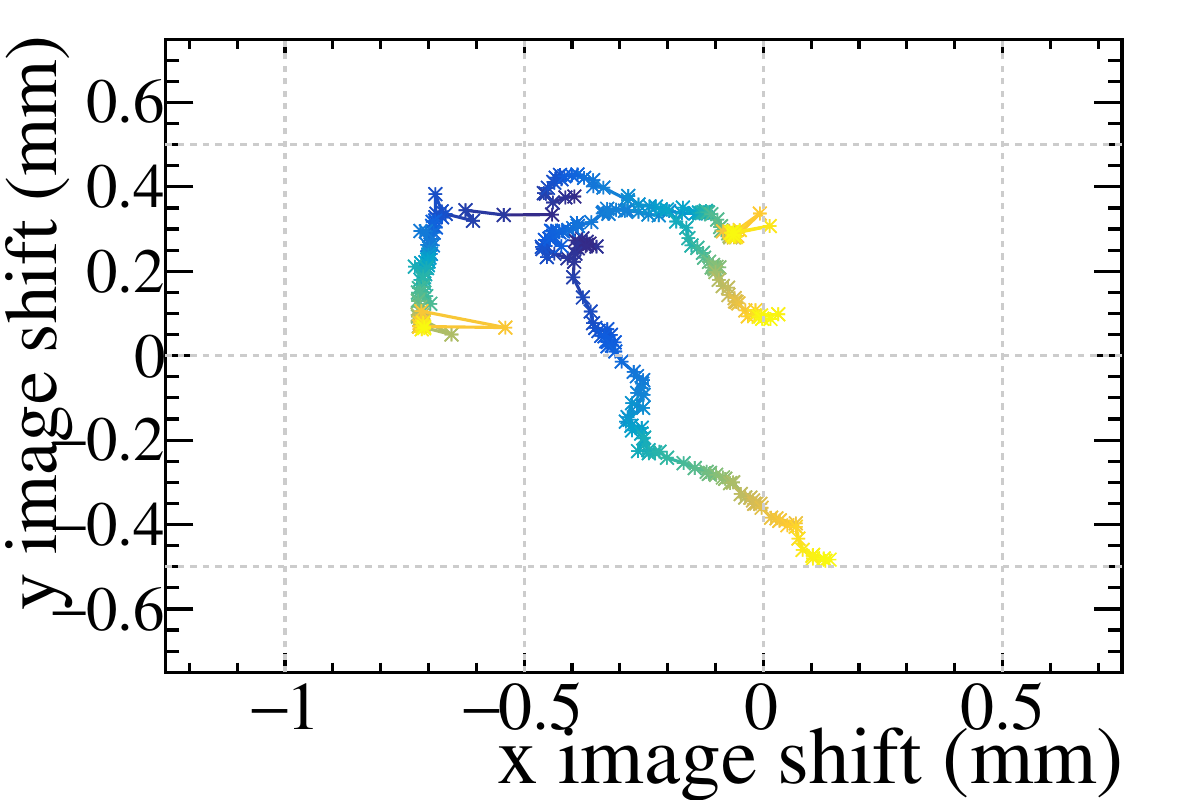}     \includegraphics[width=0.32\linewidth,page=424]{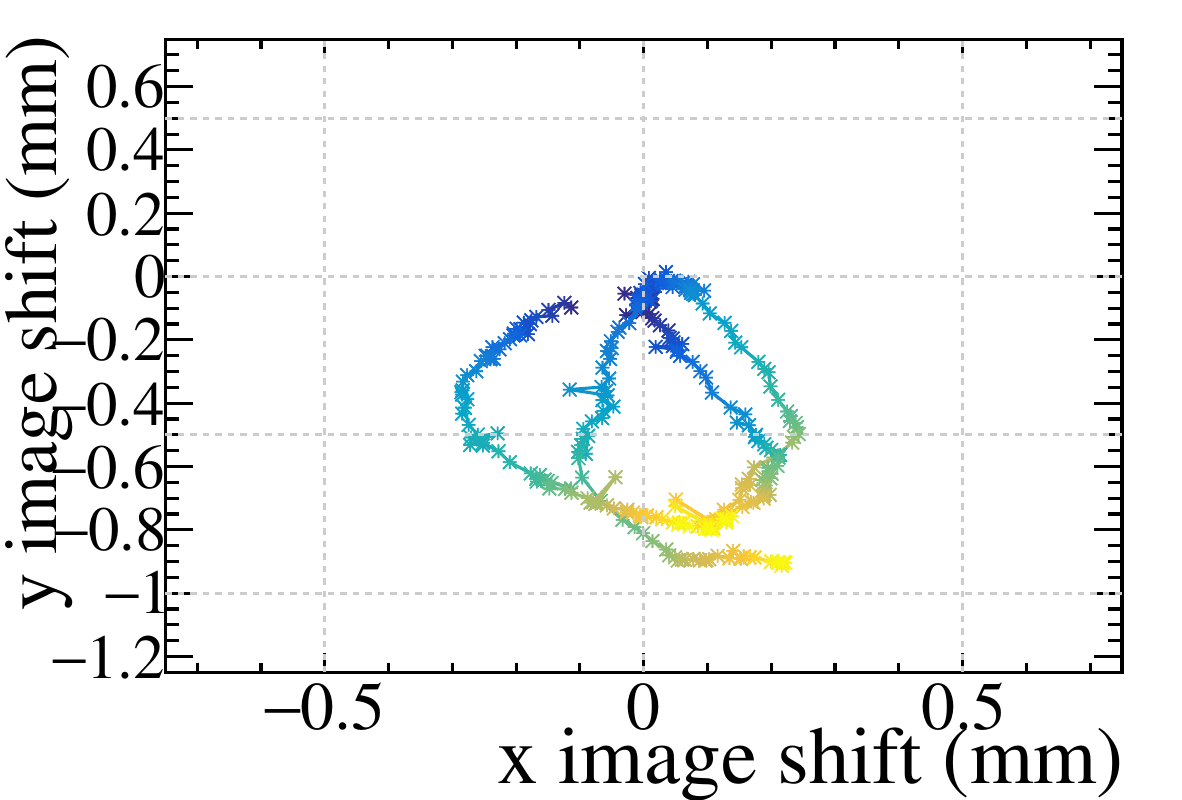} \includegraphics[width=0.32\linewidth,page=475]{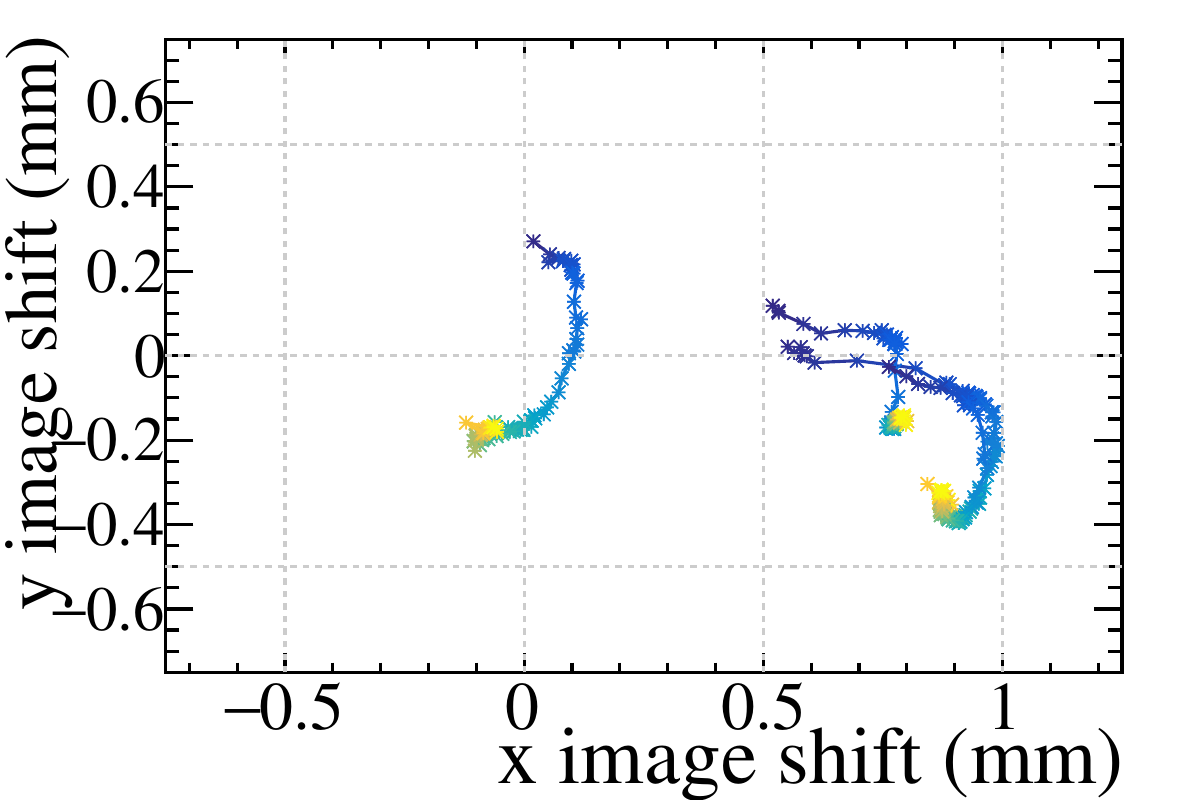}\\
    \vspace*{-0.5cm}
  \end{center}
  \caption{
    The paths taken by the centre point of the image of the photocathode of nine typical HPDs measured as displacement from the centre of the silicon chip, with four paths per HPD taken from four day-long fills of the LHC. The transitioning of the colour of the paths indicates the passage of time, with blue closest to the HPDs being switched on, and yellow the end of data taking. Dashed lines delimit pixels. A large range of behaviours is observed.}
  \label{fig:imageShifts}
\end{figure}

The position of the centre point of the images of nine HPDs' photocathodes as a function of time is shown separately for four fills of the LHC in figure~\ref{fig:imageShifts}. There is a range of behaviour, with some HPDs showing significant movements compared to the size of a pixel (0.5~mm $\times$ 0.5~mm).

\begin{figure}[tb]
  \begin{center}
    \includegraphics[width=0.49\linewidth,page=1]{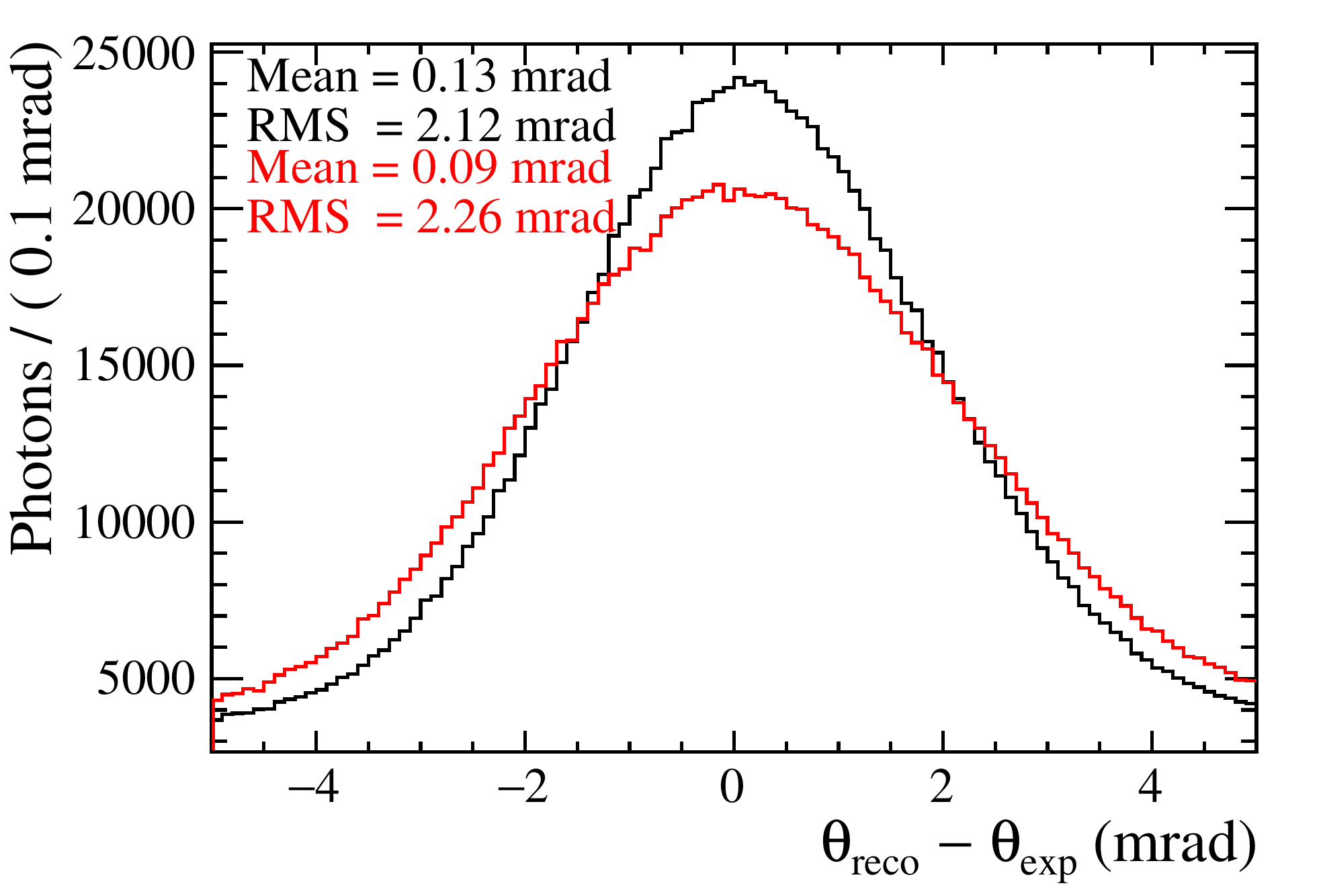}\put(-32,120){(a)} \includegraphics[width=0.49\linewidth,page=2]{figs/Fig12}\put(-32,120){(b)}
    \vspace*{-0.5cm}
  \end{center}
  \caption{
    The distribution of the difference of reconstructed and expected Cherenkov angles in (a) \richone and (b) \richtwo. The data is shown for reconstruction with corrections for the HPD photocathode image shifts applied (black) and ignored (red). A significant improvement is seen in the width of the distributions, and hence the Cherenkov angle resolutions, when applying the corrections.}
  \label{fig:imageShiftsRes}
\end{figure}

Figure~\ref{fig:imageShiftsRes} shows the change in angular resolution for both \richone and \richtwo without corrections for the HPD image shifts, and when applying corrections. A significant degradation of approximately 20\% in the resolution is observed in both RICH detectors when no corrections are applied.

\section{Cherenkov angle resolution}
\label{sec:Resolution}

The single photon resolution of both RICH detectors is measured on high momentum tracks. 
Photons emitted from the tracks used to calculated the prediction form a peak around zero in the distribution of the difference of reconstructed and expected Cherenkov angles. Photons from other tracks or random noise form a background below the peak, where in a typical $pp$ collision there is an occupancy of 1\% averaged across the detector plane in RICH1 and 0.6\% in RICH2. To obtain the resolutions a fit to this distribution is performed, modelling the correctly matched photons with a Gaussian distribution and the background as a second order polynomial. The resolution is taken to be the width of the Gaussian distribution.

\begin{figure}[tb]
  \begin{center}
    \includegraphics[width=0.49\linewidth,page=1]{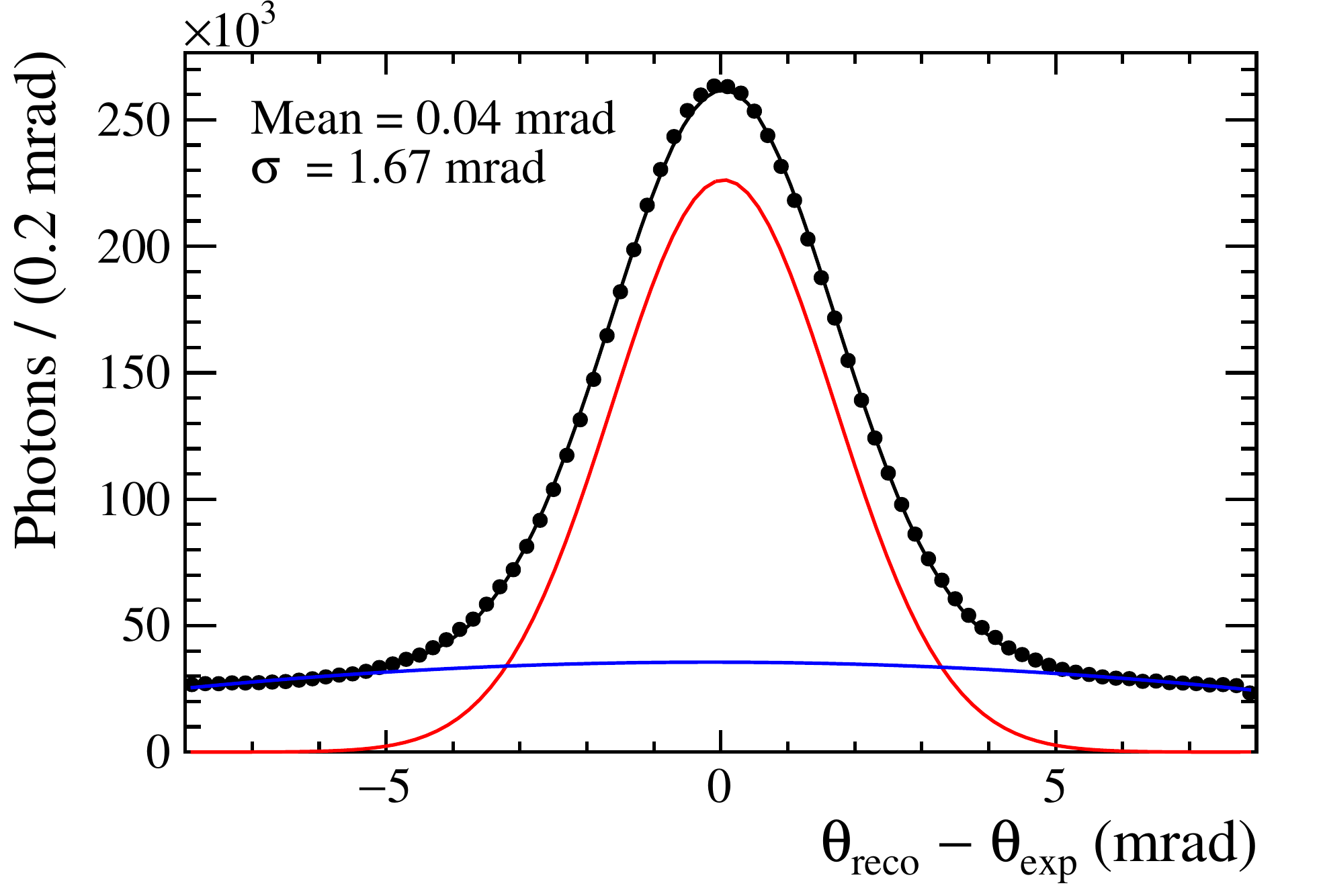}\put(-32,120){(a)}
    \includegraphics[width=0.49\linewidth,page=2]{figs/Fig13}\put(-32,120){(b)}
    \vspace*{-0.5cm}
  \end{center}
  \caption{
    Example distribution of photons detected in (a) \richone and (b) \richtwo from low track-multiplicity events. A fit to this distribution is overlaid, with a Gaussian signal component (red), polynomial background component (blue), and the total distribution (black)}
  \label{fig:lowmulti_resolution_dist}
\end{figure}

\begin{figure}[tb]
  \begin{center}
    \includegraphics[width=0.49\linewidth,page=7]{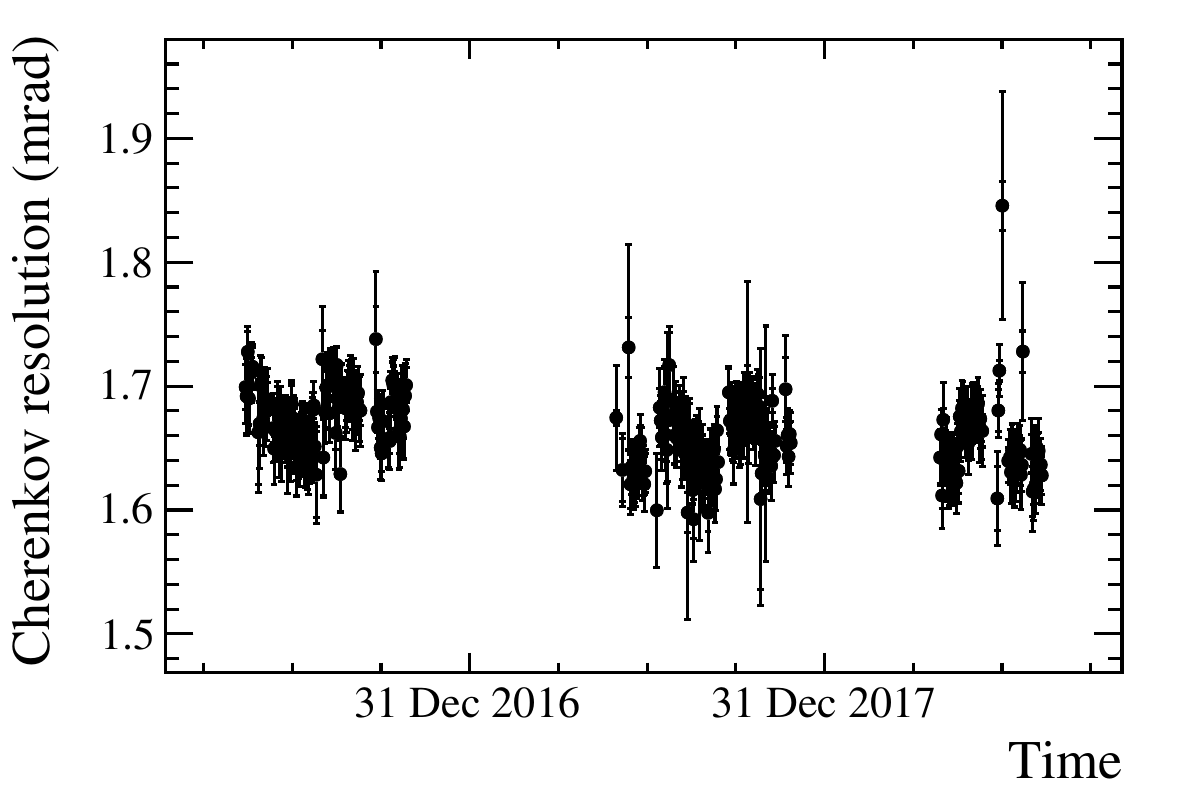}\put(-32,120){(a)}
    \includegraphics[width=0.49\linewidth,page=8]{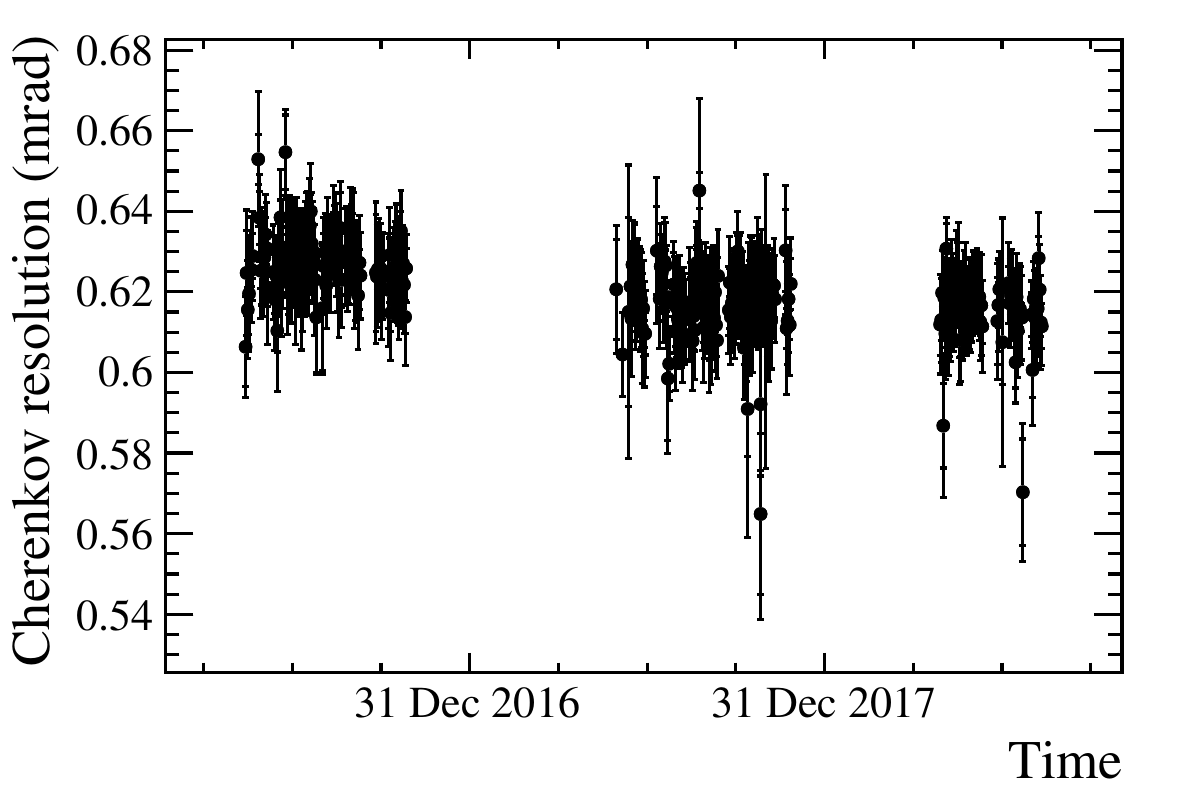}\put(-32,120){(b)}
    \vspace*{-0.5cm}
  \end{center}
  \caption{
    Resolutions for (a) \richone and (b) \richtwo measured between 2016 and 2018. Each data point is from low-multiplicity dimuon data accumulated over 24 hours. The uncertainties show individual bars for the statistical and systematic uncertainty, with the statistical uncertainty dominating.}
  \label{fig:lowmulti_resolution_timelines}
\end{figure}

\begin{figure}[tb]
  \begin{center}
    \includegraphics[width=0.49\linewidth,page=1]{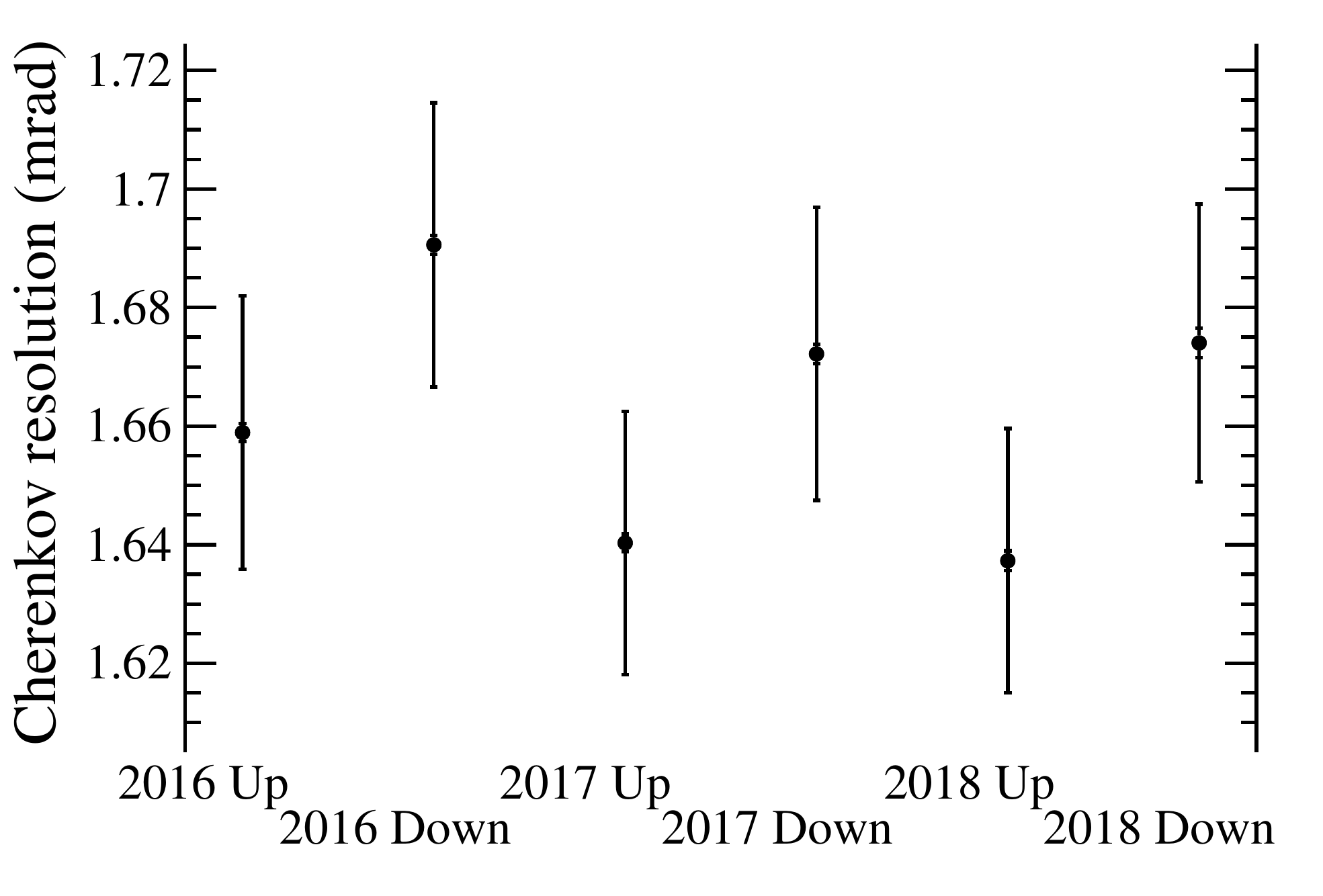}\put(-32,133){(a)}
    \includegraphics[width=0.49\linewidth,page=2]{figs/Fig14_15}\put(-32,133){(b)}\\
    \includegraphics[width=0.49\linewidth,page=5]{figs/Fig14_15}\put(-32,133){(c)}
    \includegraphics[width=0.49\linewidth,page=6]{figs/Fig14_15}\put(-32,133){(d)}
    \vspace*{-0.5cm}
  \end{center}
  \caption{
    Resolutions for (a) \richone and (b) \richtwo measured per year and per magnet polarity, and for (c) \richone and (d) \richtwo measured per year and per track charge integrated over data taken with each magnet polarity. The uncertainties are dominated by the systematic uncertainty arising from the background shape.}
  \label{fig:lowmulti_resolution_splits}
\end{figure}

The shape of the background peaks under the signal and tends to bias the resolutions obtained towards larger values. To get a reliable estimate of the absolute resolution, events with very-low charged-track multiplicity are used. This both reduces the relative amount of background and flattens the background shape under the signal peak. An example of the photon angle distribution from low-multiplicity events is shown in figure~\ref{fig:lowmulti_resolution_dist}. The systematic uncertainty from the background shape is estimated by taking the difference in resolution obtained from the nominal fit and one where the background is modelled as linear, as this has been shown to cover for a wide range of potential background shapes using simulation. The resolutions as a function of time are show in figure~\ref{fig:lowmulti_resolution_timelines}. The resolution for each year in Run~2 and in each magnet polarity are shown in figure~\ref{fig:lowmulti_resolution_splits}. The uncertainty is a combination of the uncertainty obtained from the fit and the background systematic, dominated by the latter. In \richone there does appear to be a small dependence on magnet polarity, with the resolution of the \textit{Up} polarity being slightly better than the \textit{Down}. There are several potential sources of differences between polarities in the detector, with the largest effects thought to come from differences in the magnetic distortion correction and tracking resolution.

When both magnet polarities are averaged there should be very little difference in the performance between positively and negatively charge particles. The resolutions for each charge in each Run~2 year, integrated over magnet polarities, is shown in figure~\ref{fig:lowmulti_resolution_splits}, where the resolutions determined for both charges are in excellent agreement.

Combining all run periods the Cherenkov angle resolution in \richone is \mbox{$1.662 \pm 0.023\mrad$}
and in \richtwo is 
$0.621 \pm 0.012\mrad$,
where, now, the statistical uncertainty is negligible and the systematic uncertainty is dominated by the shape of the background distribution. The measurements for both detectors are within 10\% of the expectation from simulation.

\section{Detected photon yield}
\label{sec:Yield}

\begin{figure}[tb]
  \begin{center}
    \includegraphics[width=0.49\linewidth,page=1]{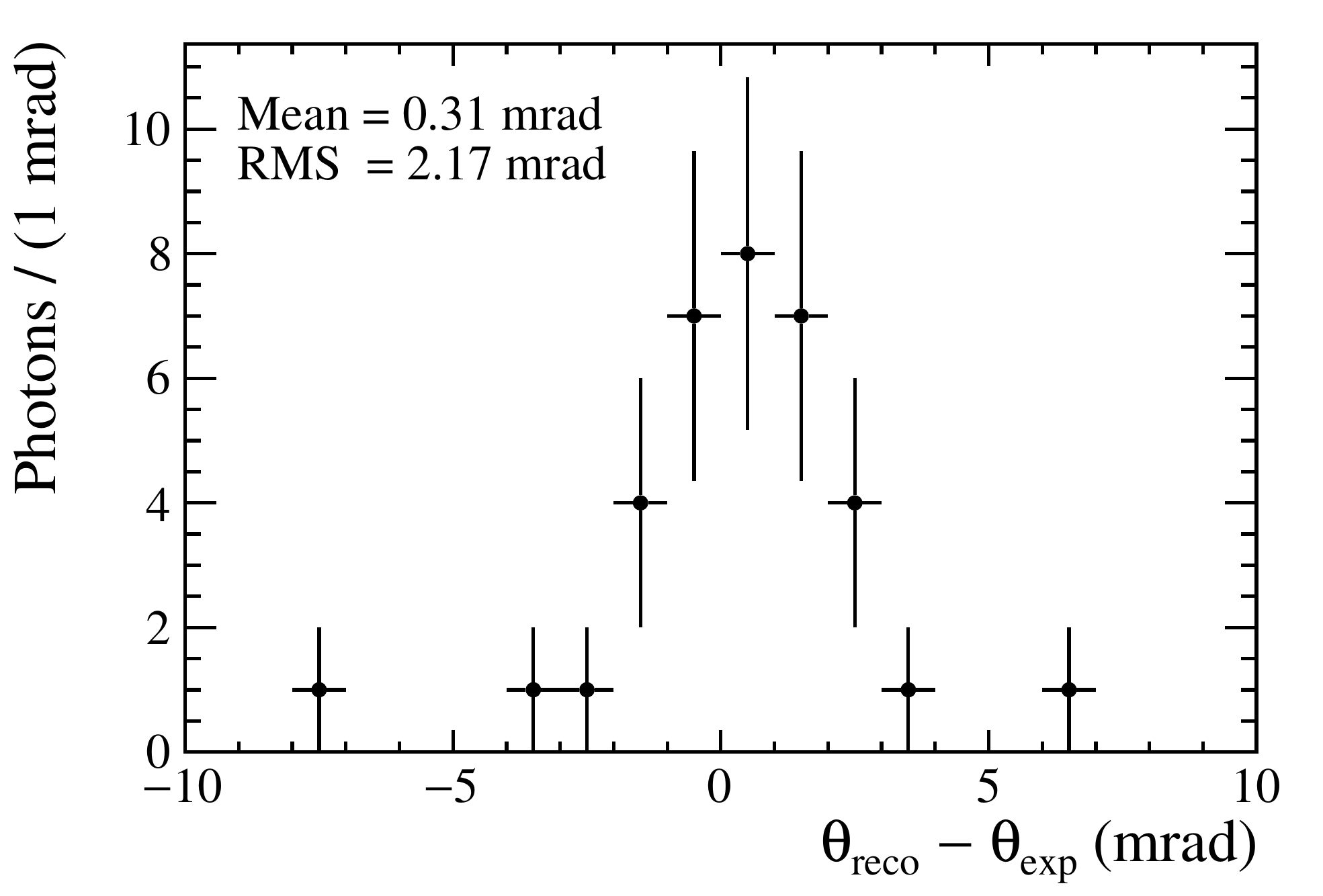}\put(-32,120){(a)}
    \includegraphics[width=0.49\linewidth,page=2]{figs/Fig16}\put(-32,120){(b)}
    \vspace*{-0.5cm}
  \end{center}
  \caption{
    Example distribution of the difference between reconstructed and expected Cherenkov angle for photons detected around a typical track in a low-multiplicity event for (a) RICH\,1 and (b) RICH\,2.}
  \label{fig:example_photon_track}
\end{figure}

The number of detected Cherenkov photons emitted per track is determined from low track-multiplicity events. The determination is performed per track, where a simple linear background subtraction is applied to the Cherenkov angle distribution of reconstructed photons. The selected tracks are positively identified as muons by the muon system, with a high momentum that produces a ring with Cherenkov angle close to its saturation value, and falls within the geometric acceptance of the detector, the same selection used to study the refractive index. An example of the distribution of the difference between the reconstructed and expected Cherenkov angles for photons around a typical track is shown in figure~\ref{fig:example_photon_track}.

The determination of the photon yield is highly dependent on the shape of the background, which can peak close to the signal. Even with the low backgrounds, simulations show that the absolute value of the number of detected photons could be biased by up to 7\,\% depending on the exact source of the background. However this bias is not a strong function of yield, and is highly correlated between measurements, so is ignored for the following relative comparisons.

\begin{figure}[tb]
  \begin{center}
    \includegraphics[width=0.49\linewidth,page=1]{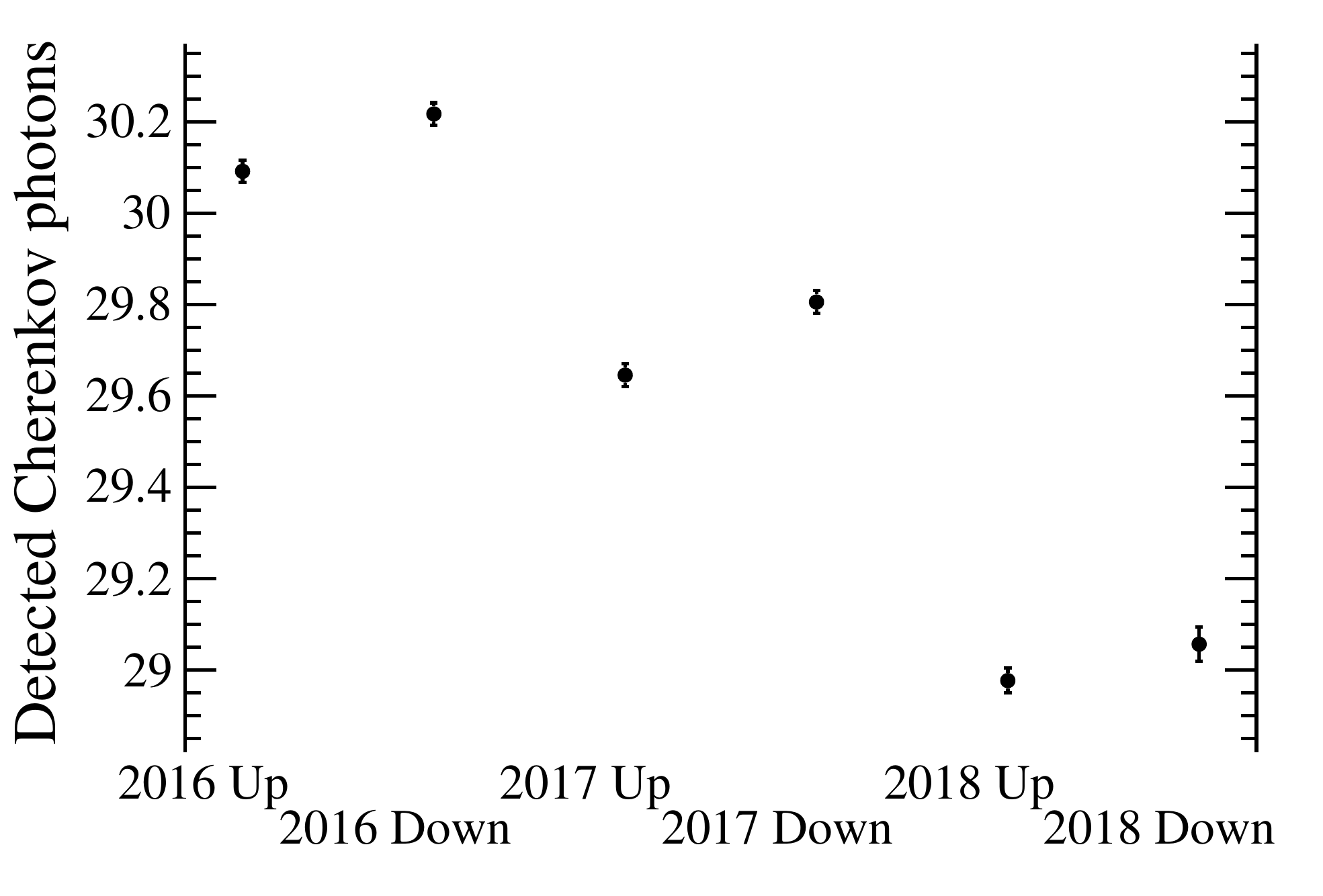}\put(-32,133){(a)}
    \includegraphics[width=0.49\linewidth,page=2]{figs/Fig17}\put(-32,133){(b)}\\
    \includegraphics[width=0.49\linewidth,page=5]{figs/Fig17}\put(-32,133){(c)}
    \includegraphics[width=0.49\linewidth,page=6]{figs/Fig17}\put(-32,133){(d)}
    \vspace*{-0.5cm}
  \end{center}
  \caption{
    Number of Cherenkov photons emitted per track for (a) \richone and (b) \richtwo measured per year and per magnet polarity, and (c) \richone and (d) \richtwo measured per year and per track charge. The uncertainty is statistical only.}
  \label{fig:photon_count}
\end{figure}

The background-subtracted mean number of detected Cherenkov photons for both \richone and \richtwo is shown in figure~\ref{fig:photon_count}. The number of detected photons slowly decreases each year as may be expected by ageing of the photon detectors. 

The number of detected photons per track for each charge, integrated over both magnet polarities, is shown in figure~\ref{fig:photon_count}. In general there is very good agreement between the number of detected photons for both charges, although a small charge asymmetry is seen for \richtwo in 2018 and a smaller asymmetry is seen in 2017. The cause of this asymmetry is not fully understood, but it is small enough to not significantly affect the performance.

The average detected photon yield for the period 2016--2018, determined from a fit to the integrated data set, in \richone is $30\pm2$ and in \richtwo is $18.5\pm1.2$, where the uncertainty is dominated by the background shape bias. This is in good agreement with expectations from simulation.

\newpage
\section{PID performance}
\label{sec:Performance}

The quality of the detector calibration, and the values of the Cherenkov angle resolution and photon yields all impact the performance of physics selections that utilise the RICH particle identification $\Delta \text{LL}$ information to identify the signal of interest and to reject backgrounds. The ultimate performance of the RICH system can be determined in terms of the ability to discriminate between the species of charged hadrons and also in the impact of those selections on the physics output of the experiment.

\subsection{Efficiency versus purity}
\label{subsec:PerformanceCurves}

The efficiency of the system for discriminating between different species of hadrons can be determined from control samples of well-identified particles obtained purely through kinematic selections. These calibration samples are selected without using the RICH information in order to not bias the results. High-purity samples of charged kaons and pions are obtained from the decay products of the $D^0$ meson identified in the decay chain $D^{*+}\to D^0(\to K^- \pi^+)\pi^+$ and the charge-conjugated decays. Samples of protons are obtained from $\Lambda^0\to p\pi^-$ decays, and the charge-conjugate decay. Details of the selection of the calibrations samples are given in ref.~\cite{pidcalibpaper}. 

The efficiency of a particular $\Delta \text{LL}$-based PID selection on a probe hadron (pion, kaon or proton) is computed by performing an unbinned maximum likelihood fit to the invariant mass distributions of the parent particle in the appropriate samples, allowing the separation of the decay candidates from any residual background. The fit is performed with and without the associated PID selection applied to the particle of interest, and the efficiency computed from the ratio of yields of candidate decays between the two fits. The PID selection is scanned over a range of loose to tight $\Delta \text{LL}$ selections, and a profile of selection efficiencies (\eg kaons identified as kaons) and associated misidentified particle leakage (\eg pions misidentified as kaons) obtained. The separation between kaons and pions, proton and pions, protons and kaons, is obtained by using the information provided by the $\Delta \text{LL}(K,\pi)$, $\Delta \text{LL}(p,\pi)$, $\Delta \text{LL}(p,K)$ distributions, respectively. 

Trigger selections vary across the different data-taking periods, and the kinematics of the $D^0$ and $\Lambda^0$ decay products are significantly different. Since the RICH performance is a function of the momentum, pseudorapidity and activity in the detectors, the latter being parametrised by the number of reconstructed tracks in the LHCb acceptance, a weighting procedure is used to equalise the distributions of these variables between different species of hadrons and across data-taking periods. The momentum range is chosen such that under at least one of the particle hypotheses compared the particle would be above the velocity threshold for generating Cherenkov radiation. Pseudorapidity ranges are chosen to match the RICH detectors' acceptances. The weighting in the number of tracks cancels most of the differences in data-taking conditions and trigger thresholds. 

\begin{figure}[p!]
  \begin{center}
    \includegraphics[width=0.49\linewidth,page=1,trim=0 0 50 0,clip]{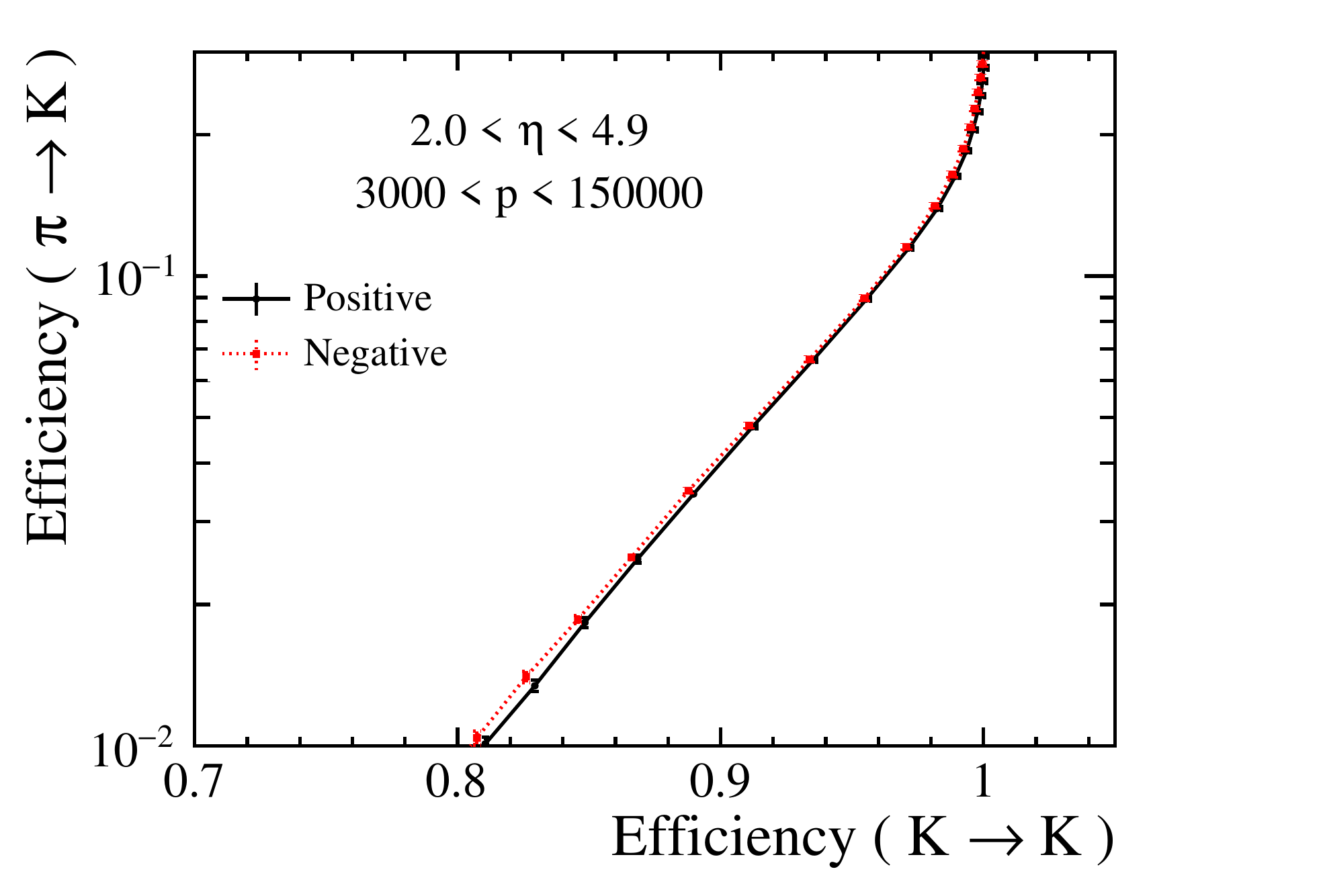}\put(-35,133){(a)}
    \includegraphics[width=0.49\linewidth,page=1,trim=0 0 50 0,clip]{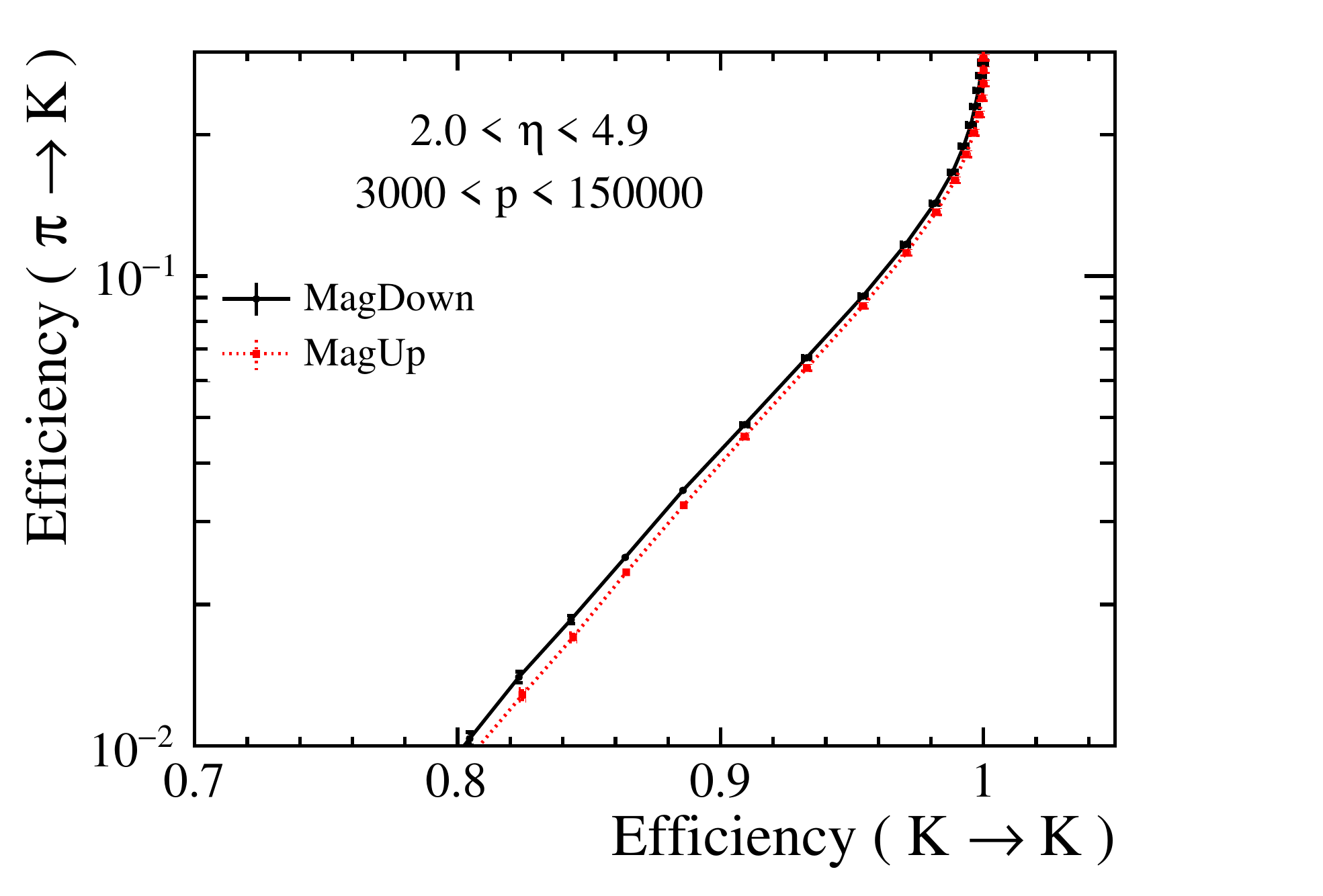}\put(-35,133){(b)}\\
    \includegraphics[width=0.49\linewidth,page=1,trim=0 0 50 0,clip]{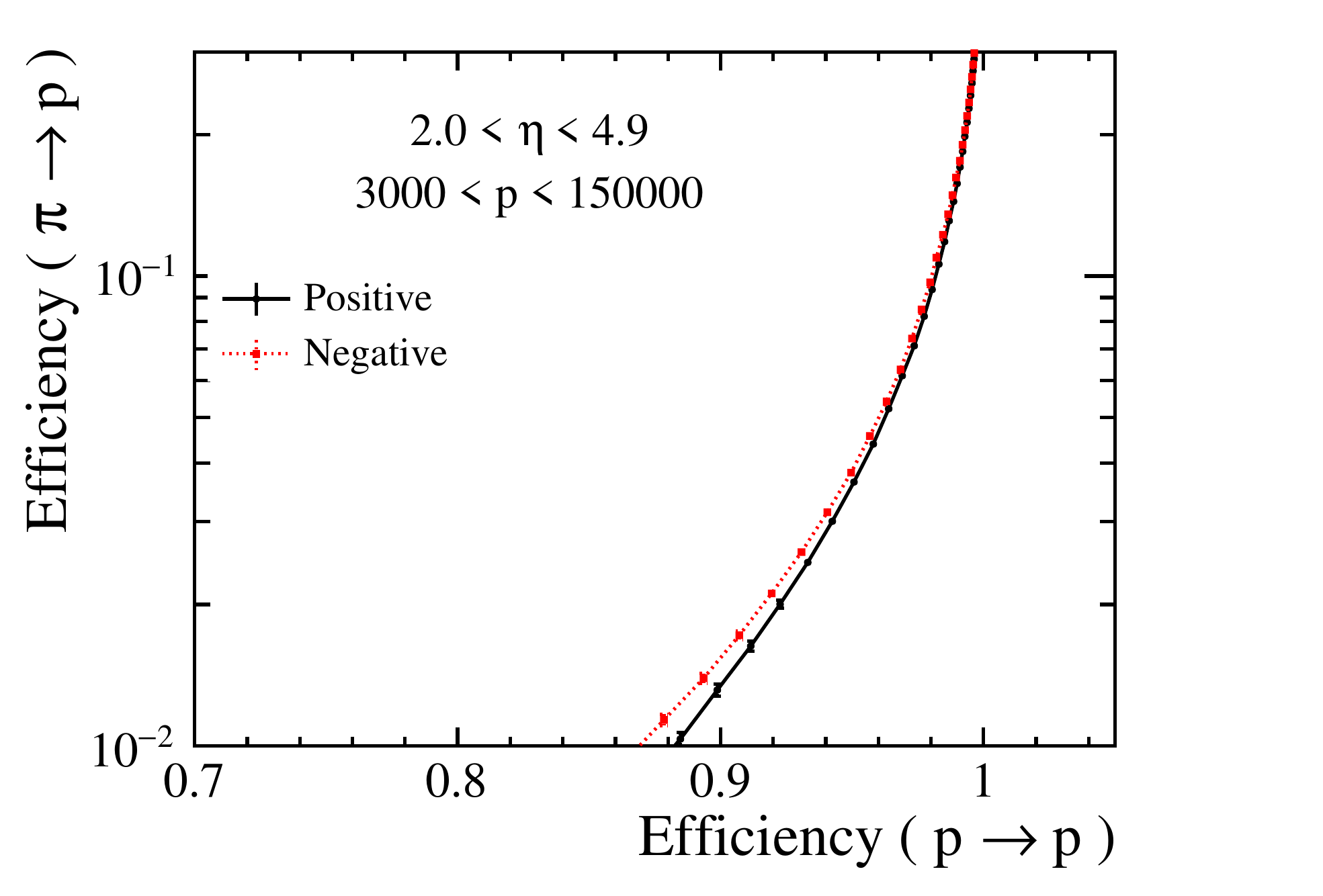}\put(-35,133){(c)}
    \includegraphics[width=0.49\linewidth,page=1,trim=0 0 50 0,clip]{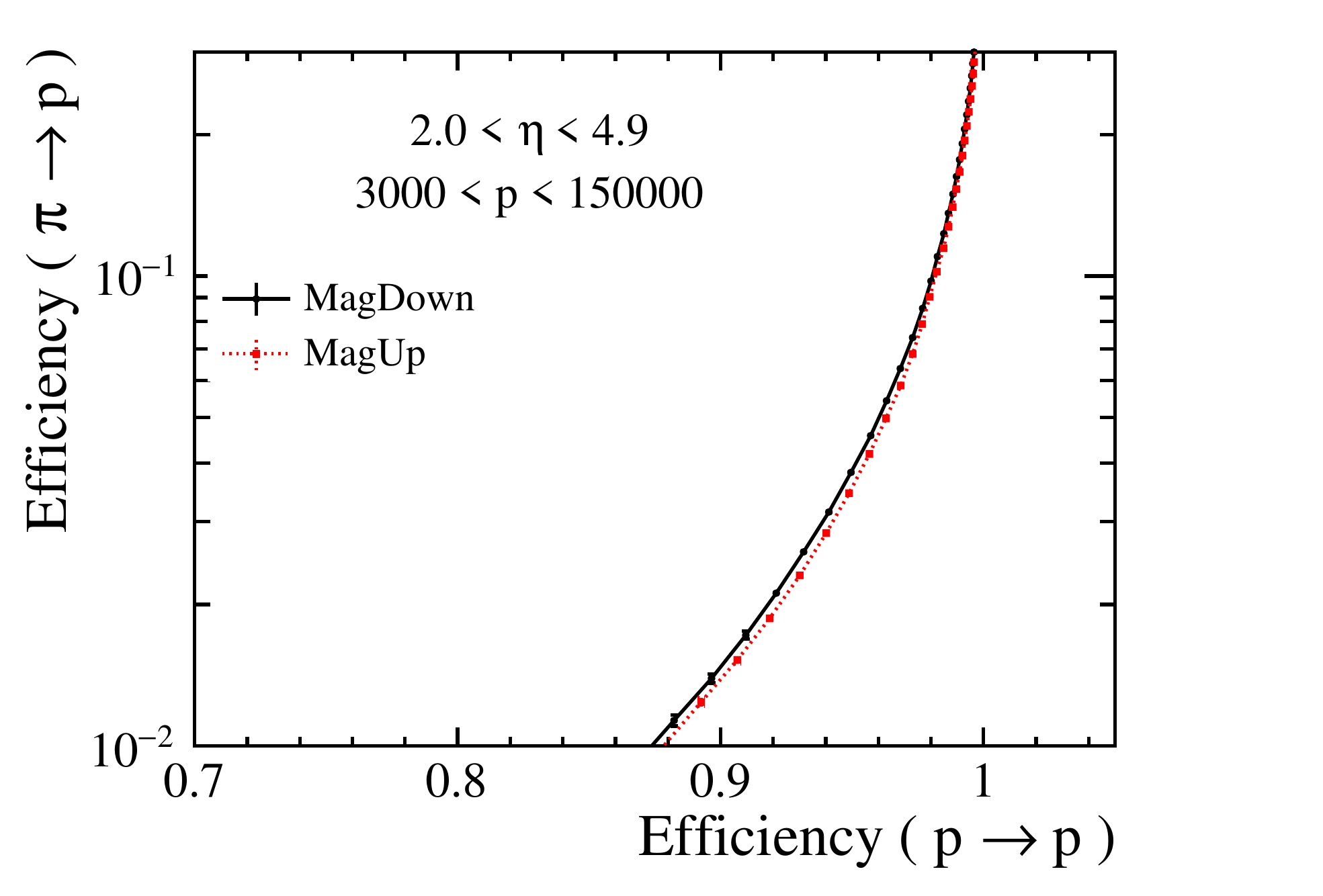}\put(-35,133){(d)}\\
    \includegraphics[width=0.49\linewidth,page=1,trim=0 0 50 0,clip]{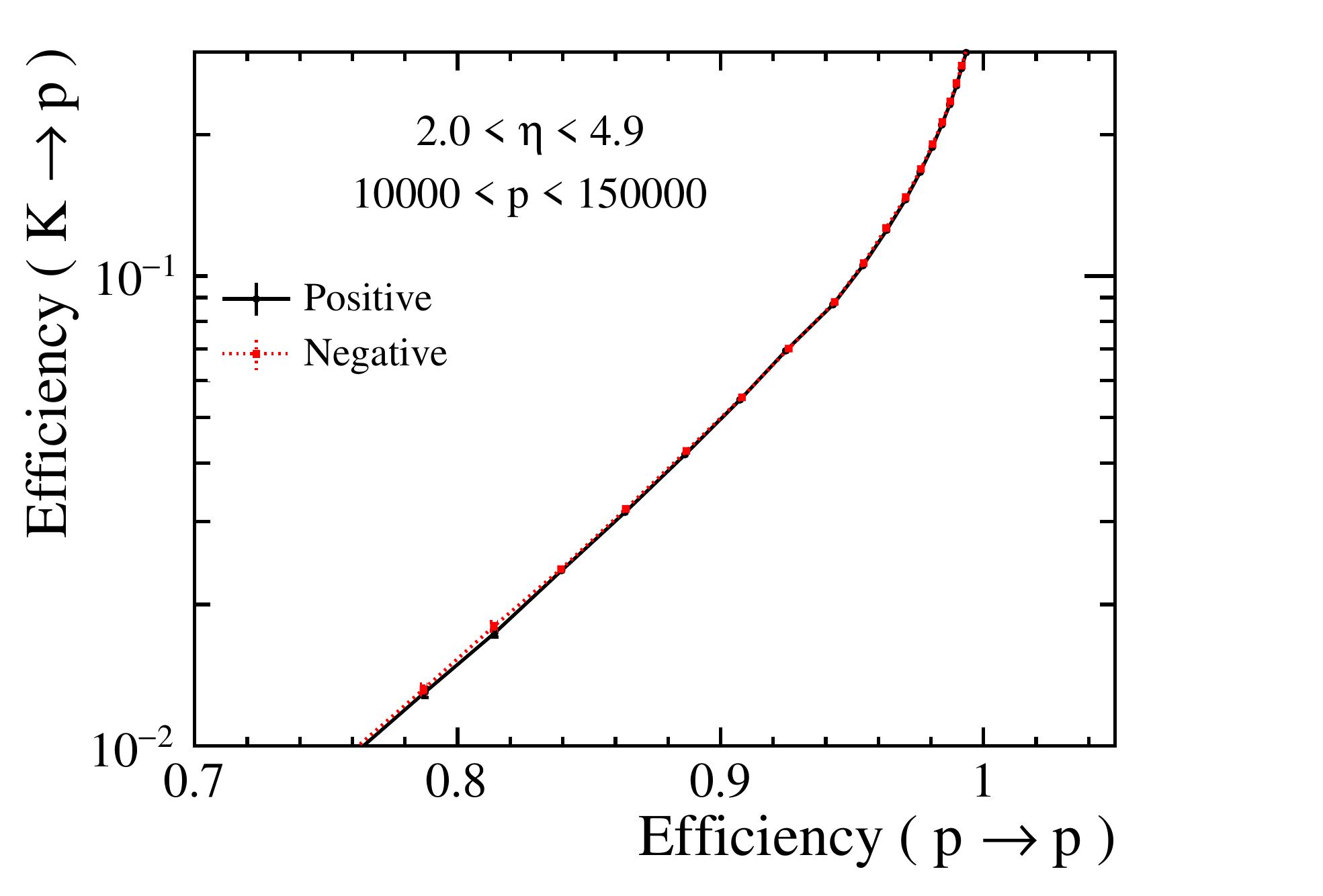}\put(-35,133){(e)}
    \includegraphics[width=0.49\linewidth,page=1,trim=0 0 50 0,clip]{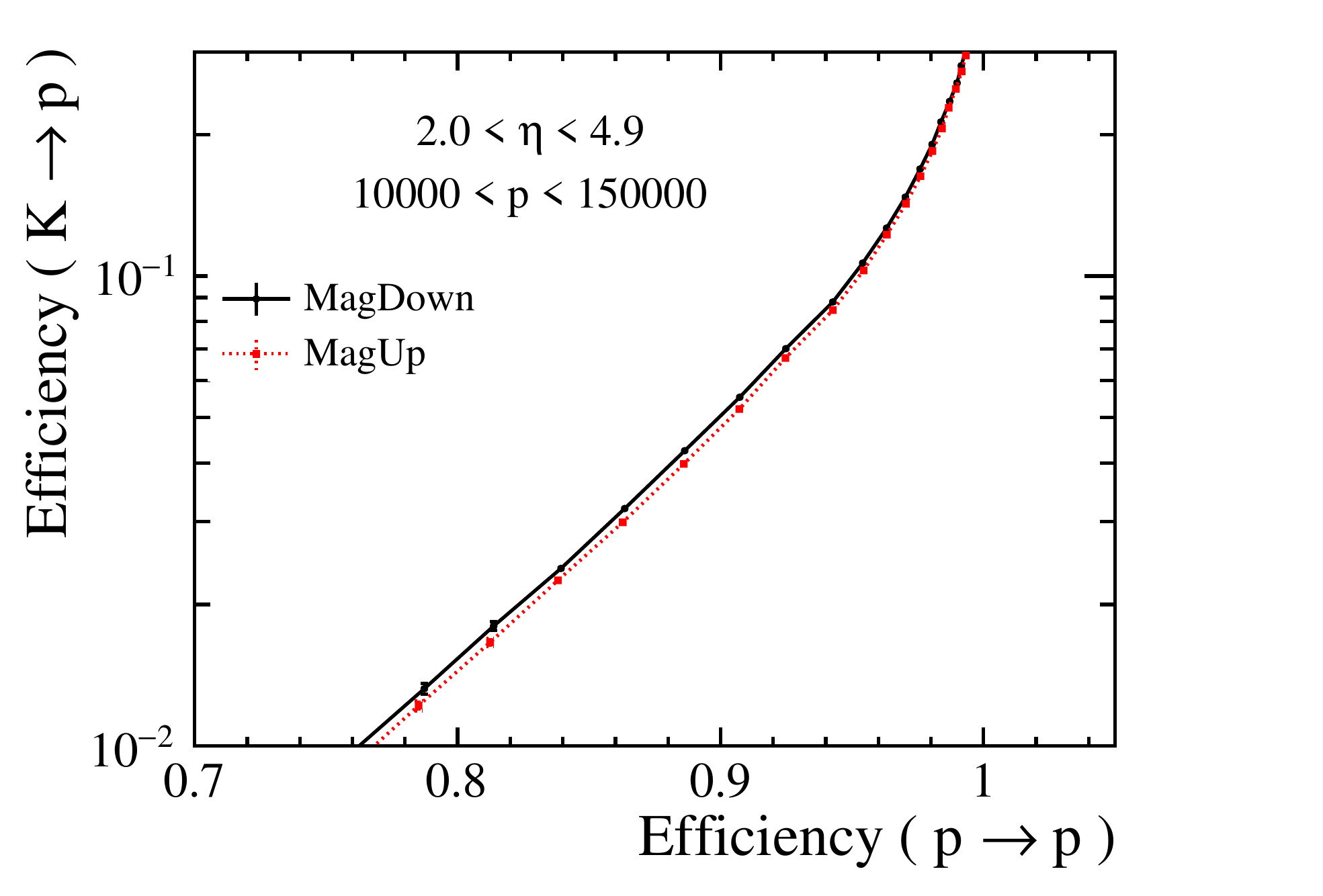}\put(-35,133){(f)}
    
    \vspace*{-0.5cm}
  \end{center}
  \caption{
    The efficiency of selecting kaons (a and b), protons (c, d, e and f), with the associated leakage from misidentifying pions (a, b, c d) and kaons (e and f) in data taken during 2018. (a, c and e)~The efficiency curves are shown for positively (black, solid) and negatively (red, dotted) charged particles. (b, d and f)~The efficiency curves are shown for \textit{Up} (red, dotted) and \textit{Down} (black, solid) magnetic field polarities. Uncertainties are statistical only, and are highly correlated between points on the same curve.}
  \label{fig:perfcharge}
\end{figure}

The performance curves for separating kaons and pions, protons and pions, and protons and kaons, split by charge, are shown in figure~\ref{fig:perfcharge} for the 2018 data-taking year and integrated over the two magnet polarities. Each datapoint corresponds to a $\Delta \text{LL}$ value, with tighter selections when moving towards the bottom-left corner of each plot. The agreement between the performance curves of positive and negative hadrons, of particular interest for \CP violation measurements, is excellent. The same behaviour is observed for all the data-taking years. 

The same performance curves, but split by magnet polarity, are shown in figure~\ref{fig:perfcharge} for the 2018 data-taking year and integrated over the hadron charges. Also here the agreement is excellent and the same behaviour is observed for all the data-taking years.

\begin{figure}[tb]
  \begin{center}
    \includegraphics[width=0.49\linewidth,page=1,trim=0 0 50 0,clip]{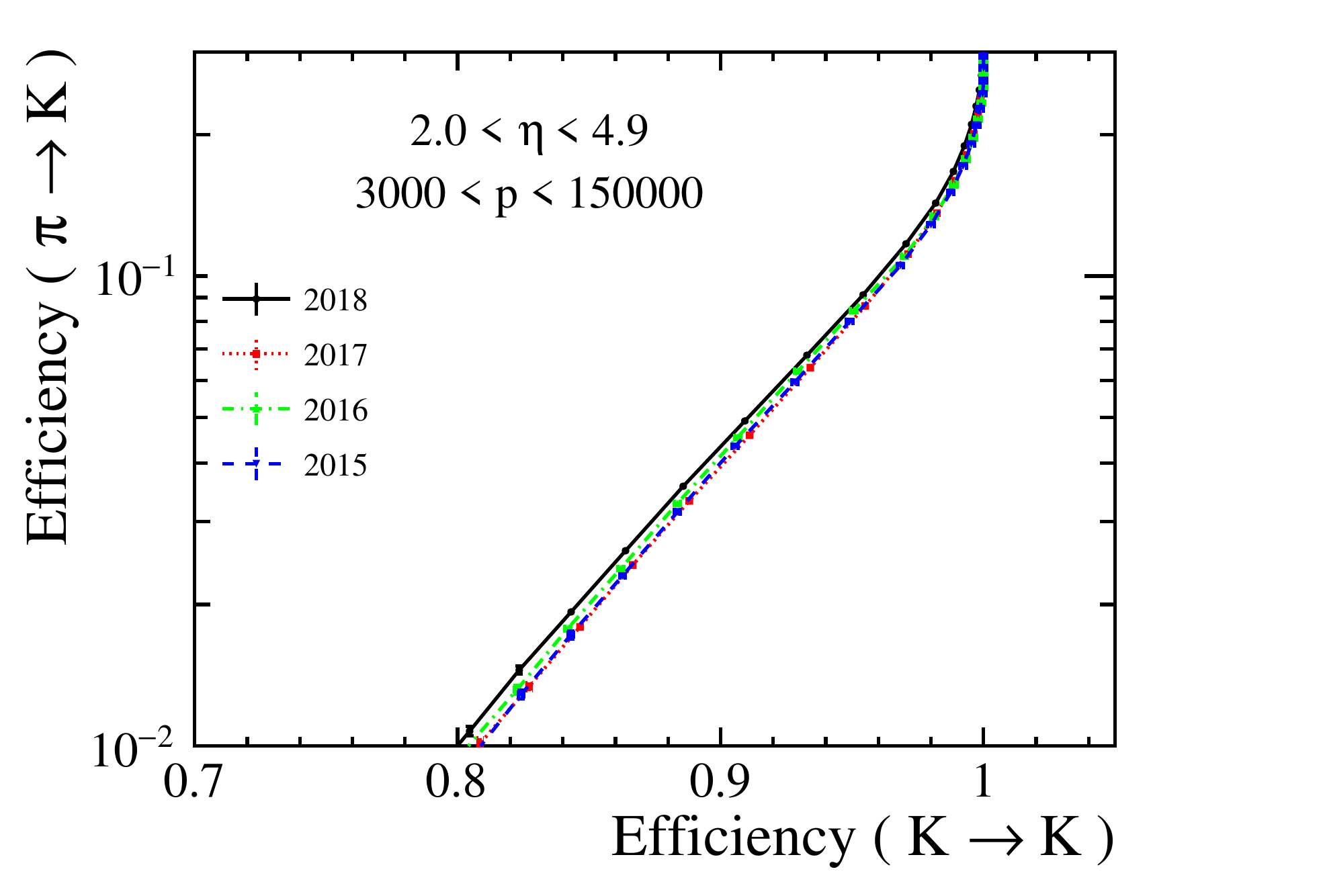}\put(-35,133){(a)}
    \includegraphics[width=0.49\linewidth,page=1,trim=0 0 50 0,clip]{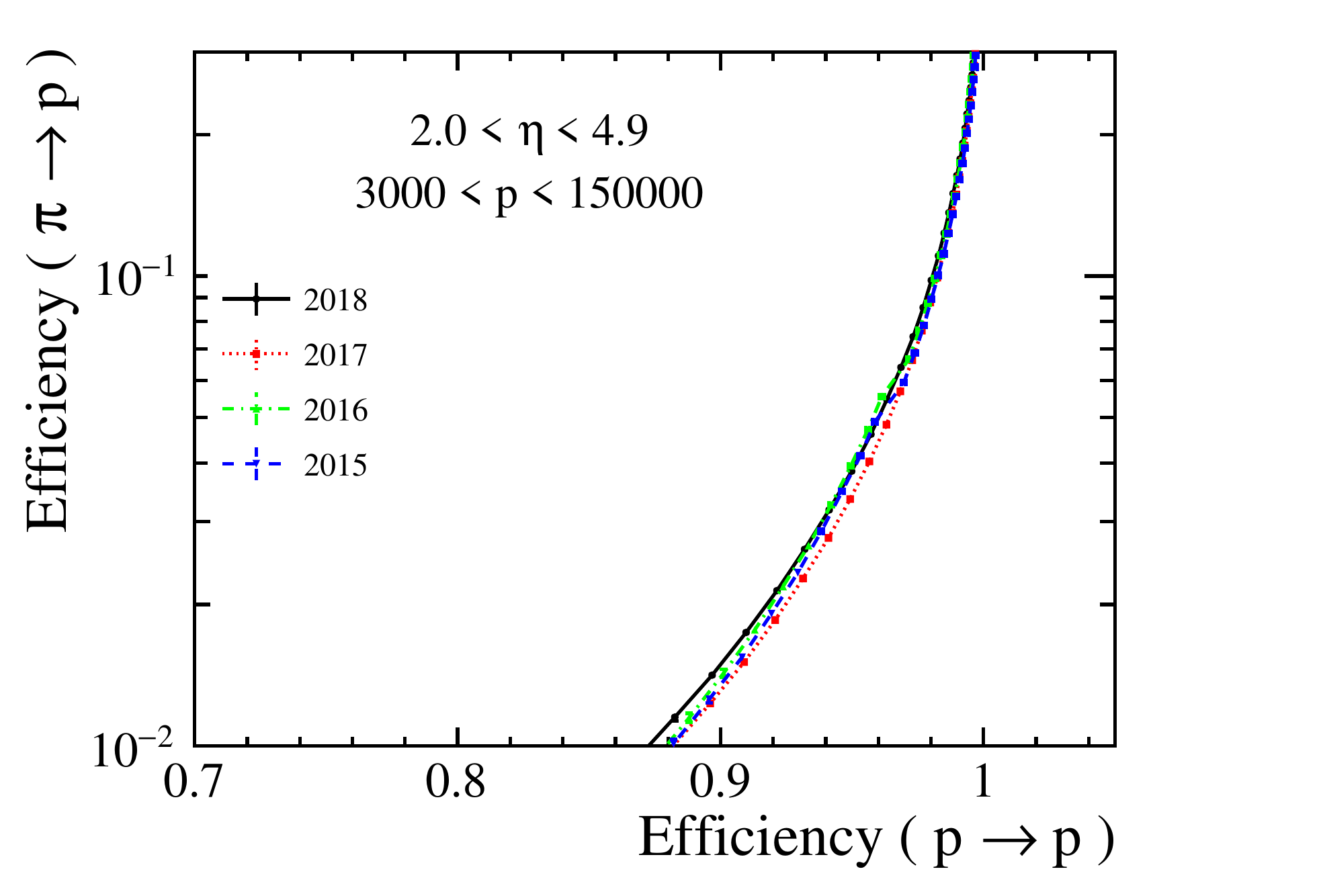}\put(-35,133){(b)}\\
    \includegraphics[width=0.49\linewidth,page=1,trim=0 0 50 0,clip]{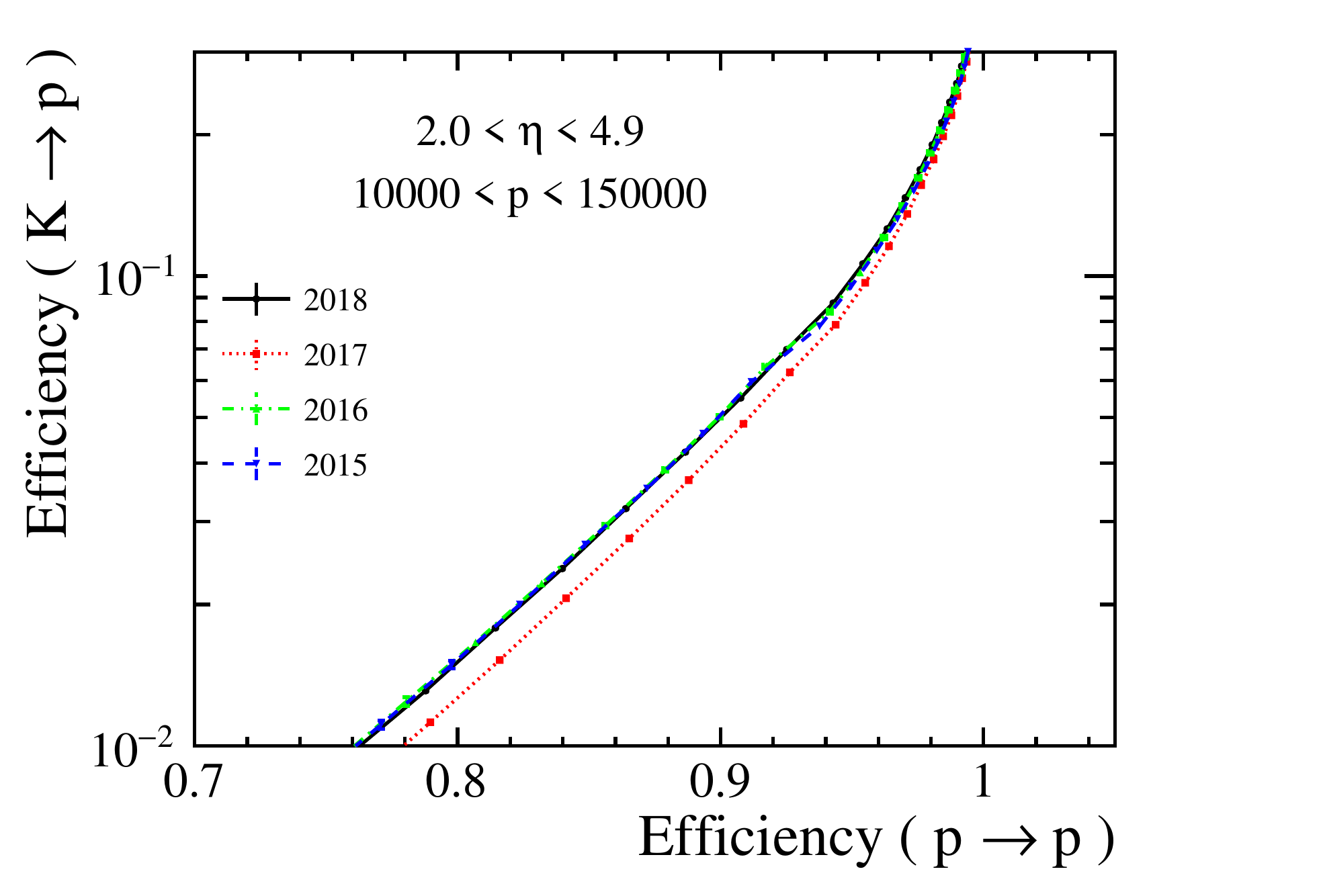}\put(-35,133){(c)}
    \vspace*{-0.5cm}
  \end{center}
  \caption{
    The efficiency of selecting kaons (a), protons (b and c), with the associate leakage from misidentifying pions (a and b) and kaons (c). The efficiency curves are shown for 2015 (blue, dashed), 2016 (green, dash-dotted), 2017 (red, dotted), 2018 (black solid). Uncertainties are statistical only, and are highly correlated between points on the same curve.}
  \label{fig:perfyears}
\end{figure}

The curves showing the stability of the PID performances across Run 2 are shown for the years 2015--2018 in figure~\ref{fig:perfyears}. Within the corresponding kinematics ranges, the probability to correctly identify each charged hadron is consistently above the 90\% for a 5\% misidentification probability. 

\begin{figure}[tb]
  \begin{center}
    \includegraphics[width=0.49\linewidth,page=1]{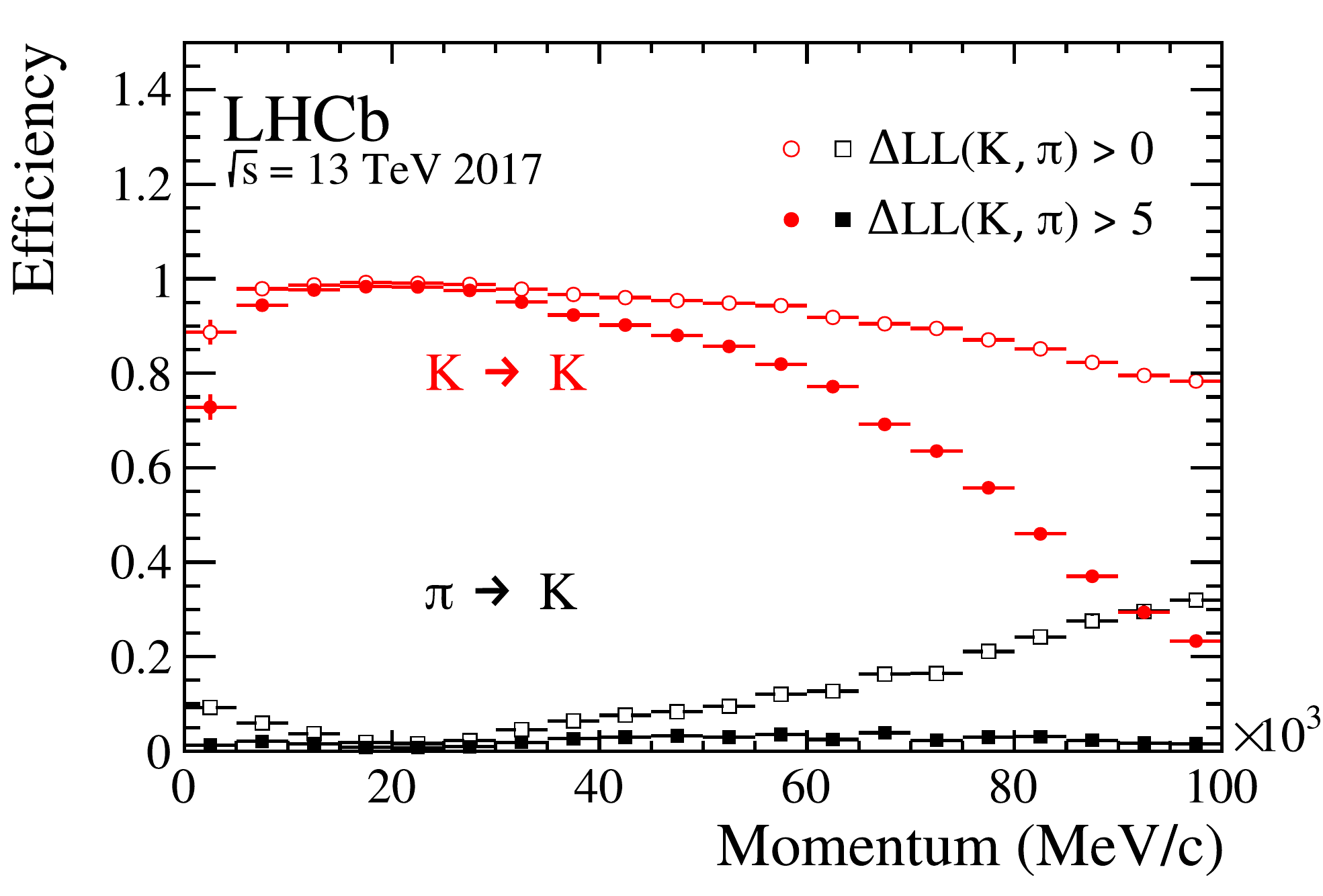}\put(-38,132){(a)}
    \includegraphics[width=0.49\linewidth,page=1]{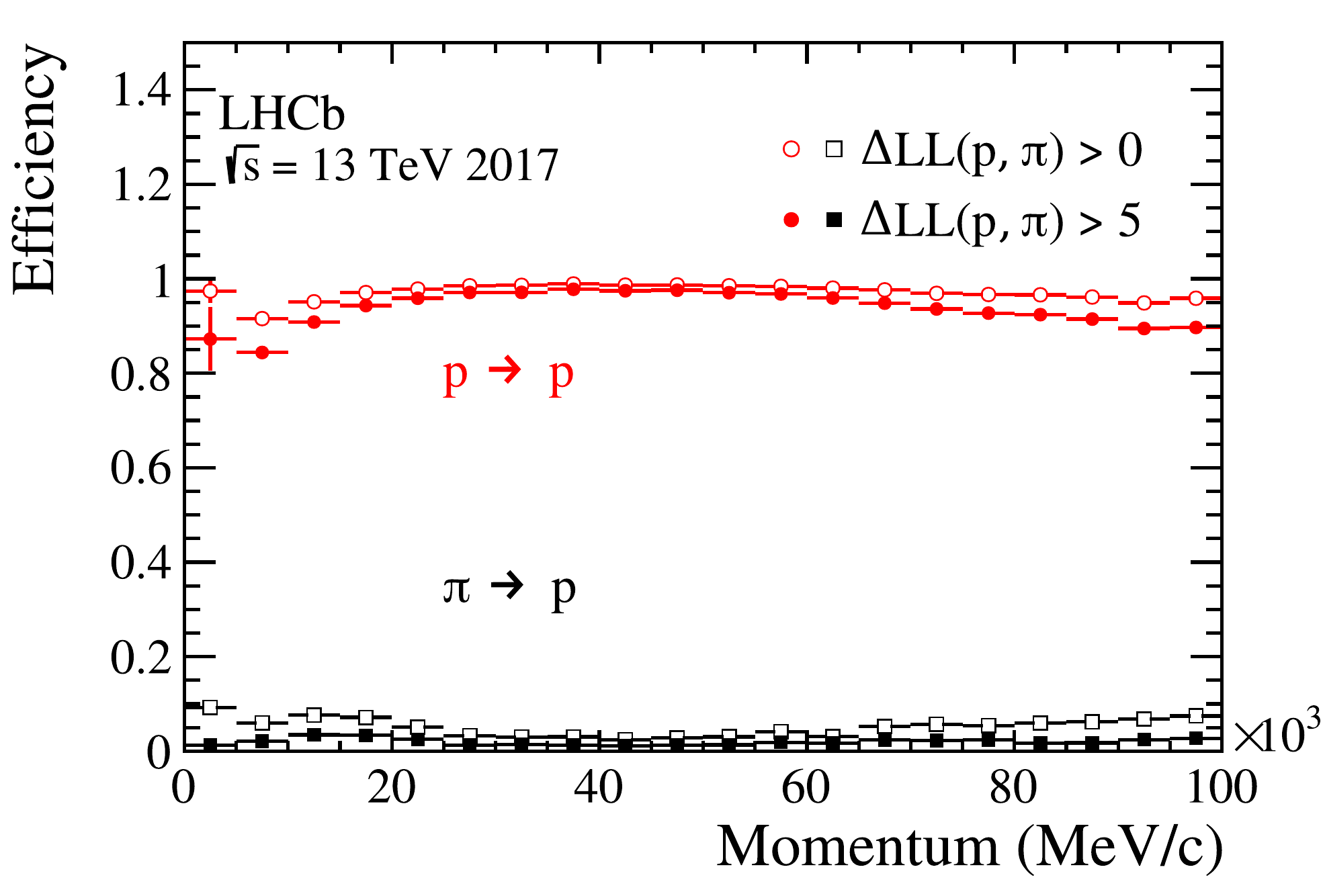}\put(-38,132){(b)}\\
    \includegraphics[width=0.49\linewidth,page=1]{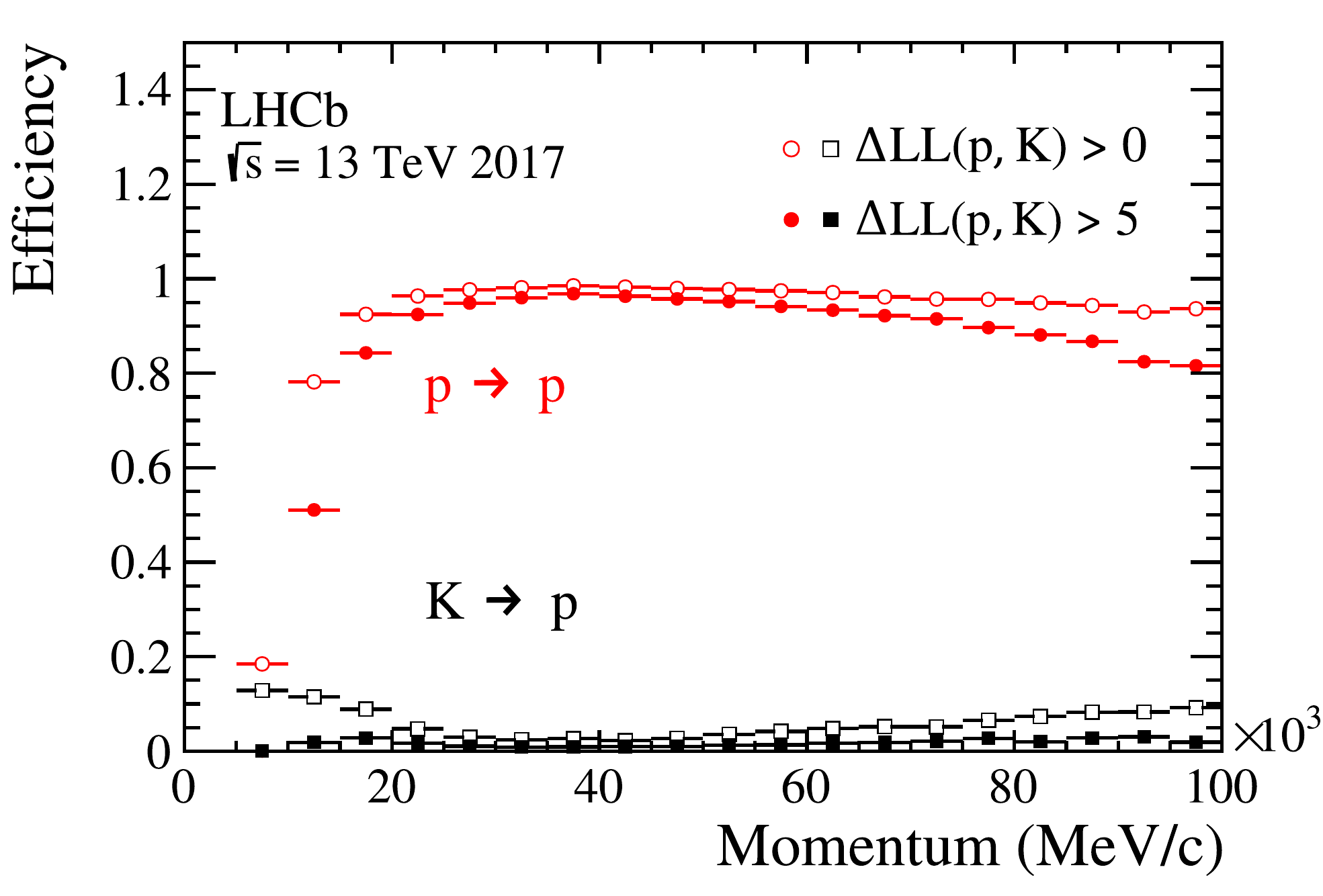}\put(-38,132){(c)}
    \vspace*{-0.5cm}
  \end{center}
  \caption{
    The efficiency of selecting kaons (a), protons (b and c), with the associate leakage from misidentifying pions (a and b) and kaons (c) as a function of momentum. Two selections are made, a loose selection (hollow circles) and a tight selection (solid circles).}
  \label{fig:perfyearsMomentum}
\end{figure}

The PID efficiency and the misidentification rate are also determined as a function of the track momentum, as shown in figure~\ref{fig:perfyearsMomentum}. In this case two selections are considered: a loose selection on the relevant $\Delta \text{LL}$ distribution, resulting in a high signal efficiency, and a tight selection on the same variable, resulting in a good background rejection.

\subsection{Physics impact}
\label{subsec:PhysicsImpact}

The performance curves shown in section \ref{subsec:PerformanceCurves} are highly illustrative, but do not tell the whole story of the performance of the hadronic particle ID within LHCb. The benefits of the application of PID on particular physics observables is dependent on the kinematics of the signal channels, and the nature of the background channels. In this section a more qualitative demonstration of the performance of the hadron PID is given through several examples of its importance for the LHCb physics programs.

\begin{figure}[tb]
  \begin{center}
    \includegraphics[width=0.49\linewidth,page=1]{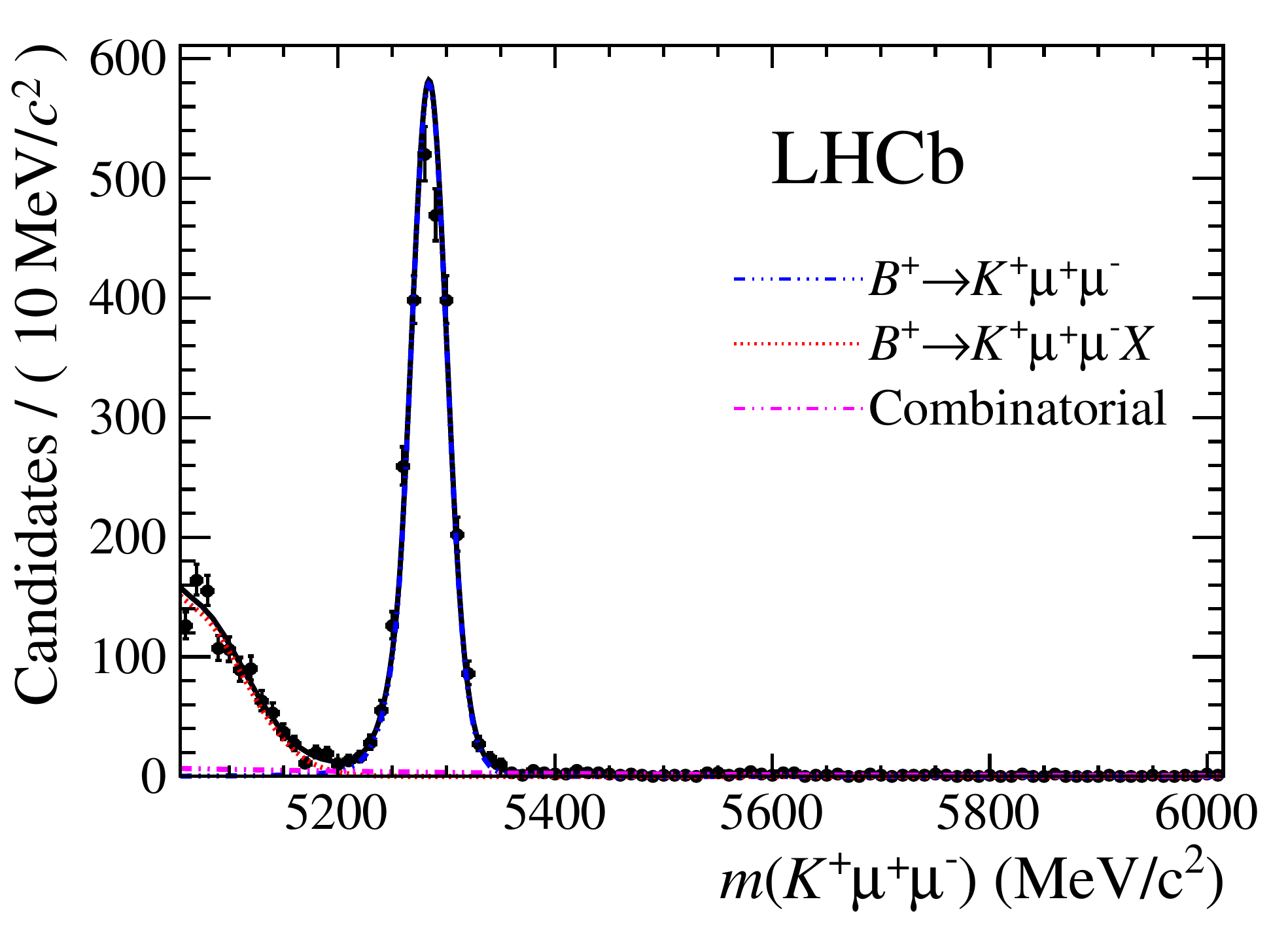}\put(-32,133){(a)}
    \includegraphics[width=0.49\linewidth,page=1]{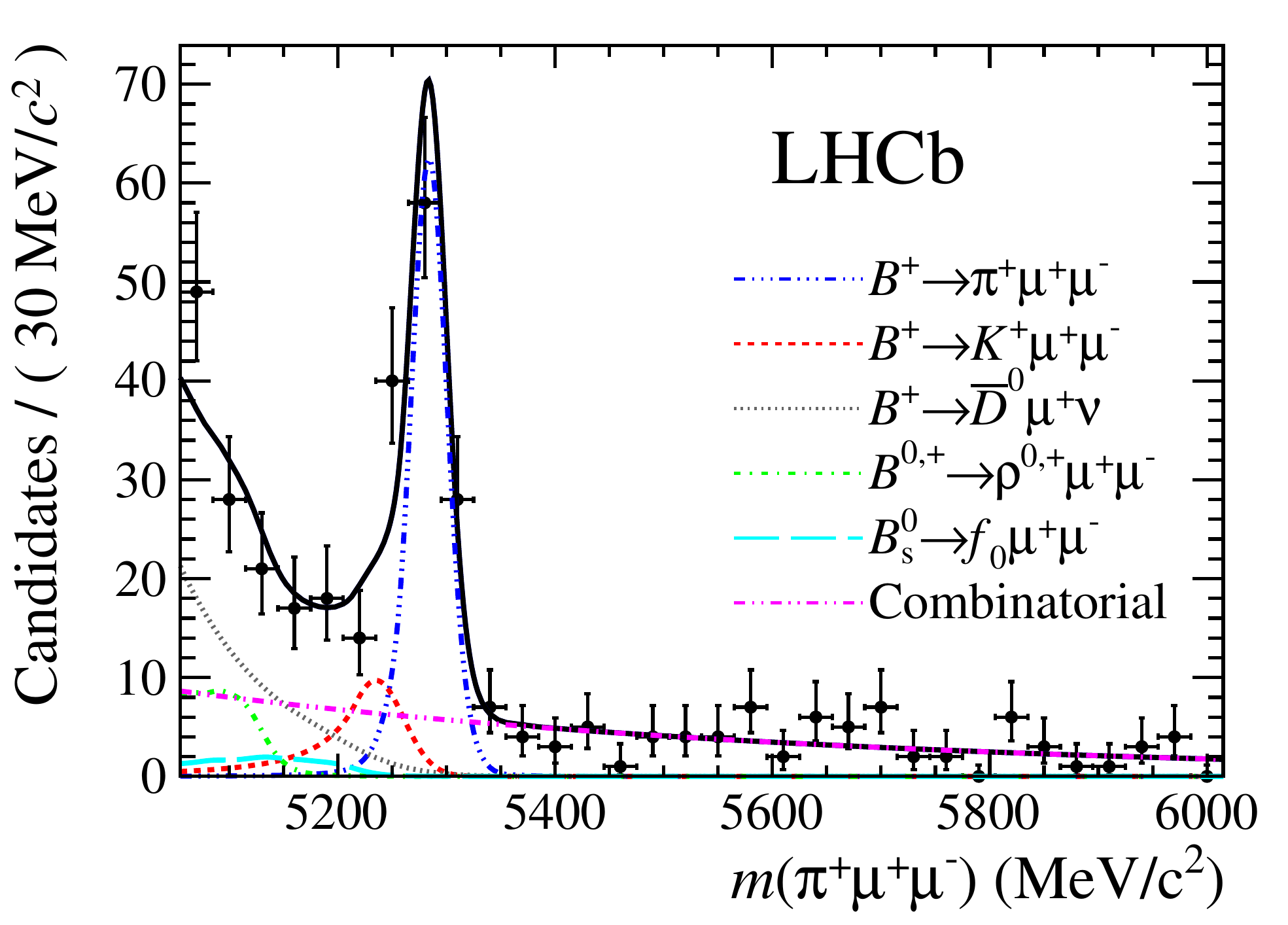}\put(-32,133){(b)}
    \vspace*{-0.5cm}
  \end{center}
  \caption{
    The effect of applying hadron PID selections to suppress the dominant kaon mode (a) to measure the CKM suppressed pion mode (b), in rare $B$ decays~\cite{LHCb-PAPER-2015-035}. Here 82~\% of pion events are retained, suppressing the kaons by a factor of 80 relative to the pions, leaving a clear peak from the pion mode.}
  \label{fig:perfexpimumu}
\end{figure}

An example of the hadron PID used in rare decay analyses is shown in figure~\ref{fig:perfexpimumu} \cite{LHCb-PAPER-2015-035}. Separate samples of $B^+\to K^+\mu^+\mu^-$ and $B^+\to \pi^+\mu^+\mu^-$ decay candidates are obtained, where the pion mode is suppressed by the ratio of CKM elements $\left|\frac{V_{td}}{V_{ts}}\right|^2$, approximately a factor of 25. The hadronic PID selection suppresses the kaon mode relative to the pion mode by a factor of 80, while retaining 82\,\% of pion candidates. A clean peak from the pion mode is seen, which would be impossible to obtain without effective hadron PID.

\begin{figure}[tb]
  \begin{center}
    \includegraphics[width=0.49\linewidth,page=1]{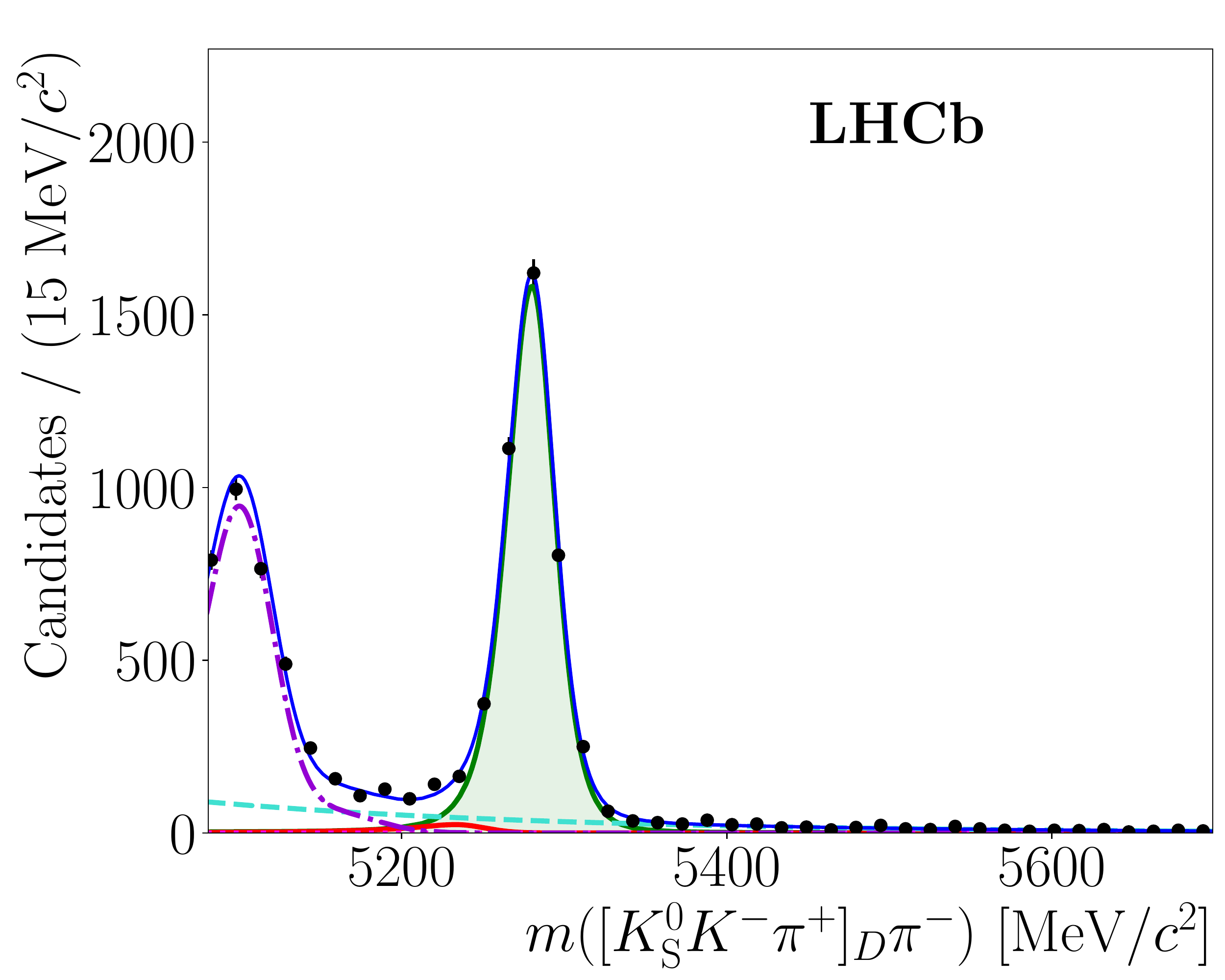}\put(-32,133){(a)}
    \includegraphics[width=0.49\linewidth,page=1]{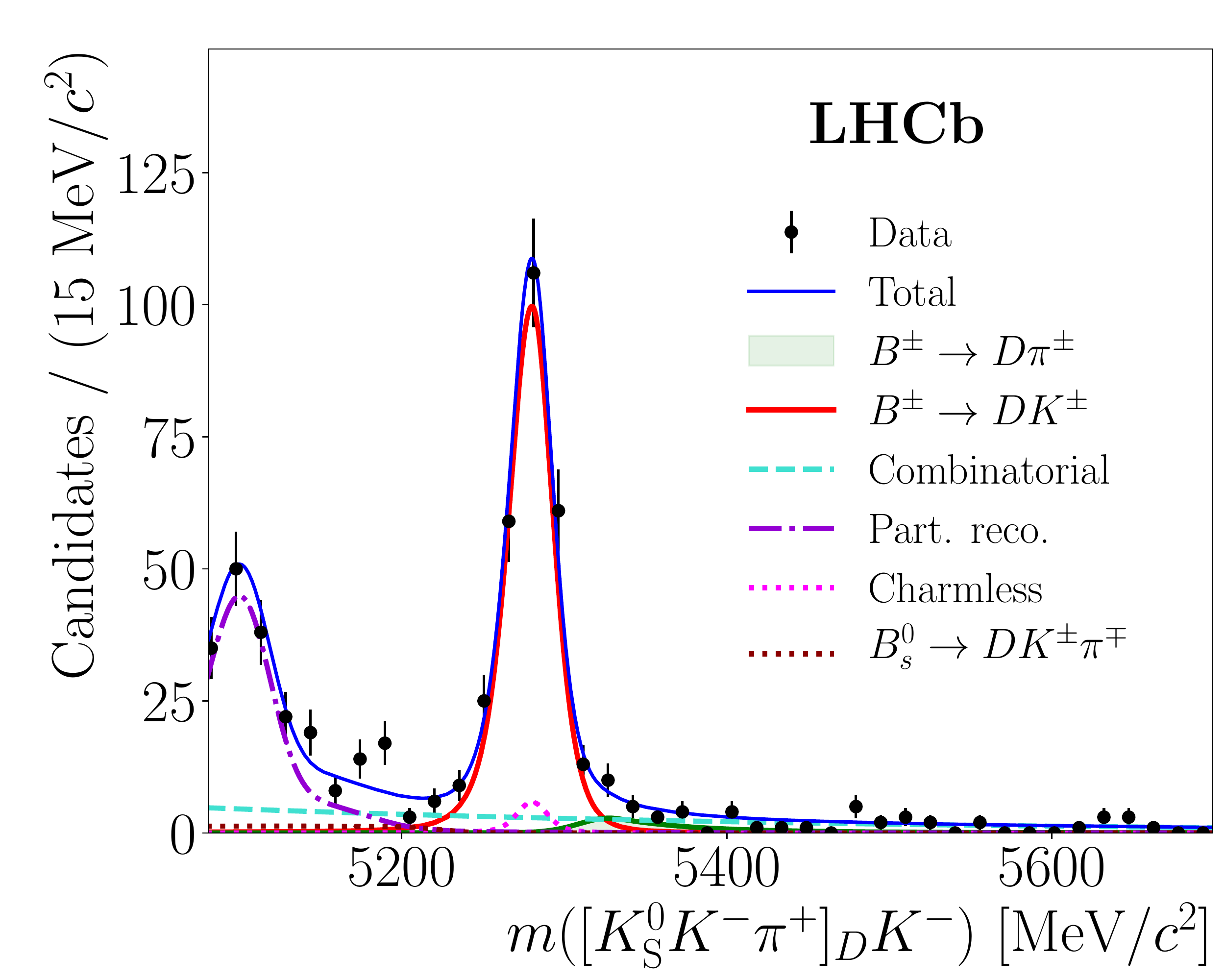}\put(-32,133){(b)}
    \vspace*{-0.5cm}
  \end{center}
  \caption{
    The effect of applying hadron PID selections to suppress the dominant pion mode (a) to measure the CKM suppressed kaon mode (b), in CP violation analyses~\cite{LHCb-PAPER-2019-044}.}
  \label{fig:perfexgamma}
\end{figure}

Measurements of the CKM angle $\gamma$ rely on distinguishing purely hadronic final states from $B$ decays. An example of isolating $B^\pm\to D^0 K^\pm$ candidates from the CKM favoured $B^\pm\to D^0 \pi^\pm$ is show in figure~\ref{fig:perfexgamma}.

\begin{figure}[tb]
  \begin{center}
    \includegraphics[width=0.49\linewidth,page=1]{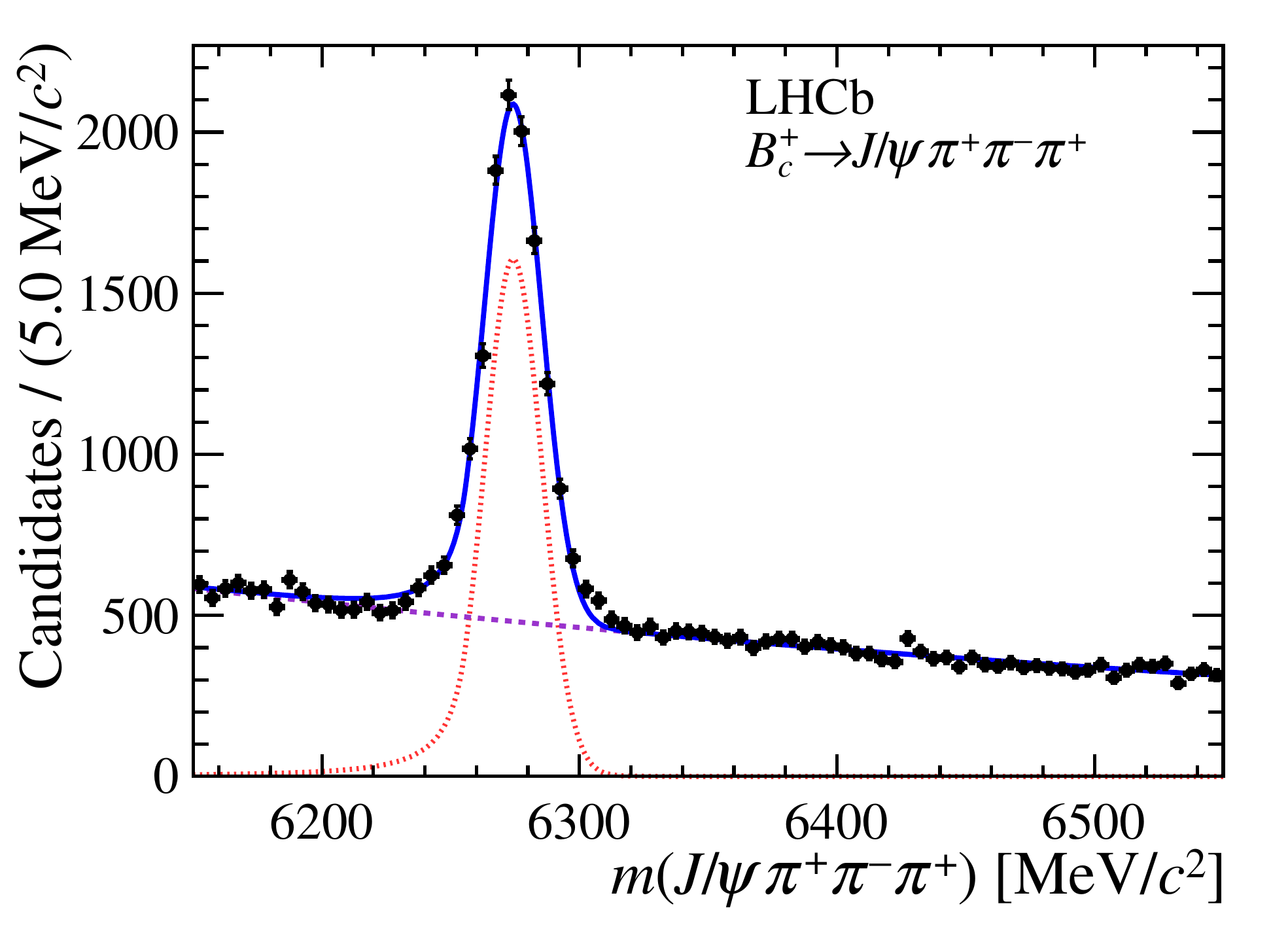}\put(-32,133){(a)}
    \includegraphics[width=0.49\linewidth,page=1]{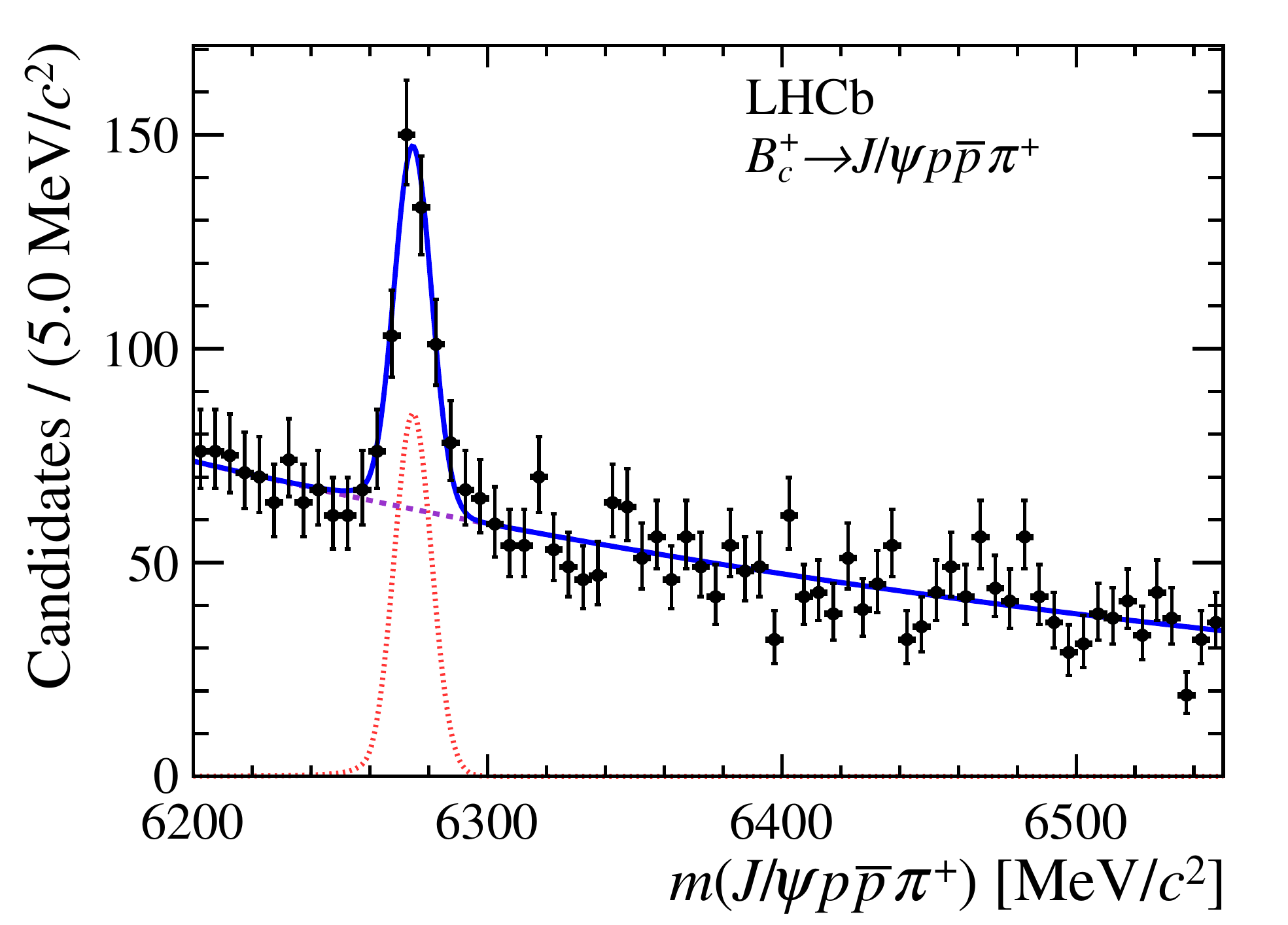}\put(-32,133){(b)}
    \vspace*{-0.5cm}
  \end{center}
  \caption{
    The effect of applying hadron PID selections to distinguish between \mbox{$B_c^+ \to J/\psi \pi^+ \pi^- \pi^+$} (a) and $B_c^+ \to J/\psi p \bar{p} \pi^+$  (b) decay modes~\cite{LHCb-PAPER-2020-003}.}
  \label{fig:perfBc}
\end{figure}

Hadron PID is also crucial in the spectroscopy program, where the capability to efficiently discriminate between long-lived hadrons is of fundamental importance in both precision measurements~\cite{LHCb-PAPER-2020-008} and new resonance searches~\cite{LHCb-PAPER-2019-014}. An application of the excellent proton identification performance is shown in figure~\ref{fig:perfBc}, where the $B_c^+$ meson invariant mass peak can be reconstructed in two decay modes only differing for the presence of a $p\overline{p}$ pair instead of a $\pi^+\pi^-$ pair in the final state. The $B_c^+ \to J/\psi p \overline{p} \pi^+$ decay mode is suppressed by a factor $\sim 15$ with respect to the $B_c^+ \to J/\psi \pi^+ \pi^- \pi^+$ decay channel~\cite{PDG2018}, but the proton selection efficiency and pion rejection provided by the RICH detectors allow clear identification of the $B_c^+ \to J/\psi p \overline{p} \pi^+$ signal. 

\clearpage
\section{Conclusions}
\label{sec:Conclusions}

The performance of the LHCb RICH detectors has been investigated. The utility of the calibration and alignment procedure has been demonstrated to give remarkable stability.

The angular resolution of the detectors has been measured to be $1.662 \pm 0.023\mrad$ and $0.621 \pm 0.012\mrad$ for \richone and \richtwo, respectively. The resolutions in \richone show small differences between magnetic field polarities due to limitations of mapping out distortions caused by the magnetic field, but otherwise the resolutions are excellent, and close to expectations from simulation.

The number of detected photons per track has been measured to be $30\pm 2$ and $18.5\pm 1.2$ for \richone and \richtwo, respectively. A slight decline in the photon yield is seen in both detectors as a function of time, as expected from the ageing of the HPD photocathodes.

The purity and efficiency of particle selections made using the information from both RICH detectors has been shown, yielding very pure samples with high efficiency. The performance has been shown to be very stable over time and independent of the track charge. The benefits of the excellent charged hadron identification have been shown in a number of physics channels demonstrating the access to a unique physics program from the LHC.

The RICH detectors are undergoing a substantial upgrade program, which will improve the performance for the much larger data sets that will be taken during Run 3 of the LHC. This will ensure the excellence of the LHCb charged hadron identification for years to come.

\section*{Acknowledgements}

\noindent We express our gratitude to our colleagues in the CERN
accelerator departments for the excellent performance of the LHC. We
thank the technical and administrative staff at the LHCb
institutes.
We acknowledge support from CERN and from the national agencies:
INFN (Italy);
MNiSW and NCN (Poland);
MEN/IFA (Romania);
STFC (United Kingdom);
NSF (USA).
We acknowledge the computing resources that are provided by CERN, IN2P3
(France), KIT and DESY (Germany), INFN (Italy), SURF (Netherlands),
PIC (Spain), GridPP (United Kingdom), CSCS (Switzerland), IFIN-HH (Romania), CBPF (Brazil),
PL-GRID (Poland) and OSC (USA).
We are indebted to the communities behind the multiple open-source
software packages on which we depend.

\bibliographystyle{unsrt}
\bibliography{main,standard,LHCb-PAPER,LHCb-CONF,LHCb-DP,LHCb-TDR}

\end{document}